\documentclass[aps,amsmath,amssymb,prd,showpacs,floatfix,preprint,superscriptaddress,nofootinbib,12pt]{article}
\usepackage{jheppub}
\usepackage{bm}
\usepackage{ulem}

\usepackage{pstricks}
\usepackage{color}

\newcommand{\p}{\partial}
\newcommand{\Tr}{\text{Tr}}

\newcommand{\J}{\mathtt{j}}

\allowdisplaybreaks[3]

\begin{document}

\title{Solvable time-like cosets and holography beyond $AdS$}

\author[a]{Soumangsu Chakraborty}
\author[b]{Mikhail Goykhman}
\affiliation[a]{Institut de Physique Th\'eorique, Universit\'e Paris-Saclay, CNRS, CEA\\
Orme des Merisiers, 91191 Gif-sur-Yvette, France}
\affiliation[b]{William I. Fine Theoretical Physics Institute, University of Minnesota, Minneapolis, MN
55455, USA}
\emailAdd{soumangsuchakraborty@gmail.com}
\emailAdd{goykhman@umn.edu}

\abstract{
We build a novel time-like coset sigma-model describing type-II superstring
theory in a charged rotating black-brane background that interpolates between a locally $AdS_3$ in the IR and
a  linear-dilaton geometry in the UV.
This allows one to perform a systematic study of holography in non-AdS backgrounds which are smoothly connected to $AdS_3$.
We construct 
massless closed string states vertex operators in the NS-NS sector,
calculate the corresponding two-point correlation functions, and discuss
holographic interpretation of our results from 4+1 dimensional boundary field theory point of view. 
Compactifying the theory on $\mathbb{T}^4$, we show that the spectrum of a single long string with unit winding agrees with the spectrum of a CFT$_2$ deformed by $T\bar{T}$.
We also calculate  correlation functions of  operators of the dual
1+1 dimensional non-conformal boundary field theory using world-sheet techniques.
}

\maketitle

\section{Introduction}

String theory on group manifolds and coset spaces supplies a rich
class of models describing strings propagating on a non-trivial curved 
background, that are  amenable to a formulation in terms of a two-dimensional
world-sheet theory, exact in string length.
This powerful feature follows from the potent combination of (super-)conformal
and affine Kac-Moody symmetries enjoyed by these theories.
The world-sheet action that respects these symmetries and describes string theory on group manifolds
is  given by the celebrated Wess-Zumino-Witten (WZW)
model, while its generalization to group cosets is described by the gauged WZW (gWZW) model \cite{Witten:1983ar,Polyakov:1983tt,Knizhnik:1984nr}.

A particular class of the (g)WZW models, solvable in the limit of small string coupling $g_s$,
yet exact in the string length $\ell_s$,  is given by the world-sheet models
with the target space based on the $SL(2,\mathbb{R})$ and $SU(2)$ group manifolds.
An enormous technical simplification due to such a choice of the target space
group manifold is afforded by the available properties of
the $SL(2,\mathbb{R})$ and $SU(2)$ primary operators, in particular, the known structure
of their correlation functions in the exact world-sheet theory. This knowledge assists one immensely
in a calculation of various observables exact in the WZW level $k$ that simultaneously
translates into the stringy, beyond gravity, accessibility of the large curvature regime: a notoriously coveted
goal of practitioners in the field.

One of the fruitful applications of the $SL(2,\mathbb{R})$ and $SU(2)$ based (g)WZW models
is found in the field  of holographic correspondence. Original examples are given by the
holographic (bulk) description of a two-dimensional conformal
field theory (CFT) \cite{Maldacena:2001km,Maldacena:2000kv,Maldacena:2000hw,Giveon:2001up,Kutasov:1999xu,Giveon:1998ns}, and holographic description of Little String Theory (LST) \cite{Seiberg:1997zk,Giveon:1999zm,Aharony:1998ub,Itzhaki:1998dd,Boonstra:1998mp,Witten:1991yr,Dijkgraaf:1991ba,Aharony:1998tt,Minwalla:1999xi,Aharony:1999dw,Narayan:2001dr,DeBoer:2003dd,Israel:2003ry,Aharony:2004xn}.
The former is given by the superconformal WZW model on $AdS_3\times\mathbb{S}^3\times \mathbb{T}^4$
(where $AdS_3\sim SL(2,\mathbb{R})$, and $\mathbb{S}^3\sim SU(2)$), 
while the latter is described by the superconformal gWZW model on
$\frac{SL(2,\mathbb{R})}{U(1)}\times SU(2)\times \mathbb{R}^5$. In the case when microscopic description of coset models is provided by a stack of $k$ coincident NS5 branes wrapping $\mathbb{R}^5$ or $\mathbb{T}^4\times \mathbb{S}^1$ and $p$ fundamental strings (F1) along $\mathbb{R}^1$ or $\mathbb{S}^1$, the $AdS_3$ background found in the near-horizon limit of the F1 strings is described by the $SL(2,\mathbb{R})$ WZW model at level $k$, while the coset $SL(2,\mathbb{R})/U(1)$ describes the linear-dilaton background, sourced by a stack of $k$ NS5-branes in its near-horizon limit.  The corresponding dual field theory interpolates between a CFT$_2$, obtained by taking the decoupling limit of the F1-strings, and the world-volume theory of the NS5-branes, namely LST, obtained by taking the decoupling limit of  the NS5 branes. The NS5-branes source the NS flux, equal to $k$,
through the $\mathbb{S}^3\sim SU(2)$ sphere surrounding the NS5-branes in the ten-dimensional
space-time. The radius of both $AdS_3$ and $\mathbb{S}^3$ behaves as $R=\sqrt{k}\,\ell_s$,
which defines the small-curvature semi-classical limit (amenable to the supergravity approximation)
at $k\rightarrow \infty$.   However, an exact solvability of the $SL(2,\mathbb{R})\times SU(2)$ WZW model
allows one to carry out calculations exactly in $k$, furnishing  an exact world-sheet description.

One can construct a more complicated class of cosets describing dynamics of NS5-F1 system.
Coset systems, described by the gWZW model, require one to impose anomaly-cancellation
conditions. These can be formulated as time-like  or null constraints (for cosets that include the time-like direction in  target space) on the gauge parameters.
Null cosets have been extensively studied in \cite{Martinec:2017ztd,Martinec:2018nco,Martinec:2020gkv,Bufalini:2021ndn,Bufalini:2022wyp}
and shown to describe micro-state geometries with  world-sheet control.
On the other hand, time-like cosets help us understand dynamics of black holes,
scattering in a black hole background,
and resolution of cosmological singularities \cite{Witten:1991yr,Elitzur:2002rt,Elitzur:2002vw,Giveon:2003ge,Goykhman:2013oja,Chakraborty:2020yka}.

In  \cite{Parnachev:2005hh,Goykhman:2013oja}  collective excitations of the LST were studied
holographically, by analyzing the pole structure of the stress-energy tensor two-point function,
described in terms of the graviton  scattering amplitudes in the bulk.\footnote{See also \cite{Polchinski:2012nh}
for related work.}
The latter were calculated exactly, in terms of the two-point correlation functions
of vertex operators on the world-sheet, describing massless closed string states
in the Neveu-Schwarz-Neveu-Schwarz (NS-NS) sector of the supergravity multiplet.
In particular, the 
diffusion mode was located, and its existence was independently verified by the supergravity
analysis. Additionally, in \cite{Goykhman:2013oja} a finite-density version of this model
was considered, by constructing the dual bulk configuration with a non-trivial gauge field profile.
Such a background was explicitly demonstrated in  \cite{Chakraborty:2020yka}
to appear in the configuration involving the near-horizon limit of a stack of NS5-branes with a large number of fundamental strings (F1).\\

In this paper, our goal is to further broaden our understanding of the asymptotically linear dilaton
backgrounds, construct spectrum of string excitations on top of these backgrounds,
calculate various correlation functions of vertex operators corresponding to massless string states,
 and comment on application of these results to holography. To this end,
we consider type-II superstring theory on  $\frac{SL(2,\mathbb{R})\times SU(2)\times U(1)_x}{U(1)}\times\mathbb{T}^4$
ten-dimensional coset space-time, and perform time-like gauging of the $U(1)$ sub-group in  denominator.
We formulate exact world-sheet action of the superconformal gWZW model on this coset,
describing type-II superstring theory in the corresponding background.
We then proceed to deriving the exact in $k$ gravity background,
consisting of non-trivial configurations of the metric, dilaton, and $B$-field.
The obtained background asymptotes to the two-dimensional linear dilaton
background times $\mathbb{S}^3\times\mathbb{S}^1\times\mathbb{T}^4$.
At the same time, the obtained geometry possesses an event horizon 
with non-vanishing angular velocities along two of the $\mathbb{S}^3$ and $\mathbb{S}^1$ directions. 

Compactifying the $ U(1)_x\sim \mathbb{S}^1$ direction and
performing the Kaluza-Klein (KK) reduction, we obtain the nine-dimensional
space-time with two non-trivial $U(1)$ gauge field profiles, non-vanishing angular velocities along two of the $\mathbb{S}^3$
directions, and an event horizon.
Such a configuration describes a holographically dual theory at finite temperature, characterized by finite values of
chemical potentials and charge densities corresponding to two $U(1)$ charges.
We provide a complete thermodynamic description of the ten- and nine-dimensional
backgrounds, by calculating the corresponding temperature, entropy, angular momenta, energy, grand potential,
and charges. We demonstrate that for the particular choice of the coset gauging parameters
our results reproduce the known results in the literature, while largely extending
those results to a more general case of a rotating charged asymptotically linear-dilaton background with 
a black brane event horizon.

We subsequently discuss collective excitations of the boundary theory by inspecting the structure
of two-point correlation functions of the stress-energy tensor and the currents corresponding
to two conserved $U(1)$ charges.
These can be determined holographically in terms of the two-point functions of the vertex
operators of the world-sheet theory, describing massless excitations of the 
NS-NS sector of the type-II supergravity background, involving the metric, dilaton, $B$-field,
and gauge fields. We construct the corresponding  vertex operators
by performing the coset BRST quantization of the world-sheet theory,
ensuring the $U(1)$ symmetry at the quantum level. In particular, we determine
the tachyon vertex operator, and study its asymptotic behavior. 
We then construct the most general massless NS-NS physical states,
and calculate their two-point correlation functions.

The string theory model that we study in this paper, at zero temperature,  interpolates between local $AdS_3\times \mathbb{S}^3\times \mathbb{T}^4$ in the IR to a linear-dilaton times $\mathbb{S}^3\times \mathbb{S}^1\times \mathbb{T}^4$ in the UV. The dual boundary field theory interpolates between a certain vacua of LST in the UV to a CFT$_2$ in the IR. This allows us to systematically study holography in asymptotically non-$AdS$ space-time. Furthermore, it has been argued that such models are intimately related to solvable models in string theory that go by the name of `single-trace' $T\bar{T}$ deformation \cite{Giveon:2017nie,Chakraborty:2018vja,Chakraborty:2019mdf,Chakraborty:2020swe,Chakraborty:2020yka,Chakraborty:2020cgo}. We show that the spectrum of a single winding one long string in the interpolating coset background under investigation, exactly matches with the spectrum obtained in the case of $T\bar{T}$ deformed CFT$_2$ \cite{Smirnov:2016lqw,Cavaglia:2016oda} for the particular sign of the irrelevant coupling that gives rise to a real-valued spectrum. For a generic winding $\mathrm{w}>1$ sector, the spectrum agrees with the $\mathbb{Z}_\mathrm{w}$ twisted sector of a symmetric orbifold theory where the block of the symmetric orbifold is obtained by $T\bar{T}$ deformation of a CFT$_2$.  Using world-sheet string theory techniques we compute the two and three-point functions of operators of the boundary theory in momentum space. As one moves away from the IR conformal fixed point, the operators develop a momentum dependent ambiguity in the normalization of the operators, although it is not clear to us how to fix this normalization issue. The `dimensions' \footnote{Strictly speaking, as one moves away from the conformal fixed point, there is no notion of dimension of operators in the CFT sense. What we mean by `dimension' is in the sense discussed in  \cite{Guica:2021pzy,Guica:2021fkv}. } of the operators at a generic point on the renormalization group flow develop a momentum dependence which resonates with the idea of momentum dependent dimensions discussed in the context of non-local CFTs \cite{Guica:2021pzy,Guica:2021fkv}.

The rest of this paper is organized as follows.
In section~\ref{sec:review ads3 and s3} we review basic setup of
superstring theory on $AdS_3\times \mathbb{S}^3\times \mathbb{T}^4$
and introduce our conventions. In section~\ref{sec:coset construction}
we construct the main ten-dimensional coset target space-time background that we are interested in in this paper.
We develop thermodynamics of this background in section~\ref{sec:thermodynamics}.
In section~\ref{sec:two charge} we perform the Kaluza-Klein reduction of the $U(1)_x$
circle and construct the nine-dimensional background, characterized by two
non-trivial $U(1)$ gauge field profiles.
We construct the spectrum of physical string excitation states on top of the background fields
configuration in section~\ref{sec:BRST}, focusing specifically on the NS-NS sector.
In section~\ref{sec:holography} we explore the holographic interpretation of our results,
by considering two-point correlation functions of the dual boundary theory calculated
in terms of the world-sheet vertex operators in the bulk.
We discuss our results and comment on future directions in section~\ref{sec:discussion}.
Appendix~\ref{app:conventions} is dedicated to review of relevant background
concerning (superconformal) gauged WZW models.
Additional ancillary material is provided in appendices~\ref{sec:GHY in string frame},~\ref{sec: sl2r u1 over u1 thermodynamics}.

\section{Review of superstrings on $AdS_3\times \mathbb{S}^3\times \mathbb{T}^4$}
\label{sec:review ads3 and s3}

In this section, we will provide a brief summary
of the type-II superstring theory on the $AdS_3\times \mathbb{S}^3\times \mathbb{T}^4$
background, focusing on the low-lying states in the Neveu-Schwarz-Neveu-Schwarz (NS-NS) sector. In particular, we will
set up conventions related to the $SL(2,\mathbb{R})$ and $SU(2)$
Wess-Zumino-Witten (WZW) models, that we will be following for the rest of the paper.
We will also review the affine Kac-Moody symmetry algebra of the target space-time, discuss
its representation space, normalizability of the  string states, and the corresponding vertex operators. 
A complementary condensed glossary of our notations can be found in appendix~\ref{app:conventions}.
For a detailed discussion of string theory on $AdS_3$ we refer the reader to 
\cite{Maldacena:2001km,Maldacena:2000kv,Maldacena:2000hw,Giveon:2001up,Kutasov:1999xu,Giveon:1998ns}.

The world-sheet superstring theory on $AdS_3\times \mathbb{S}^3\times \mathbb{T}^4$
is given by the superconformal
 WZW model on the $SL(2,\mathbb{R})\times SU(2)\times U(1)^4$ group manifold.
This world-sheet sigma-model is constrained by the left- and right-moving $SL(2,\mathbb{R})\times SU(2)$
supersymmetric affine Kac-Moody algebra, at supersymmetric level $k$. In the remainder of this section,
we will discuss the affine symmetry algebra and the vertex operators of the
$SL(2,\mathbb{R})\sim AdS_3$ and $SU(2)\sim\mathbb{S}^3$ subspaces separately.

\subsection{Symmetries and vertex operators of $AdS_3$}

In this subsection, we will discuss the $SL(2,\mathbb{R})$ superconformal WZW model,
describing the world-sheet string theory on $AdS_3$.
The classical supersymmetric WZW sigma-model consists of the bosonic
sector, given by the $SL(2,\mathbb{R})_{k+2}$ WZW model
at the level $k+2$,  and  three pairs of free (anti)-holomorphic two-dimensional
Majorana-Weyl fermions $\psi^a$, $\tilde \psi^a$ $a=1,2,3$. Bosonic degrees of freedom comprise the
 (anti-)holomorphic Kac-Moody
currents $j_a$, $\tilde j_a$, $a=1,2,3$ at the  level $k+2$.
The fermionic sector can be bosonized, and
described in terms of the $SL(2,\mathbb{R})_{-2}$  WZW model at the level $-2$.
Combining the bosonic
and the fermionic degrees of freedom
we obtain
the total bosonic Kac-Moody currents forming the algebra at level $k$.

The exact (in the sense of world-sheet path integral) effective action for the bosonic field $g\in SL(2,\mathbb{R})$ is
given by the $SL(2,\mathbb{R})_{k}$
WZW  action, 
\begin{equation}
\label{slwzw}
S_{sl}[g] = k\,\mathtt{S}[g]\,,
\end{equation}
where the $k=1$ action $\mathtt{S}[g]$ is given by (\ref{WZWgen}).
The action \eqref{slwzw} is invariant under the $SL(2,\mathbb{R})_L\times SL(2,\mathbb{R})_R$ affine symmetry algebra.
The global part of this
 symmetry algebra is generated by the Kac-Moody currents $j^a$ and $\tilde{j}^a$, $a=1,2,3$, given by
\begin{equation}\label{slgcurrents}
j^a=\frac{k+2}{2} {\rm Tr} \left(\partial g g^{-1}\, \tau^a \right), \ \ \ \ \
\tilde{j}^a=\frac{k+2}{2} {\rm Tr} \left(g^{-1}\bar{\partial} g\,  \tau^a \right).
\end{equation}
where
\begin{equation}
\tau ^{1,2} = \sigma^{1,2}\,,\qquad  \tau^{3} = i\,\sigma^3\,,
\end{equation}
and $\sigma^a$ are Pauli matrices given by
\begin{equation}\label{pauli}
\sigma^1=\begin{pmatrix}
0& 1\\ 1&0
\end{pmatrix}, \ \ \ \ \sigma^2=\begin{pmatrix}
0& -i\\ i&0
\end{pmatrix}, \ \ \ \ \sigma^3=\begin{pmatrix}
1& 0\\ 0&-1
\end{pmatrix}\,.
\end{equation}
At the same time, the total Kac-Moody currents, incorporating 
contributions both from the bosonic and the fermionic degrees of freedom,
form the (anti-)holomorphic affine algebra $\J^a$, $\tilde{\J}^a$, $a=1,2,3$ at level $k$,
with the currents defined as\footnote{See eq. (\ref{tt j def}) in appendix~\ref{app:effective wzw action} for details.}
\begin{equation}
\mathtt{j}^a = j^a - \frac{i}{k} f^a_{\;\; bc} \,\psi^b\psi^c\,,\qquad
\tilde{\mathtt{j}}^a = \tilde j^a - \frac{i}{k} f^a_{\;\; bc} \,\tilde\psi^b\tilde\psi^c\,.
\end{equation}

Using the Sugawara construction,  we can write the (anti-) holomorphic
components of the stress-energy tensor of the world-sheet sigma model.
Focusing on the contributions from the bosonic degrees of freedom, we obtain\footnote{The sign $\supset$
means that the terms to the right of this sign constitute some of the contributions to the expression to the left of this sign.
In the case of (\ref{slst}), we focus on the contributions to the stress-energy
tensor from the bosonic d.o.f., and omit writing explicitly contributions due to the fermions.}
\begin{equation}\label{slst}
\begin{aligned}
T_{sl}\supset\frac{1}{k}\, j^a\,j^a\,,\qquad
\tilde{T}_{sl}\supset\frac{1}{k}\,\tilde{j}^a\,\tilde{j}^a\,.
\end{aligned}
\end{equation}
The quadratic Casimir of $SL(2,\mathbb{R})_k$ is given by
\begin{eqnarray}\label{slcasimir}
c_2=(j^3)^2+\frac{1}{2}\{j^+,j^-\},
\end{eqnarray}
where we defined
\begin{eqnarray}\label{jpm}
j^{\pm}=j^1\pm i j^2.
\end{eqnarray}
Representations of $SL(2,\mathbb{R})$ are labeled by
the number $j$ related to the quadratic Casimir as $c_2=-j(j+1)$.

The spectrum of string states on $AdS_3$ can be constructed 
in terms of the unitary representations of $SL(2,\mathbb{R})$. The representations relevant for studying string theory on $AdS_3$ can be largely classified into two groups: the principal discrete series representation\footnote{The discrete series states are often referred to in the literature as the short strings. The short strings can be thought of as bound states of strings that are trapped deep inside the bulk of $AdS_3$.} (lowest and highest weight) denoted by $\mathcal{D}_j^{\pm}$ and the principal continuous series representation denoted by $\mathcal{C}_j^\alpha$.   The lowest weight principal discrete series states (incoming states) are built on states of the form $|j,j\rangle$ which are annihilated by $j^+$. Thus one can write 
\begin{equation}\label{Dp}
\mathcal{D}_j^+=\left\{|j,m\rangle   ; \;\;  j\in \mathbb{R}, \;\; m=j,j+1,j+2,\dots \right\},
\end{equation}
where $m$ is the $j^3$ eigenvalue of $|j,m\rangle$. The higher weight discrete series states (outgoing states), $\mathcal{D}^-_j$, are obtained by taking the charge conjugate of the $\mathcal{D}_j^+$ states,
\begin{equation}\label{Dm}
\mathcal{D}_j^-=\left\{|j,m\rangle ;  \;\; j\in \mathbb{R}, \;\; m=-j, -j-1,-j-2,\dots \right\},
\end{equation}
 Normalizability of the states requires 
\begin{equation}\label{kbound}
-\frac{1}{2}<j<\frac{k-1}{2}\,.
\end{equation}
The principal continuous series\footnote{The continuous series states are often referred to in the literature as the long strings. Unlike the short strings, the long strings form the scattering states. They live very close to the boundary of $AdS_3$ and have a continuous momentum.} representation corresponds to states which are delta-function normalizable, given by
\begin{equation}\label{C}
\mathcal{C}_j^\alpha=\left\{|j,\alpha,m\rangle; \;\; j\in-\frac{1}{2}+i \mathbb{R}, \;\; m=\alpha,\alpha\pm1,\alpha\pm2,\dots \right\}\,,
\end{equation}
where $|j,\alpha,m\rangle$ are eigenstates of $j^3$ with eigenvalue $m$. Without loss of generality, one can set $0\leq \alpha<1$.
String theory on $AdS_3$ also contains operators that violate the bound \eqref{kbound}.
In terms of the AdS/CFT correspondence, such operators give rise to local operators of the boundary theory (see, {\it e.g.},
\cite{Kutasov:1999xu} for a detailed discussion of the normalizability of the states in $AdS_3$).

Next, let us review construction of the $SL(2,\mathbb{R})$ primaries. The vertex operators of $SL(2,\mathbb{R})$ can always be locally decomposed into a product of vertex operators of $U(1)_y$ and $\frac{SL(2,\mathbb{R})}{U(1)}$. Such a decomposition of vertex operators is known as the parafermionic decomposition.  Let us also assume that $U(1)_y$ is generated by $\J^3$ and is  parametrized by the scalar field $y$, satisfying the OPE $y(z)y(w)\simeq -\log(z-w)$. In terms of the field $y$, the current $\J^3$ takes the form
\begin{equation}
\J^3=-i\sqrt{\frac{k}{2}}\partial y.
\end{equation}
Thus, primary operator $V_{j;m,\bar{m}}$ of string theory on $AdS_3$, obtained by diagonalizing $\J^3,\tilde{\J}^3$ with eigenvalues $m,\bar{m}$ respectively, can be represented as
\begin{equation}\label{vertoparaf}
V_{j;m,\bar{m}}=e^{i\sqrt{\frac{2}{k}}(my+\bar{m}\bar{y})}\Psi_{j;m,\bar{m}}\,,
\end{equation}
where $\Psi_{j;m,\bar{m}}$ is a primary vertex operator on $\frac{SL(2,\mathbb{R})}{U(1)}$.
Using the OPE of $V_{j;m,\bar{m}}$ with the stress-energy tensor,
\begin{equation}\label{tvope}
T_{sl}(z)V_{j;m,\bar{m}}(w)\simeq -\frac{j(j+1)}{k} \frac{1}{(z-w)^2} V_{j;m,\bar{m}}(w)\,,
\end{equation}
one can read of the dimension of the vertex operator $V_{j;m,\bar{m}}$ as 
\begin{equation}\label{dimsl}
\Delta_{sl}=-\frac{j(j+1)}{k}\,.
\end{equation}
This implies that that the dimensions of $\Psi_{j;m,\bar{m}}$ are given by
\begin{equation}\label{deltapsi}
\Delta_\Psi=-\frac{j(j+1)}{k}+\frac{m^2}{k}\,, \ \ \ \ \bar{\Delta}_\Psi=-\frac{j(j+1)}{k}+\frac{\bar{m}^2}{k}\,.
\end{equation}
The vertex operator $V_{j;m,\bar{m}}$ satisfies the following OPE relations
with the generators of the $SL(2,\mathbb{R})$ current algebra:
\begin{equation}
\label{jvope}
\begin{aligned}
&\J^3(z)V_{j;m,\bar{m}}(w)\simeq \frac{m}{z-w} V_{j;m,\bar{m}}(w)\,,\\
&\J^{\pm}(z) V_{j;m,\bar{m}}(w)\simeq \frac{m\mp j}{z-w}V_{j;m\pm1,\bar{m}}(w)\,.
\end{aligned}
\end{equation}
Switching  chirality of the currents, one can write a similar set of OPE algebra of $V_{j;m,\bar{m}}$ with the
right-moving (anti-holomorphic) $SL(2,\mathbb{R})$ currents.

It is often useful to transform the vertex operators from the momentum space $(m,\bar{m})$
to the the position space $(x,\bar{x})$. Let us denote the position space vertex operators by $\Phi_j(z,\bar{z};x,\bar{x})$. Here, the position space coordinates $(x,\bar{x})$ can be thought of as  coordinates of the  boundary of $AdS_3$ on which the dual CFT$_2$ lives. The position space representations are related to the momentum space ones via the following transformation:
\begin{equation}\label{phitov}
V_{j;m,\bar{m}}=\int_{\partial AdS_3} d^2x~x^{j+m}\bar{x}^{j+\bar{m}} \Phi_j(z,\bar{z};x,\bar{x})\,.
\end{equation}
One can invert the transformation \eqref{phitov} to calculate $\Phi_h(z,\bar{z};x,\bar{x})$:
\begin{equation}\label{vtophi}
\Phi_{h}(z,\bar{z};x,\bar{x})=\sum_{m,\bar{m}} x^{-m-h}\bar{x}^{-\bar{m}-h}V_{h-1;m,\bar{m}}\,,
\end{equation}
where $(h,h)=(j+1,j+1)$ is the spin of  $\Phi_h(z,\bar{z};x,\bar{x})$ under space-time $SL(2,\mathbb{R})_L\times SL(2,\mathbb{R})_R$ symmetry.

The operator $\Phi_j$ satisfies the following reflection symmetry:
\begin{equation}\label{reflection}
\Phi_{j+1}(z,\bar{z};x,\bar{x})=\frac{2j+1}{\pi}\int d^2x'~ \frac{\Phi_{-j}(z,\bar{z};x',\bar{x}')}{|x-x'|^{4(j+1)}}\,.
\end{equation}
Thus for real $j$ one can restrict, without loss of generality, to $j>-1/2$, while for $j=-1/2+is$ \eqref{reflection} relates wave function with $s>0$ (out-going scattering waves) to those with $s<0$ (in-coming scattering waves).

In the semiclassical approximation, {\it i.e.},
in the limit $k\to\infty$, the two-point function of the operator
$\Phi_h(z;x)$ can be determined via the world-sheet conformal invariance and the invariance under the global $SL(2,\mathbb{R})_L\times SL(2,\mathbb{R})_R$ symmetry of the space-time theory \cite{Kutasov:1999xu} 
\begin{equation}\label{phi2ptfn}
\langle \Phi_{h}(z_1;x_1)  \Phi_{h'}(z_2;x_2)  \rangle=\delta(h-h')\frac{B(h)}{|z_1-z_2|^{4\Delta_{sl}}|x_1-x_2|^{4h}}~,
\end{equation}
where 
\begin{equation}\label{bh}
B(h)=\frac{k}{\pi}X^{2h-1}\gamma\left(1-\frac{2h-1}{k}\right),  \ \ \ \ \gamma(x)=\frac{\Gamma(x)}{\Gamma(1-x)}\,,
\end{equation}
and $X$ is an arbitrary constant whose value can be adjusted by shifting the radial coordinate or rescaling the transverse field theory coordinates. In the momentum basis the two-point function takes the form
\begin{equation}\label{VV2ptfn}
\langle V_{j;m,\bar{m}}(z)V_{j';-m,-\bar{m}}(w)\rangle{=}\frac{\pi\delta(j{-}j')}{|z{-}w|^{4\Delta_{sl}}}
\frac{\Gamma\left(1{-}\frac{2j{+}1}{k}\right)\Gamma({-}2j{-}1)\Gamma(j{-}m{+}1)
\Gamma(j{+}\bar{m}{+}1)}{\Gamma\left(1{+}\frac{2j{+}1}{k}\right)\Gamma(2j{+}2)\Gamma({-}j{-}m)\Gamma({-}j{+}\bar{m})}.
\end{equation}

The structure of
three-point correlation function of the primary world-sheet operator $\Phi_h$ is
fixed by the conformal symmetry of the world-sheet and the space-time theories.
For the operator $\Phi_h$ normalized so that its two-point function is given by (\ref{phi2ptfn}), one obtains \cite{Maldacena:2001km}
\begin{align}
\langle \Phi_{h_1}(z_1,x_1)&\Phi_{h_2}(z_2,x_2)\Phi_{h_3}(z_3,x_3)\rangle=\frac{C_{sl}(h_1,h_2,h_3)}{|z_{12}|^{2(\Delta_{h_1}+\Delta_{h_2}-\Delta_{h_3})}|z_{23}|^{2(\Delta_{h_2}+\Delta_{h_3}-\Delta_{h_1})}|z_{31}|^{2(\Delta_{h_3}+\Delta_{h_1}-\Delta_{h_2})}}\notag\\
&\times \frac{1}{|x_{12}|^{2(h_1+h_2-h_3)}|x_{23}|^{2(h_2+h_3-h_1)}|x_{31}|^{2(h_3+h_1-h_2)}}\label{phiphiphi}
\end{align}
where $\Delta_{h}=\Delta_{sl}$ and $z_{ij}=z_i-z_j$ and $x_{ij}=x_i-x_j$.
The OPE coefficient $C_{sl}(h_1,h_2,h_3)$ can be computed from the conformal Ward identity, giving
\begin{equation}\label{Csl}
C_{sl}(h_1,h_2,h_3){=}{-}\frac{G(1{-}h_1{-}h_2{-}h_3)G(h_3{-}h_1{-}h_2)G(h_3{-}h_3{-}h_1)
G(h_1{-}h_2{-}h_3)}{2\pi^2X^(h_1{+}h_2{+}h_3-1)\gamma\left(\frac{k{+}1}{k}\right)G({-}1)G(1{-}2h_1)
G(1{-}2h_2)G(1{-}2h_3)}
\end{equation}
where 
\begin{eqnarray}
G(h)=k^{\frac{h(k+1-h)}{2k}}\Gamma_2(-h|1,k)\Gamma_2(k+1+h|,1,k)
\end{eqnarray}
with $\Gamma_2$ defined as
\begin{eqnarray}
\log \Gamma_2(x|1,\omega)=\lim_{\epsilon\to 0}\frac{\partial}{\partial \epsilon}\left[\sum_{m,n=0}^\infty(x+n+m\omega)^{-\epsilon}-\sum_{m,n=0, (m,n)\neq (0,0)}^\infty(n+m\omega)^{-\epsilon}\right].
\end{eqnarray}

String theory on $AdS_3$ also contains operators obtained by a spectral flow automorphism of the symmetry algebra:
\begin{equation}
\begin{aligned}\label{specjsl}
&\J^3(z)\to \J^3(z)-\frac{k\hat{\omega}}{2z}\,, \\
&\J^{\pm}(z)\to z^{\pm \hat{\omega}}\,\J^{\pm}(z)\,,
\end{aligned}
\end{equation}
where $\hat{\omega}\in \mathbb{Z}$ is the spectral flow parameter. Note that since we are working in the universal cover of $SL(2,\mathbb{R})$, the left- and the right-moving spectral flow parameters are identical. Under spectral flow automorphism the Virasoro generators, obtained by mode expansion of the stress-energy tensor \eqref{slst}, transform as
\begin{eqnarray}
L_n\to L_n +\hat{\omega} j^3_{n}-\frac{k\hat{\omega}^2}{4}\delta_{n,0} \qquad \forall n\in \mathbb{Z}\,,
\end{eqnarray}
where $j^3_n$ are the Kac-Moody modes of the current $j^3$.

For each spectrally flowed sector $\hat{\omega}\in \mathbb{Z}$, string theory on $AdS_3$ contains twist operators given by
\begin{eqnarray}\label{twistfield}
t_{sl}^{\hat{\omega}}(z,\bar{z})=t^{\hat{\omega}}(z)\, \tilde{t}^{\hat{\omega}}(\bar{z})
=e^{-\hat{\omega}\, (\int \J^3(z)+\int \tilde{\J}^3(\bar{z}) )}=e^{i\,\hat{\omega}
\,\sqrt{\frac{k}{2}}\,(y(z)+\bar{y}(\bar{z}))}\,.
\end{eqnarray}
The twist field $t_{sl}^{\hat{\omega}}$ acts on the spectrally un-flowed ({\it i.e.}, $\hat{\omega}=0$) vertex operators \eqref{vertoparaf} to generate operators in the spectrally flowed sector:
\begin{equation}
\begin{aligned}\label{twistVope}
t_{sl}^{\hat{\omega}}(z,\bar{z}) V_{j;m,\bar{m}}(w,\bar{w})&\simeq (z-w)^{-\hat{\omega}m}\,
(\bar{z}-\bar{w})^{-\hat{\omega}\bar{m}}\,V^{\hat{\omega}}_{j;m,\bar{m}}(w,\bar{w})\\
&\sim (z-w)^{-(m-\bar{m})\hat{\omega}}\, |z-w|^{2\bar{m}\hat{\omega}}\, V^{\hat{\omega}}_{j;m,\bar{m}}(w,\bar{w})\,,
\end{aligned}
\end{equation}
where
\begin{equation}\label{spectfV}
V^{\hat{\omega}}_{j;m,\bar{m}}=e^{i\,\sqrt{\frac{2}{k}}\,\left[\left(m+\frac{k\hat{\omega}}{2}\right)\,y+\left(\bar{m}+\frac{k\hat{\omega}}{2}\right)\,\bar{y}\right]}\Psi_{j;m,\bar{m}}~.
\end{equation}
Note that the operator $\Psi_{j;m,\bar{m}}$ remains unchanged under spectral flow. In particular, its world-sheet dimension $\Delta_\Psi$, given by \eqref{deltapsi}, does not change under spectral flow. Demanding mutual locality of the spectrally flowed vertex operators with respect to the twist fields imposes the constraint 
\begin{equation}
m-\bar{m}\in \mathbb{Z}~.
\end{equation}
It is easy to read off the world-sheet dimension of the vertex operator \eqref{spectfV} as 
\begin{equation}
\label{deltaslsf}
\Delta_{sl}=\Delta_\Psi-\frac{\left(m+\frac{k\hat{\omega}}{2}\right)^2}{k}
=-\frac{j(j+1)}{k}-m\hat{\omega}-\frac{k\hat{\omega}^2}{4}\,.
\end{equation}
Switching the chiralities, one obtains the right-moving dimension as
\begin{equation}
\label{deltabarslsf}
\bar{\Delta}_{sl}=\bar{\Delta}_\Psi-\frac{\left(\bar{m}+\frac{k\hat{\omega}}{2}\right)^2}{k}=-\frac{j(j+1)}{k}-\bar{m}\hat{\omega}-\frac{k\hat{\omega}^2}{4}\,.
\end{equation}
The OPEs of the spectrally flowed vertex operator \eqref{spectfV} with the  $SL(2,\mathbb{R})$ currents take the form
\begin{equation}
\label{jvopesf}
\begin{aligned}
&\J^3(z)V^{\hat{\omega}}_{j;m,\bar{m}}(w)\simeq \frac{m+\frac{k\hat{\omega}}{2}}{z-w} V_{j;m,\bar{m}}(w)\,,\\
&\J^{\pm}(z) V^{\hat{\omega}}_{j;m,\bar{m}}(w)\simeq \frac{m\mp j}{z-w}V_{j;m\pm1,\bar{m}}(w)\,.
\end{aligned}
\end{equation}

\subsection{Symmetries and vertex operators of $\mathbb{S}^3$}

Superstring theory on $\mathbb{S}^3$ is described 
by the superconformal $SU(2)$ WZW model that
is composed of the bosonic $SU(2)_{k-2}$ WZW model at level $k-2$ and three pairs of free
(anti-)holomorphic fermions that can be described by $SU(2)_2$ WZW model at level $2$. The bosonic and the fermionic
sectors combine to give rise to the $SU(2)_k$ affine symmetry algebra at level $k$.

For $g'\in SU(2)$, the exact effective $SU(2)_k$ WZW  action is given by
\begin{equation}\label{suwzw}
S_{su}= k\,\mathtt{S}[g']
\end{equation}
where the level-one WZW action is defined in (\ref{WZWgen}).
The action \eqref{suwzw} is invariant w.r.t.
$SU(2)_L\times SU(2)_R$ affine symmetry algebra at level $k$.  The global part of the $SU(2)_L\times SU(2)_R$ symmetry algebra is generated by
the  Kac-Moody currents $j^{\prime a}$ and $\tilde{j}^{\prime a}$, $a=1,2,3$, given by
\begin{equation}\label{sugcurrents}
j^{\prime a}=\frac{k}{2}\, {\rm Tr} \left(\partial g' g'^{-1} \sigma^a \right), \ \ \ \ \ \tilde{j}^{\prime a}=
\frac{k}{2} \, {\rm Tr} \left(g'^{-1}\bar{\partial} g'  \sigma^a \right).
\end{equation}
These currents get combined with contributions
from the fermions, rendering the $SU(2)_k$ Kac-Moody algebra
\begin{equation}\label{suope}
\J^{\prime a}(z)\, \J^{\prime b}(w) = \frac{k}{2}\,\frac{\delta^{ab}}{(z-w)^2} + \frac{i\,\epsilon^{abc}}{z-w}\,\J'_c(w)+\dots\,.
\end{equation}
The world-sheet stress-energy tensor according to the Sugawara construction takes the form,
\begin{equation}\label{sust}
\begin{aligned}
T_{su}\supset\frac{1}{k}\,j^{\prime a}\,j^{\prime a}\,,\qquad
\tilde{T}_{su}\supset\frac{1}{k}\,\tilde{j}^{\prime a}\,\tilde{j}^{\prime a} \,,
\end{aligned}
\end{equation}
where we omitted contributions from the fermions.
The quadratic Casimir of $SU(2)_k$ is given by
\begin{equation}\label{sucasimir}
c'_2=(j^{\prime 3})^2+\frac{1}{2}\{j^{'+},j^{'-}\}\,,
\end{equation}
where 
\begin{eqnarray}\label{jsupm}
j^{'\pm}=j^{\prime 1}\pm i j^{\prime 2}\,.
\end{eqnarray}
All representations of $SU(2)$ can be parametrized by $j'$, that is related to the  Casimir as $c'_2=j'(j'+1)$.

The group manifold $SU(2)_k$, being compact,  possesses only discrete series representation given by
\begin{equation}\label{repsu}
\mathcal{D}_{j'}'=\left\{|j';m',\bar{m}'\rangle: \;\; j'=0,\frac{1}{2},1,\dots,\frac{k}{2};\;\; m',\bar{m}'=-j',j'+1,\dots,j'\right\}\,,
\end{equation}
where $m',\bar{m}'$ are  eigenvalues of $\J^{\prime 3}$ and $\tilde{\J}^{\prime 3}$. 

Analogously to the case of $SL(2,\mathbb{R})$, the primary vertex operators of $SU(2)$ 
can be described in terms of the parafermionic decomposition into $U(1)_{y_{su}}$ and $\frac{SU(2)}{U(1)}$. The subgroup $U(1)_{y_{su}}$ is generated by $\J^{\prime 3}$ and is  parametrized by a  scalar field $y_{su}$,
satisfying the OPE $y_{su}(z)y_{su}(w)\simeq-\log(z-w)$. Thus one can write 
\begin{equation}\label{j3psu}
\J^{\prime 3}=i\sqrt{\frac{k}{2}}\,\partial y_{su}\,.
\end{equation}
The primary operators, $V'_{j';m',\bar{m}'}$,  can be written as
\begin{equation}\label{vertoparafsu}
V'_{j';m',\bar{m}'}=e^{i\sqrt{\frac{2}{k}}(m'y_{su}+\bar{m}'\bar{y}_{su})}\Psi'_{j';m',\bar{m}'}\,,
\end{equation}
where $\Psi'_{j';m',\bar{m}'}$ is a primary vertex operator on $\frac{SU(2)}{U(1)}$.  The vertex operator \eqref{vertoparafsu} transforms
in the spin $(j',j')$ representation of $SU(2)$.

The $SU(2)$ vertex operator $V'_{j';m',\bar{m}'}$ satisfies the following OPE with the stress-energy tensor:
\begin{equation}\label{tvopesu}
T_{su}(z)V'_{j';m',\bar{m}'}(w)\simeq \frac{\Delta_{su}}{(z-w)^2} V_{j';m',\bar{m}'}(w)\,,
\end{equation}
where we defined its scaling dimension as
\begin{equation}\label{dimsu}
\Delta_{su}=\frac{j(j+1)}{k}\,.
\end{equation}
It follows that the dimension of $\Psi'_{j';m',\bar{m}'}$ is given by
\begin{equation}\label{deltapsip}
\Delta_{\Psi'}=\frac{j(j+1)}{k}-\frac{m'^2}{k},\ \ \ \ \bar{\Delta}_{\Psi'}=\frac{j(j+1)}{k}-\frac{\bar{m}'^2}{k} \,.
\end{equation}

The OPEs of the primary vertex operator $V'_{j';m',\bar{m}'}$ with the $SU(2)$ currents are given by
\begin{equation}
\label{jvopesu}
\begin{aligned}
&\J^{\prime 3}(z)V'_{j';m',\bar{m}'}(w)\simeq \frac{m'}{z-w} V'_{j';m',\bar{m}'}(w)\,,\\
&\J^{'\pm}(z) V'_{j';m',\bar{m}'}(w)\simeq \frac{\sqrt{j'(j'+1)-m'(m'\pm1)}}{z-w}V'_{j';m'\pm1,\bar{m}'}(w)\,.
\end{aligned}
\end{equation}
Switching the chirality of the $SU(2)$ currents, one can analogously write the OPE relations
of $V'_{j';m',\bar{m}'}$ with the right-moving $SU(2)$ currents.

Similarly to the vertex operators in $AdS_3$,
the $SU(2)$ vertex operators can be equivalently described in two different bases, related by
\begin{equation}\label{phiptovp}
\Phi_{j'}'(z,\bar{z};u,\bar{u})=\sum_{m',\bar{m}'
=-j'}^{j'}\sqrt{C_{2j'}^{j'+m'}C_{2j'}^{j'+\bar{m}'}}u^{j'+m'}\bar{u}^{j'+\bar{m}'}V'_{j';m',\bar{m}'}(z,\bar{z})\,,
\end{equation}
where 
\begin{equation}\label{coeffC}
C_{q}^{p}=\frac{q!}{p! \, (q-p)!}\,.
\end{equation}

The two-point functions of the $SU(2)$ primaries are given by
\begin{equation}
\begin{aligned}\label{su2ptfn}
&\langle \Phi_{j_1'}'(z,\bar{z};u,\bar{u})\Phi_{j_2'}'(w,\bar{w};v,\bar{v})\rangle= \frac{\delta_{j_1'j_2'}}{|z-w|^{4\Delta_{su}}|u-v|^{4j_1'}}\,, \\
&\langle (V'_{j_1';m_1',\bar{m}_1'}(z,\bar{z}))^\dagger V'_{j_2';m_2',\bar{m}_2'}(w,\bar{w}) \rangle=\frac{\delta_{j_1'j_2'}\delta_{m_1'm_2'}\delta_{\bar{m}_1'\bar{m}_2'}}{|z-w|^{4\Delta_{su}}}\,,
\end{aligned}
\end{equation}
where the conjugate operator is given by 
\begin{equation}\label{herconj}
(V'_{j';m',\bar{m}'})^\dagger=(-1)^{2j-m'-\bar{m}'}V'_{j';-m',-\bar{m}'}\,.
\end{equation}
The three-point function of $V'_{j';m',\bar{m}'}$ is given by \cite{Fateev:1985mm}
\begin{equation}\label{VVV}
\langle V'_{j_1';m_1',\bar{m}_1'}(z_1),V'_{j_2';m_2',\bar{m}_2'}(z_2)V'_{j_3';m_3',\bar{m}_3'}(z_3)\rangle=\frac{
C_{su}(j_i',m_i',\bar m_i')}{|z_{12}|^{2(\Delta_{j'_1}+\Delta_{j'_2}-\Delta_{j'_3})}|z_{23}|^{2(\Delta_{j'_2}+\Delta_{j'_3}-\Delta_{j'_1})}|z_{31}|^{2(\Delta_{j'_3}+\Delta_{j'_1}-\Delta_{j'_2})}}
\end{equation}
where $\Delta_{j'}=\Delta_{su}$ and
\begin{equation}\label{Csu}
C_{su}(j_i',m_i',\bar m_i')=\begin{bmatrix}
j'_1& j'_2 &j'_3\\ 
m'_1 & m'_2 &m'_3
\end{bmatrix}\begin{bmatrix}
j'_1& j'_2 &j'_3\\ 
\bar{m}'_1 & \bar{m}'_2 &\bar{m}'_3
\end{bmatrix} \rho(j'_1,j'_2,j'_3) \ \ 
\end{equation}
with the first two factors are the Wigner $3j$-symbols, and 
\begin{align}
&\frac{\rho^2(j'_1,j'_2,j'_3)}{(2j'_1+1)(2j'_2+1)(2j'_3+1)}=\frac{\Gamma\left(\frac{k+1}{k}\right)\Gamma\left(1-\frac{2j'_1+1}{k}\right)\Gamma\left(1-\frac{2j'_2+1}{k}\right)\Gamma\left(1-\frac{2j'_3+1}{k}\right)}{\Gamma\left(\frac{k-1}{k}\right)\Gamma\left(1+\frac{2j'_1+1}{k}\right)\Gamma\left(1+\frac{2j'_2+1}{k}\right)\Gamma\left(1+\frac{2j'_3+1}{k}\right)}\notag\\
&\times \frac{\Pi^2(j'_1+j'_2+j'_3+1)\Pi^2(j'_1+j'_2-j'_3)\Pi^2(j'_2+j'_3-j'_1)\Pi^2(j'_3+j'_1-j'_2)}{\Pi^2(2j'_1)\Pi^2(2j'_2)\Pi^2(2j'_3)}
\end{align}
where $\Pi$ denotes
\begin{equation}
\Pi(j')=\prod_{n=1}^{j'}\frac{\Gamma\left(1+\frac{n}{k}\right)}{\Gamma\left(1-\frac{n}{k}\right)}\,.
\end{equation}

String theory on $\mathbb{S}^3$ contains states obtained by spectral flow automorphism of the $SU(2)$ algebra:
\begin{equation}
\begin{aligned}\label{specjsu}
&\J^{\prime 3}(z)\to \J^{\prime 3}(z)-\frac{k\hat{\omega}'}{2z}\,,
  \ \ \ &\tilde{\J}^{\prime 3}(\bar{z})\to \tilde{\J}^{\prime 3}(z)-\frac{k\hat{\bar{\omega}}'}{2\bar{z}}\,, \\
&\J^{'\pm}(z)\to z^{\pm \hat{\omega}'}\J^{'\pm}(z)\,,
 \ \ \ &\tilde{\J}^{'\pm}(\bar{z})\to \bar{z}^{\pm \hat{\bar{\omega}}'}\tilde{\J}^{'\pm}(\bar{z})\,,
\end{aligned}
\end{equation}
where $\hat{\omega}',\hat{\bar{\omega}}\in\mathbb{Z}$ are the left- and right-moving $SU(2)$ spectral flow parameters. Unlike spectral flow automorphism of $SL(2,\mathbb{R})$, the left- and the right-moving $SU(2)$ spectral flow parameters are different and independent of each other. Under spectral flow automorphism, the world-sheet Virasoro generators transform as
\begin{equation}\label{virsu}
\begin{aligned}
&L_n'\to L_n' -\hat{\omega}' j^{'3}_{n}+\frac{k\hat{\omega}^{'2}}{4}\delta_{n,0} \,,\\
&\tilde{L}_n'\to \tilde{L}_n' -\hat{\bar{\omega}}' \tilde{j}^{'3}_{n}+\frac{k\hat{\bar{\omega}}^{'2}}{4}\delta_{n,0}\,, \\
\end{aligned}
\end{equation}
where $j^{'3}_n$ and $\tilde{j}^{'3}_n$ are  the Kac-Moody modes of the currents $j^{\prime 3}$ and $\tilde{j}^{\prime 3}$.

In the parafermionic notation, the spectrally flowed vertex operators are given by
\begin{equation}\label{sfvsu}
{V'}^{\hat{\omega}',\hat{\bar{\omega}}'}_{j';m',\bar{m}'}=e^{i\,\sqrt{\frac{2}{k}}\,\left[\left(m'+\frac{k\hat{\omega}'}{2}\right) y_{su}+\left(\bar{m}+\frac{k\hat{\bar{\omega}}'}{2}\right) \bar{y}_{su}\right]} \Psi'_{j';m',\bar{m}'}\,.
\end{equation}
Its (anti-)holomorphic scaling dimensions are  given by
\begin{equation}
\label{deltasfsu}
\begin{aligned}
&\Delta_{su}=\frac{j'(j'+1)}{k}+m'\hat{\omega}'+\frac{k\hat{\omega}'^2}{4}\,, \\
 & \bar{\Delta}_{su}=\frac{j'(j'+1)}{k}+\bar{m}'\hat{\bar{\omega}}'+\frac{k\hat{\bar{\omega}}'^2}{4}\,.
\end{aligned}
\end{equation}

The OPEs of the spectrally flowed vertex operator \eqref{sfvsu} with the global $SU(2)$ generators take the form
\begin{equation}
\begin{aligned}
\label{sfvjpope}
& \J^{\prime 3}(z) \, {V'}^{\hat{\omega}',\hat{\bar{\omega}}'}_{j';m',\bar{m}'}(w) =
\frac{m'+\frac{k}{2}\hat{\omega}'}{z-w}\,{V'}^{\hat{\omega}',\hat{\bar{\omega}}'}_{j';m',\bar{m}'}\,,\\
&\tilde \J^{\prime 3}(\bar z) \, {V'}^{\hat{\omega}',\hat{\bar{\omega}}'}_{j';m',\bar{m}'}(\bar{w})
= \frac{\bar m'+\frac{k}{2}\hat{\bar{\omega}}'}{\bar z-\bar{w}}\,{V'}^{\hat{\omega}',\hat{\bar{\omega}}'}_{j';m',\bar{m}'}\,.
\end{aligned}
\end{equation}
The OPE of ${V'}^{\hat{\omega}',\hat{\bar{\omega}}'}_{j';m',\bar{m}'}$ with $\J^{'\pm}$ and $\tilde{\J}^{'\pm}$ remain unchanged
under spectral flow.

\section{Coset construction}
\label{sec:coset construction}

The main model we are going to work with in this paper is given by the
 type-II superstring theory
on the ten-dimensional space-time,
\begin{equation}
\label{big coset}
\frac{SL(2,\mathbb{R})\times SU(2)\times U(1)}{U(1)} \times U(1)^4\,,
\end{equation}
given by a product of the six-dimensional coset space-time and the four-dimensional flat compact space
given by the torus $\mathbb{T}^4 = U(1)^4$. The radii of $\mathbb{T}^4$ are free parameters of our model.
In particular, we can take the limit of infinite radii, and consider the theory on the
six-dimensional coset space times the $\mathbb{R}^4$
manifold.

Our primary interest will be the case of asymmetric anomaly-free time-like gauging
of the $U(1)$ sub-group in the coset sub-space of (\ref{big coset}).
The goal of this section is to construct the world-sheet action, obtain the gauge currents, and derive
the corresponding  supergravity target space-time background geometry.
We will demonstrate that the target space-time geometry has a black hole horizon,
while its asymptotic region is given by the two-dimensional linear dilaton background times
$\mathbb{S}^3\times \mathbb{S}^1\times \mathbb{T}^4$.\footnote{
We will denote indices of vectors in ten-dimensional space-time 
with capital-case Latin letters, and indices of the six-dimensional
coset space-time with lower-case Greek letters. \textit{E.g.}, the space-time coordinates
are given by $X^M = (X^\mu, z^1,z^2,z^3,z^4)$, where $X^\mu$
are coordinates of the six-dimensional coset space-time, and $z^i=(z^1,z^2,z^3,z^4)$
are coordinates of the torus $\mathbb{T}^4$. Throughout this paper, we will generally
assume that all the coordinates are dimensionless, and therefore, \textit{e.g.}, the
ten-dimensional metric is given by $ds_{10}^2 = \ell_s^2\,G_{MN}dX^MdX^N$,
where $\ell_s$ is the string length.}

We start by considering sigma-model on the group manifold
\begin{equation}
\label{sl2r su2 and u1}
{\cal G}\times U(1)^4 \qquad\textrm{where}\qquad {\cal G} = SL(2,\mathbb{R})\times SU(2)\times U(1)_x\,,
\end{equation}
that is described by the action consisting of the WZW terms for the field $G\in {\cal G}$ and the Polyakov 
terms for the abelian components.
Supersymmetric version of this model is obtained by supplementing
the bosonic WZW/Polyakov sector with the free Majorana-Weyl fermions.
The latter are split into (anti-)holomorphic sectors, and are labeled by indices
in adjoint representation of the space-time symmetry group:
\begin{equation}
\psi^a,\;\tilde\psi^a \in sl(2,\mathbb{R})\,,\quad
\psi^{\prime b},\;\tilde\psi^{\prime b} \in su(2)\,,\quad
\psi^x,\;\tilde\psi^x \in u(1)_x\,,\quad
\psi^{i},\;\tilde\psi^{i} \in u(1)^4\,.
\end{equation}
The world-sheet fermions are then described by the action
\begin{equation}
\begin{aligned}
S_ f {=} \frac{k}{2\pi}\int d^2z\left(\psi^a \p \psi^a {+}\tilde\psi^a\bar\p\tilde\psi^a
{+}\psi^{\prime b} \p \psi^{\prime b} {+}\tilde\psi^{\prime b}\bar\p\tilde\psi^{\prime b}
{+}\psi^x \p \psi^x {+}\tilde\psi^x\bar\p\tilde\psi^x
{+}\psi^i \p \psi^i {+}\tilde\psi^i\bar\p\tilde\psi^i
 \right)
\end{aligned}
\end{equation}
where sums over repeated indices $a=1,2,3$, $b=1,2,3$, $i=1,2,3,4$ are implied.

As mentioned above,
 bosonic sector of the model is described by the sum of 
WZW terms for the $SL(2,\mathbb{R})$ and $SU(2)$ sectors, plus the Polyakov terms for the $U(1)_x$
and $U(1)^4$ sub-spaces. Due to the nature of the coset (\ref{big coset}),
it is convenient to combine
the group elements
\begin{equation}
\label{sl2r su2 and u1 defs}
g\in SL(2,\mathbb{R})\,,\qquad g'\in SU(2)\,,\qquad e^{i\sqrt{\frac{2}{k}}\,x} \in U(1)_x
\end{equation}
into one 
block-diagonal element of the sub-group ${\cal G}$ of (\ref{sl2r su2 and u1}), that we denote as
\begin{equation}
G=\textrm{diag}\left[g,\; g',\; e^{i\sqrt{\frac{2}{k}}\,x}\right]\,.
\end{equation}
The WZW action for the field $G$ on the group manifold (\ref{sl2r su2 and u1}) is then given by
\begin{equation}
\label{WZWgen G component}
S_{\textrm{WZW}}[g,g',x] = k\,\mathtt{S}[g] - k\,\mathtt{S}[g'] +\frac{1}{2\pi}\,\int d^2z\,\partial x\bar\partial x\,,
\end{equation}
where the level-one WZW action $\mathtt{S}$ is defined in Appendix~\ref{app:conventions}.
In terms of the field $G$ the action (\ref{WZWgen G component}) can be equivalently re-written as
\begin{equation}
\label{WZWgen G} 
\begin{aligned}
S_{\textrm{WZW}}'[G] \equiv S_{\textrm{WZW}}[g,g',x] &={k\over 4\pi} \int d^2z\,\Tr \left(P\,G^{-1}\p G \, G^{-1}\bar\p G\right)\\
&+{i k \over 24\pi} \int _{{\cal B}}\,\Tr\left( P\,G^{-1}d G
\wedge G^{-1}d G\wedge G^{-1}d G\right)\,,
\end{aligned}
\end{equation}
where we introduced the projection operator
\begin{equation}
P = \textrm{diag}\left[\,1,\;1,\;-1,\;-1,\;-1\,\right]\,,
\end{equation}
and  prime in the WZW action notation in l.h.s. of (\ref{WZWgen G}) indicates the presence of this projection operator.

Together with the Polyakov terms for the $U(1)^4$ sector,
\begin{equation}
S_{\mathbb{T}^4} = \frac{1}{2\pi}\,\int d^2z \,\p z^i\,\bar\p z^i\,,
\end{equation}
total action of the supersymmetric sigma model on (\ref{sl2r su2 and u1}) is given by
\begin{equation}
\label{total susy wzw action}
S = S_{\textrm{WZW}}'  + S_{\mathbb{T}^4} + S_f\,.
\end{equation}
We view the action (\ref{total susy wzw action}) as exact (in $k$, that also means exact in $\ell_s$)
effective action of the world-sheet
theory. In particular, the level $k$ of the WZW terms that we wrote down for the $SL(2,\mathbb{R})$
and $SU(2)$ sectors are exact in quantum theory.\footnote{See Appendix~\ref{app:effective wzw action}
for a review of the corresponding formalism.}

\subsection{Gauging and gauge currents}

Next, we are going to gauge the $U(1)$ subgroup of the group ${\cal G}$.
Let us start with the derivation in the bosonic sector, where
an action of the gauged $U(1)$ sub-group of the target space symmetry group ${\cal G}$ is defined by 
the following transformation laws
\begin{equation}
\label{U1 def}
\begin{aligned}
&g\sim e^{\frac{a_1}{\sqrt{k}}\,\xi\,\sigma_3}\,g\,e^{\frac{b_1}{\sqrt{k}}\,\xi\,\sigma_3}\,,\\
&g'\sim e^{i\,\frac{a_2}{\sqrt{k}}\,\xi\,\sigma_3}\,g'\,e^{i\,\frac{b_2}{\sqrt{k}}\,\xi\,\sigma_3}\,,\\
&x_L\sim x_L +a_3\,\xi\,,\qquad x_R\sim x_R + b_3\,\xi\,.
\end{aligned}
\end{equation}
Here $\xi$ parametrizes the gauged $U(1)$ group, and $x_{L,R}$ are (anti-)holomorphic
components of the world-sheet field $x$.
We are going to be working with the following parametrization of the
$SL(2,\mathbb{R})$ and $SU(2)$ group elements
\begin{equation}
\label{groups parametrizations}
\begin{aligned}
g &= e^{\alpha\sigma_3}e^{\theta\sigma_1}e^{\beta\sigma_3}\,,\\
g' &= e^{i\alpha'\sigma_3}e^{i\theta'\sigma_1}e^{i\beta'\sigma_3}\,,
\end{aligned}
\end{equation}
in terms of which
the action (\ref{WZWgen G component}) is given by
(see sections~\ref{sec:su2wzw},~\ref{sec:sl2rwzw})
\begin{equation}
\label{WZW action in parametrization}
\begin{aligned}
S_{\textrm{WZW}}[g,g',x] &= \frac{k}{2\pi}\int d^2z\,\left(
\p\alpha\bar\p\alpha+\p\beta\bar\p\beta+\p\theta\bar\p\theta
+2\cosh(2\theta)\bar\p\alpha\p\beta\right.\\
&+\left.\p\alpha'\bar\p\alpha'+\p\beta'\bar\p\beta'+\p\theta'\bar\p\theta'
+2\cos(2\theta')\bar\p\alpha'\p\beta'+\p x\bar\p x
\right)\,,
\end{aligned}
\end{equation}
while the transformations (\ref{U1 def}) are given by
\begin{equation}
\label{u1 transform in parametrization}
\begin{aligned}
&\alpha\sim\alpha+\frac{a_1}{\sqrt{k}}\,\xi\,,\quad
\beta\sim\beta+\frac{b_1}{\sqrt{k}}\,\xi\,,\quad
\alpha'\sim\alpha'+\frac{a_2}{\sqrt{k}}\,\xi\,,\quad
\beta'\sim\beta'+\frac{b_2}{\sqrt{k}}\,\xi\,,\\
&x\sim x + (a_3+b_3)\xi\,,\quad \theta\sim\theta\,,\quad \theta'\sim\theta'\,.
\end{aligned}
\end{equation}

The $U(1)$ transformation rules (\ref{U1 def}) can be compactly reformulated as 
\begin{equation}
G\sim e^{\xi \,T_L}\,G\, e^{\xi\,T_R}\,,
\end{equation}
where the left/right-moving generators of the corresponding $u(1)$ algebra are given by
the following block-diagonal matrices
\begin{equation}
\label{TL and TR defs}
\begin{aligned}
T_L &= \textrm{diag} \left[\, \frac{a_1}{\sqrt{k}}\,\sigma_3,\; i\,\frac{a_2}{\sqrt{k}}\,\sigma_3,\; ia_3\,\sqrt{\frac{2}{k}}\, \right]\,,\\
T_R &= \textrm{diag} \left[\, \frac{b_1}{\sqrt{k}}\,\sigma_3,\; i\,\frac{b_2}{\sqrt{k}}\,\sigma_3,\; ib_3\,\sqrt{\frac{2}{k}}\, \right]\,.
\end{aligned}
\end{equation}
Then introducing the compensator fields 
\begin{equation}
U = e^{-u\,T_L}\,,\qquad V= e^{-v\,T_L}\,,
\end{equation}
obeying the $U(1)$ transformation rules
\begin{equation}
u\sim u+\xi\,,\qquad v\sim v+\xi\,,
\end{equation}
we can construct $U(1)$-invariant field $UGV$. The corresponding locally $U(1)$-invariant WZW action
can be written using the Polyakov-Wiegmann identity as
\begin{equation}
\label{UGV action}
\begin{aligned}
&S'_{\textrm{WZW}}[UGV]=S'_{\textrm{WZW}}[G]+S'_{\textrm{WZW}}[U]+S'_{\textrm{WZW}}[V]\\
&+\frac{k}{2\pi}\,\int d^2z\,
\textrm{Tr}\left[P\left(G^{-1}\bar\p G\p VV^{-1}+U^{-1}\bar\p U \p G G^{-1}
+U^{-1}\bar\p U G \p VV^{-1}G^{-1}\right)\right]\,,
\end{aligned}
\end{equation}
where $S'_{\textrm{WZW}}[G]$ is given by (\ref{WZWgen G component}), (\ref{WZWgen G}), while
\begin{equation}
\label{U and V action}
\begin{aligned}
S'_{\textrm{WZW}}[U] &= \frac{k}{4\pi}\,\textrm{Tr}(P\,T_L^2)\,\int d^2z\,\p u\bar\p u\,,\\
S'_{\textrm{WZW}}[V] &= \frac{k}{4\pi}\,\textrm{Tr}(P\,T_R^2)\,\int d^2z\,\p v\bar\p v\,.
\end{aligned}
\end{equation}
Using (\ref{TL and TR defs}) we obtain
\begin{equation}
\textrm{Tr}(P\,T_L^2) = \frac{2}{k}\,(a_1^2+a_2^2+a_3^2)\,,\qquad
\textrm{Tr}(P\,T_R^2) = \frac{2}{k}\,(b_1^2+b_2^2+b_3^2)\,.
\end{equation}
Notice that the action (\ref{UGV action}), while being invariant w.r.t. local  $U(1)$ gauge
transformations (\ref{U1 def}), cannot be an action of the gauged WZW model. This is because it includes
Polyakov kinetic terms (\ref{U and V action}) for
the fields $u$, $v$, making these fields dynamical rather than auxiliary. 
One obvious way to eliminate these kinetic terms is given by the choice of null gauging\footnote{
In case of null gauging, $a_1$ and $b_1$ are imaginary-valued, and so are $\alpha$ and $\beta$
parameters of the $SL(2,\mathbb{R})$ group element (\ref{groups parametrizations}).
Gauge-fixing, \textit{e.g.}, $\alpha = \beta = i\, t/2$ then gives Lorentzian time $t$.
In case of time-like gauging, $a_1$, $b_1$, $\alpha$, $\beta$ are real-valued,
but the target space-time metric of the considered gWZW model is Lorentzian.}
\begin{equation}
a_1^2+a_2^2+a_3^2 = 0\,,\qquad
b_1^2+b_2^2+b_3^2 = 0\,.
\end{equation}
We are not going to pursue such a direction in this section. Instead we are going to consider
the time-like gauging
\begin{equation}
\label{time-like gauging}
a_1^2+a_2^2+a_3^2 = 1\,,\qquad
b_1^2+b_2^2+b_3^2 = 1\,,
\end{equation}
that ensures the anomaly-cancellation condition $\textrm{Tr}(P\,T_L^2) =\textrm{Tr}(P\,T_R^2)$,
thereby allowing to eliminate kinetic terms for the fields $u$, $v$ by considering the action
\begin{equation}
\label{gWZW compact}
S_{\textrm{gWZW}}[G] = S'_{\textrm{WZW}}[UGV] - \frac{1}{2\pi}\int d^2z\,\p w\bar\p w\,.
\end{equation}
Here we introduced the $U(1)$-invariant field $w = u - v$ described by the Polyakov action
with the `wrong sign'. In the time-like gauging (\ref{time-like gauging}) one easily derives
\begin{equation}
S'_{\textrm{WZW}}[U]+S'_{\textrm{WZW}}[V] - \frac{1}{2\pi}\int d^2z\,\p w\bar\p w
=\frac{1}{\pi}\int d^2z\,A\tilde A\,,
\end{equation}
where we denoted the auxiliary gauge field potentials as
\begin{equation}
\label{A tilde A in terms of u and v}
A = -\p v\,,\qquad \tilde A = -\bar\p u\,.
\end{equation}
Furthermore, introducing the $U(1)_{L,R}$ gauge currents
\begin{equation}
\begin{aligned}
\label{general currents}
J = k\,\textrm{Tr} \left(P\,T_L\,\p G\, G^{-1}\right)\,,\qquad
\tilde J = k\,\textrm{Tr} \left(P\,T_R\, G^{-1}\,\bar\p G\right)\,,
\end{aligned}
\end{equation}
we can re-write the action (\ref{gWZW compact}) as
\begin{equation}
\label{gWZW result}
S_{\textrm{gWZW}}[G] = S'_{\textrm{WZW}}[G] +\frac{1}{2\pi}\int d^2z\,
\left(A\tilde J+\tilde A J + (M+2)\,A\tilde A\right)\,,
\end{equation}
where we denoted
\begin{equation}
M = k\,\textrm{Tr}\left(P\,T_L\,G\,T_R\,G^{-1}\right)\,.
\end{equation}
Eliminating the non-dynamic fields $A$, $\tilde A$ via e.o.m., we obtain the action
\begin{equation}
\label{gWZW result1}
S_{\textrm{gWZW}}[G] = S'_{\textrm{WZW}}[G] -\frac{1}{2\pi}\int d^2z\,
\frac{J\tilde J}{M+2}\,,
\end{equation}
and a non-trivial dilaton field
\begin{equation}
\label{general dilaton}
\Phi = \Phi_0-\frac{1}{2}\,\log\left(\frac{M}{2}+1\right)\,,
\end{equation}
where $\Phi_0$ is the  dilaton background.

Let us denote
\begin{equation}
\Delta = \frac{M}{2} + 1\,.
\end{equation}
Using the parametrization (\ref{groups parametrizations}) we obtain
\begin{equation}
\label{Delta result}
\Delta = 1+a_1b_1\cosh(2\theta) + a_2b_2\cos(2\theta')+a_3b_3\,.
\end{equation}
The dilaton background (\ref{general dilaton}) can then be determined as
\begin{equation}
\label{dilaton result}
\Phi = \Phi_0-\frac{1}{2}\,\log\left(1+a_1b_1\cosh(2\theta) + a_2b_2\cos(2\theta')+a_3b_3\right)\,.
\end{equation}
For the currents (\ref{general currents}) we obtain
\begin{equation}
\label{J tJ in terms of constituents b}
\begin{aligned}
J= 2\left(-i\frac{a_1}{\sqrt{k}}\,j^3-i\frac{a_2}{\sqrt{k}}j^{\prime 3}+a_3\p x\right)\,,\quad
\tilde J= 2\left(-i\frac{b_1}{\sqrt{k}}\,\tilde j^3-i\frac{b_2}{\sqrt{k}}\tilde j^{\prime 3}+b_3\bar\p x\right)\,,
\end{aligned}
\end{equation}
where we defined (recall that $\sigma^a$ are Pauli matrices)
\begin{equation}
\label{u1 currents of sl2r and su2}
\begin{aligned}
j^3&= i\frac{k}{2}\,\textrm{Tr}\left(\p g g^{-1}\sigma^3\right)\,,\qquad
&&\tilde j^3=i\frac{k}{2}\,\textrm{Tr}\left(g^{-1}\bar \p g \sigma^3\right)\,,\\
j^{\prime 3}&=\frac{k}{2}\,\textrm{Tr}\left(\p g' g^{\prime -1}\sigma^3\right)\,,\qquad
&&\tilde j^{\prime 3}=\frac{k}{2}\,\textrm{Tr}\left(g^{\prime-1}\bar \p g' \sigma^3\right)\,.
\end{aligned}
\end{equation}
In the parametrization (\ref{groups parametrizations}) we obtain
\begin{equation}
\label{u1 currents}
\begin{aligned}
j^3&=i k\,\left(\p\alpha+\cosh(2\theta)\,\p\beta\right)\,,\qquad
&&\tilde j_3=i k\,\left(\bar\p\beta+\cosh(2\theta)\,\bar\p\alpha\right)\,,\\
j^{\prime 3}&=ik\,\left(\p\alpha'+\cos(2\theta')\,\p\beta'\right)\,,\qquad
&&\tilde j_3'=ik\,\left(\bar\p\beta'+\cos(2\theta')\,\bar\p\alpha'\right)\,.
\end{aligned}
\end{equation}
We will also use the notations
\begin{equation}
\label{jx and jw defs}
\begin{aligned}
j^x &= i\,\p x\,,\;\; \tilde j^x = i\, \bar\p x\,,\quad
j^w &=  i\,\p w\,,\;\; \tilde j^w = i\, \bar\p w\,,\quad
j^i &=  i\,\p z^i\,,\;\; \tilde j^i = i\, \bar\p z^i\,.
\end{aligned}
\end{equation}
The current components $j^3$ and $j^{\prime 3}$, and their anti-holomorphic 
counterparts, belong to the $sl(2,\mathbb{R})$ and $su(2)$ Kac-Moody algebras
at levels $k+2$ and $k-2$, satisfying the OPEs\footnote{In our conventions, both
$j^3$ and $j^{\prime 3}$ have real eigenvalues. An alternative convention for the $sl(2,\mathbb{R})$ currents,
related to ours by redefinitions $j^3\rightarrow i j^3$, $j_3\rightarrow - i j_3$ would
result in the $\eta_{ab}=\textrm{diag}\{1,\,1,\,-1\}$ Cartan metric in (\ref{jj jpjp b OPE}) and its fermionic counterpart
(\ref{psi psi ope}).}
\begin{equation}
\label{jj jpjp b OPE}
\begin{aligned}
j^a(z)\, j^b(w) &= \frac{k+2}{2}\,\frac{\delta^{ab}}{(z-w)^2} + \frac{\epsilon^{abc}}{z-w}\,j_c(w)+\dots\,,\\
j^{\prime a}(z)\,j^{\prime b}(w) &=  \frac{k-2}{2}\,\frac{\delta^{ab}}{(z-w)^2} + \frac{i\,\epsilon^{abc}}{z-w}\,j_c'(w)+\dots\,,
\end{aligned}
\end{equation}
where
$\epsilon^{abc}$
is the Levi-Civita symbol, $\epsilon^{123} = 1$,
and similarly for the anti-holomorphic components.
The currents $j^{x}$, $j^i$, and $j^w$ are abelian Kac-Moody currents at levels $1$, $1$, and $-1$ respectively,\footnote{
These follow from the free field OPE
\begin{equation}
\langle \p x(z)\p x(w)\rangle = - \frac{1}{2}\,\frac{1}{(z-w)^2} \,,\qquad
\langle \p w(z)\p w(w)\rangle =  \frac{1}{2}\,\frac{1}{(z-w)^2} \,.
\end{equation}
More comments on the sign of the $\langle \p w\p w\rangle$ two-point
function will be given in section~\ref{sec: coset brst current}.
}
\begin{equation}
\label{jx jx and others opes}
\langle j^x(z)\,j^x(w)\rangle = \frac{1}{2}\,\frac{1}{z-w}\,,\quad
\langle j^i(z)\,j^j(w)\rangle = \frac{1}{2}\,\frac{\delta^{ij}}{z-w}\,,\quad
\langle j^w(z)\,j^w(w)\rangle = - \frac{1}{2}\,\frac{1}{z-w}\,,
\end{equation}
and similarly for the anti-holomorphic sector.

In the supersymmetric theory (\ref{total susy wzw action})
one additionally has contributions to the gauge $U(1)_L\times U(1)_R$ currents (\ref{J tJ in terms of constituents b})
coming from the fermionic sector. The total gauge currents can be compactly reformulated 
in terms of the 
total bosonic currents of supersymmetric Kac-Moody algebra,\footnote{See Appendix~\ref{app:effective wzw action}
for a review.}
\begin{equation}
\J^a = j^a + \frac{1}{k}\,\epsilon^a_{\;\; bc}\,\psi^b\,\psi^c\,,\qquad
\J^{\prime a} = j^{\prime a} - \frac{i}{k}\,\epsilon^a_{\;\; bc}\,\psi^{\prime b}\,\psi^{\prime c}\,,
\end{equation}
that in particular implies
\begin{equation}
\J^3 = j^3+\frac{2}{k}\,\psi^1\,\psi^2\,,\qquad
\J^{\prime 3} = j^{\prime 3}-\frac{2i}{k}\,\psi^{\prime 1}\,\psi^{\prime 2}\,.
\end{equation}
Here the  $sl(2,\mathbb{R})$ and $su(2)$ fermions satisfy the OPEs
\begin{equation}
\label{psi psi ope}
\psi^a(z)\psi^{b}(w) = \frac{k}{2}\,\frac{\delta^{ab}}{z-w}\,,\qquad
\psi^{\prime a}(z)\psi^{\prime b}(w) = \frac{k}{2}\,\frac{\delta^{ab}}{z-w}\,.
\end{equation}
For the total 
bosonic currents,
we have the following OPEs of  the supersymmetric Kac-Moody algebras
$sl(2,\mathbb{R})$ and $su(2)$
\begin{equation}
\label{jj jpjp OPE}
\begin{aligned}
\J^a(z)\, \J^b(w) &= \frac{k}{2}\,\frac{\delta^{ab}}{(z-w)^2} + \frac{\epsilon^{abc}}{z-w}\,\J_c(w)+\dots\,,\\
\J^{\prime a}(z)\,\J^{\prime b}(w) &=  \frac{k}{2}\,\frac{\delta^{ab}}{(z-w)^2} + \frac{i\,\epsilon^{abc}}{z-w}\,\J_c'(w)+\dots\,,
\end{aligned}
\end{equation}
both at level $k$.
The supersymmetric theory counterpart of (\ref{J tJ in terms of constituents b}) is then given by
\begin{equation}
\label{J tJ in terms of constituents}
\begin{aligned}
\mathtt{J}= -2i\left(\frac{a_1}{\sqrt{k}}\,\J^3+\frac{a_2}{\sqrt{k}}\J^{\prime 3}+a_3\,j^ x\right)\,,\quad
\tilde{\mathtt{J}}= -2i\left(\frac{b_1}{\sqrt{k}}\,\tilde{\J}^3+\frac{b_2}{\sqrt{k}}\tilde{\J}^{\prime 3}+b_3\tilde j^ x\right)\,.
\end{aligned}
\end{equation}
Taking into account anomaly-cancellation conditions (\ref{time-like gauging}),
we then obtain
\begin{equation}
\label{JJ cor}
\langle \mathtt{J}(z)\mathtt{J}(w)\rangle = -2\,\frac{1}{(z-w)^2}\,,\qquad
\langle \tilde{\mathtt{J}}(\bar z)\tilde{\mathtt{J}}(\bar w)\rangle = -2\,\frac{1}{(\bar z-\bar w)^2}\,.
\end{equation}

Supersymmetric partners of the gauge currents (\ref{J tJ in terms of constituents})
are given by the  fermions
\begin{equation}
\label{psi tpsi in terms of constituents}
\begin{aligned}
\psi= -2i\left(\frac{a_1}{\sqrt{k}}\,\psi^3+\frac{a_2}{\sqrt{k}}\psi^{\prime 3}+a_3\psi^x\right)\,,\quad
\tilde{\psi}= -2i\left(\frac{b_1}{\sqrt{k}}\,\tilde{\psi}^3+\frac{b_2}{\sqrt{k}}\tilde{\psi}^{\prime 3}+b_3\tilde\psi^x\right)\,,
\end{aligned}
\end{equation}
Using (\ref{psi psi ope}) in combination with the OPEs
\begin{equation}
\label{psi x ope}
\psi^x (z) \psi^x(w) = \frac{1}{2}\,\frac{1}{z-w}\,,\qquad
\tilde \psi^x (\bar z) \tilde \psi^x(\bar w) = \frac{1}{2}\,\frac{1}{\bar z-\bar w}\,,
\end{equation}
we notice that the gauge fermions (\ref{psi tpsi in terms of constituents})
that satisfy the OPEs
\begin{equation}
\label{psipsi cor}
\langle \psi(z)\psi(w)\rangle = -2\,\frac{1}{z-w}\,,\qquad
\langle \tilde{\psi}(\bar z)\tilde{\psi}(\bar w)\rangle = -2\,\frac{1}{\bar z-\bar w}\,.
\end{equation}

\subsection{Supergravity background}
\label{sec:background}

Combining (\ref{WZW action in parametrization}), (\ref{Delta result}), (\ref{J tJ in terms of constituents}),
(\ref{u1 currents}) in (\ref{gWZW result1}) we can write down the full world-sheet gauged WZW action
for the considered parametrization (\ref{groups parametrizations}).
This action is invariant w.r.t. gauge transformation (\ref{u1 transform in parametrization}).
We can fix the associated redundancy in the degrees of freedom by the gauge choice
\begin{equation}
\alpha = \frac{\nu}{2}\,y\,,\qquad \beta = \frac{1}{2}\,y\,,
\end{equation}
where $\nu$ is an arbitrary parameter.
For the sake of calculating supergravity background geometry, we
set all the fermionic d.o.f. to zero.
The resulting action is then given by
\begin{equation}
S_{\textrm{gWZW}}[G] = \frac{1}{2\pi}\int d^2z\,({\cal G}_{MN}+{\cal B}_{MN})\,\p X^M\bar\p X^N\,,
\end{equation}
where we defined the world-sheet field defining coordinates in ten-dimensional target space
\begin{equation}
X^M=(X^\mu,\; z^1,\; z^2, \; z^3,\; z^4)\,,\qquad X^\mu = \left(y,\;\theta,\;\alpha',\;\theta',\;\beta',\;x\right)\,,
\end{equation}
with the non-vanishing components of the metric given by the six-dimensional coset sub-space components
\begin{equation}
\label{G solution}
\begin{aligned}
{\cal G}_{yy} &{=} \frac{k}{4}  \left(1{+}\nu ^2{+}2 \nu  \cosh (2 \theta )
{-}\frac{2 a_1 b_1}{\Delta }\, (\nu{+}\cosh (2 \theta ) ) (1{+}\nu  \cosh (2 \theta ))\right) \,,\\
{\cal G}_{y\alpha'}&={\cal G}_{\alpha'y}=
-\frac{k }{2 \Delta }\,\left(a_1 b_2 \cos (2 \theta') 
(\nu+\cosh (2 \theta ) )+a_2 b_1 (1+\nu  \cosh (2 \theta ))\right)\,,\\
{\cal G}_{y\beta'}&={\cal G}_{\beta'y}=
-\frac{k}{2 \Delta }\, (a_1 b_2 (\nu+\cosh (2 \theta ) )+a_2
b_1 \cos (2 \theta') (1+\nu  \cosh (2 \theta )))\,,\\
{\cal G}_{yx}&={\cal G}_{xy} = -\frac{\sqrt{k} }{2 \Delta }\,( a_3 b_1+a_1 b_3 \nu+(a_1 b_3
+a_3 b_1 \nu )\cosh (2 \theta ) )\,,\\
{\cal G}_{\theta\theta} &={\cal G}_{\theta'\theta'} =k\,,\\
{\cal G}_{\alpha'\alpha'} &= {\cal G}_{\beta'\beta'} =k\left(1-\frac{2}{\Delta }\, a_2 b_2  \cos (2 \theta')\right)\,,\\
{\cal G}_{\alpha'\beta'} &= {\cal G}_{\beta'\alpha'} =\frac{k }{\Delta }\,\left(\cos (2 \theta') (1+a_3 b_3+a_1 b_1
\cosh (2 \theta ))-a_2 b_2 \right)\,,\\
{\cal G}_{\alpha'x} &= {\cal G}_{x\alpha'} =-\frac{\sqrt{k} }{\Delta }\,(a_2 b_3+a_3 b_2 \cos (2 \theta'))\,,\\
{\cal G}_{\beta'x} &= {\cal G}_{x\beta'} =-\frac{\sqrt{k}}{\Delta }\, (a_3 b_2+a_2 b_3 \cos (2 \theta'))\,,\\
{\cal G}_{xx} &=1-\frac{2 a_3 b_3}{\Delta }\,,
\end{aligned}
\end{equation}
as well as the components of the metric on the torus $\mathbb{T}^4$, given by 
${\cal G}_{z^1z^1}={\cal G}_{z^2z^2}={\cal G}_{z^3z^3}={\cal G}_{z^4z^4}=1$.
Non-trivial components of the $B$-field are given by
\begin{equation}
\label{B solution}
\begin{aligned}
{\cal B}_{y\alpha'}=-{\cal B}_{\alpha'y}&=
-\frac{k }{2 \Delta }\,\left(a_1 b_2 \cos (2 \theta') 
(\nu+\cosh (2 \theta ) )-a_2 b_1 (1+\nu  \cosh (2 \theta ))\right)\,,\\
{\cal B}_{y\beta'}=-{\cal B}_{\beta'y}&=
-\frac{k}{2 \Delta }\, (a_1 b_2 (\nu+\cosh (2 \theta ) )-a_2
b_1 \cos (2 \theta') (1+\nu  \cosh (2 \theta )))\,,\\
{\cal B}_{yx}=-{\cal B}_{xy} &= -\frac{\sqrt{k} }{2 \Delta }\,( -a_3 b_1+a_1 b_3 \nu+(a_1 b_3
-a_3 b_1 \nu )\cosh (2 \theta ) )\,,\\
{\cal B}_{\alpha'\beta'} = -{\cal B}_{\beta'\alpha'} &=-k\,\left(\cos(2\theta')+\frac{1}{\Delta}\,
a_2b_2\sin(2\theta')^2\right)\,,\\
{\cal B}_{\alpha'x} = -{\cal B}_{x\alpha'} &=\frac{\sqrt{k} }{\Delta }\,(-a_2 b_3+a_3 b_2 \cos (2 \theta'))\,,\\
{\cal B}_{\beta'x} = -{\cal B}_{x\beta'} &=\frac{\sqrt{k}}{\Delta }\, (a_3 b_2-a_2 b_3 \cos (2 \theta'))\,.
\end{aligned}
\end{equation}
Together with the dilaton (\ref{dilaton result}),
expressions (\ref{G solution}), (\ref{B solution}) represent the classical supergravity background, where
the parameter $\nu$ is arbitrary and no physical quantities are expected to depend on it.
One can explicitly verify that provided the anomaly cancellation
conditions (\ref{time-like gauging}) are satisfied, this background obeys the classical equations of motions
\begin{align}
\label{g eom 1}
R_{MN}+2\nabla_M\p_N\phi  - \frac{1}{4}\,{\cal H}_{MLR}{\cal H}_N^{\;\;LR}&= 0\,,\\
\label{g eom 2}
4(\partial\Phi)^2-4\nabla^M\p_M\Phi-R+\frac{1}{12}{\cal H}_{MNL}{\cal H}^{MNL}&=0\,,\\
\label{g eom 3}
\nabla_L {\cal H}^L_{\;\;MN}-2\p_L\Phi {\cal H}^L_{\;\;MN }&= 0\,,
\end{align}
that can be derived from the corresponding bulk gravity action in the string frame
\begin{equation}
\label{bulk gravity action}
S = \frac{1}{2\kappa_0^2}\,V_{\mathbb{T}^4}\,
\int d^{6}x\,\sqrt{-\det\, {\cal G}}\,e^{-2\Phi}\,\left(R+4(\p\Phi)^2-\frac{1}{12}\,{\cal H}_{\mu\nu\lambda}
{\cal H}^{\mu\nu\lambda}\right)\,,
\end{equation}
where\footnote{Recall that the $B$-field is defined up to gauge transformations
\begin{equation}
\label{B gauge transform}
\delta B_{MN}=\partial_M\Lambda_N-\partial_N\Lambda_M\,,
\end{equation}
that leaves the tensor (\ref{H def}) invariant.}
\begin{equation}
\label{H def}
{\cal H}_{\mu\nu\lambda} = \p_\mu {\cal B}_{\nu\lambda} + \p_\lambda {\cal B}_{\mu\nu}+\p_\nu {\cal B}_{\lambda\mu}\,,
\end{equation}
and the dimensionful volume of the four-dimensional compactified  torus (in  string frame) is given by
(here $R_i$ are dimensionless radii of the circles of
$\mathbb{T}^4$, where $z^i\sim z^i +2\pi \ell_s R_i$, $i=1,\dots,4$)
\begin{equation}
V_{\mathbb{T}^4} = (2\pi\ell_s)^4\,\prod_{i=1}^4 R_i\,.
\end{equation}
The corresponding metric on target space-time is given by
\begin{equation}
ds_{10}^2 = ds_{6}^2+\ell_s^2\,\left((dz^1)^2+\dots+(dz^4)^2\right)\,,
\qquad ds_{6}^2 =\ell_s^2\, {\cal G}_{\mu\nu}\,dX^\mu \, dX^\nu\,,
\end{equation}
where we restored the string length $\ell_s$, taking into account that in our conventions
 the space-time coordinates $X^M$ are dimensionless.
 The metric in Einstein frame,  $\tilde {\cal G}_{MN}$, is related to the metric
 in  string frame, ${\cal G}_{MN}$, via
 \begin{equation}
 \label{Einstein frame metric def}
{\cal G}_{MN} = e^{\frac{\Phi-\Phi_0}{2}}\,  \tilde {\cal G}_{MN} \,.
 \end{equation}
 In  Einstein frame normalization of the gravity action is given by $1/(2\kappa^2)$,
 where
 \begin{equation}
 \label{kappa and kappa0}
 \kappa=\kappa_0e^{\Phi_0}=(8\pi G_{N})^{1/2}\,,
 \end{equation}
 and $G_{N}$
 is the Newton's constant.\footnote{See, \textit{e.g.}, \cite{Polchinski:1998rq}.}
 Notice that due to (\ref{Einstein frame metric def}) the volume of the four-torus in the Einstein frame is given by
 $\tilde V_{\mathbb{T}^4} =e^{\Phi_0-\Phi}\,  V_{\mathbb{T}^4} $.

 To study the background  (\ref{G solution}), (\ref{B solution}),  (\ref{dilaton result}) it is convenient to choose a
 different coordinate frame. Let us begin by rescaling the $U(1)_x$ coordinate as follows,
 \begin{equation}
 \label{def of x prime}
 x = \sqrt{k}\,x'\,.
 \end{equation}
 Then every component of the metric and the $B$-field will be proportional to $k$, and therefore 
 dependence of these tensors on the WZW level can be compactly expressed through an overall common factor.
 Next, we rescale the time coordinate $y$ as follows,\footnote{
 While this choice of the time coordinate can appear \textit{ad hoc} at this point,
 the rationale behind it will be explained below, when we will show that
 in such a frame one obtains adjusted asymptotic value $\hat{\cal G}_{tt}/k = -1$.}
 \begin{equation}
 \label{y to t change}
 y = \frac{2a_1b_1}{a_1-b_1\nu}\,t\,.
 \end{equation}
We also introduce a new radial coordinate $\rho(\theta)$ according to the rule
 (recall that $\Delta(\theta)$ was defined in (\ref{Delta result}))
 \begin{equation}
 \label{theta to rho coordinate transform}
 \Delta = \rho^2 +2a_3b_3\quad\Rightarrow\quad\theta = \frac{1}{2}\,\cosh^{-1}\left(
 1+\frac{\rho^2-\rho_+^2}{a_1b_1}\right)\,,
 \end{equation}
 where we defined
 \begin{equation}
 \label{rho pm def}
 \rho_\pm^2 = 1\pm a_1b_1-a_3b_3+a_2b_2\cos(2\theta')\,.
 \end{equation}
For the choice of parameters
\begin{equation}
\label{positive a1b1}
a_1b_1>0\quad\Leftarrow\quad a_1=\sqrt{1-a_2^2-a_3^2}\,,\quad  b_1=\sqrt{1-b_2^2-b_3^2}
\end{equation}
the coordinate transformation (\ref{theta to rho coordinate transform})
therefore implies the range of validity of the $\rho$ coordinate given by
\begin{equation}
\rho \geq \rho_+\,.
\end{equation}
In the new coordinates we then obtain
the dilaton
\begin{equation}
\Phi = \Phi_0-\frac{1}{2}\,\log(\rho^2+2a_3b_3)\,.
\end{equation}
and the metric
\begin{equation}
\label{gen interval in rho coordinate}
ds_6^2 = \ell_s^2\,\left(\hat {\cal G}_{tt}\,dt^2{+}
{\cal G}_{\rho\rho}\,d\rho^2{+}2{\cal G}_{\rho\theta'}\,d\rho\,d\theta'
{+} {\cal G}_{ab}\,(d X^a{+}R^a\,dt)\,(d X^b{+}R^b\,dt)\right)\,,
\end{equation}
Here we have defined
\begin{equation}
\label{hat X def}
X^a = \left(\alpha',\;\beta',\; x'\right)\,,
\end{equation}
and used ${\cal G}_{ab}$,
$a,b=\alpha',\beta',x'$ to denote projection of the ${\cal G}_{\mu\nu}$
tensor on the three-dimensional sub-space spanned by the vector (\ref{hat X def}).
We have also re-arranged the terms in (\ref{gen interval in rho coordinate}) in such a way that the off-diagonal
metric elements ${\cal G}_{t\alpha'}$, ${\cal G}_{t\beta'}$, ${\cal G}_{tx'}$
are expressed in terms of the vector
\begin{equation}
R^a =  {\cal G}^{ab}\,{\cal G}_{tb}\,,
\end{equation}
where $ {\cal G}^{ab}$ is the inverse metric on the three-dimensional sub-space 
spanned by (\ref{hat X def}).
We then obtain the adjusted metric component
\begin{equation}
\begin{aligned}
\hat {\cal G}_{tt} &= {\cal G}_{tt}- {\cal G}_{ab}\,R^a\,R^b\\
&={\cal G}_{tt} - {\cal G}^{ab}\,{\cal G}_{ta}\,{\cal G} _{tb}\,.
\end{aligned}
\end{equation}
Introducing
\begin{equation}
\ell_\pm^2 = a_2b_2\cos(2\theta') - a_3b_3 \pm \sqrt{(a_2^2+a_3^2)(b_2^2+b_3^2)}
\end{equation}
we can express
\begin{equation}
\hat {\cal G}_{tt} = -k\,\frac{(\rho^2-\rho_+^2)(\rho^2-\rho_-^2)}{(\rho^2-\ell_+^2)(\rho^2-\ell_-^2)}
\end{equation}
We therefore observe that $\rho^2=\rho_+^2$ is the outer horizon of the metric.
Notice that $\rho_+^2 \geq \ell_+^2$, $\rho_-^2\geq \ell_-^2$, and $\ell_+^2\geq \ell_-^2$.\footnote{This
is easiest to see by taking $a_1=\cos\psi$, $b_1 = \cos\psi'$.}
Therefore the apparent singularities of $\hat {\cal G}_{tt}$ at $\rho^2=\ell_\pm^2$
are hidden behind the outer horizon $\rho^2 = \rho_+^2$.

Let us define
\begin{equation}
\begin{aligned}
R^{\alpha'} &= {\cal R}^{\alpha'}-\frac{a_2 b_1\nu}{a_1-b_1\nu}\,,\\
R^{\beta'} &= {\cal R}^{\beta'}-\frac{a_1 b_2}{a_1-b_1\nu}\,,\\
R^{x'} &= {\cal R}^{x'}-\frac{a_1b_3+a_3b_1\nu}{a_1-b_1\nu}\,,
\end{aligned}
\end{equation}
where we separated explicitly asymptotic values at $\rho\rightarrow\infty$, leaving us with ${\cal R}^a = {\cal O}(1/\rho)$,
where
\begin{equation}
\begin{aligned}
 {\cal R}^{\alpha'}&=\frac{a_2\,(b_2^2+b_3^2-a_3b_3+a_2b_2\cos(2\theta')-\rho^2)}{(\rho^2-\ell_+^2)
 (\rho^2-\ell_-^2)}\,,\\
 {\cal R}^{\beta'}&=\frac{b_2\,(-a_2^2-a_3^2+a_3b_3-a_2b_2\cos(2\theta')+\rho^2)}{(\rho^2-\ell_+^2)
 (\rho^2-\ell_-^2)}\,,\\
{\cal R}^{x'}&=\frac{a_3b_2^2-b_3a_2^2+(a_3-b_3)(a_2b_2\cos(2\theta')-2a_3b_3-\rho^2)}{(\rho^2-\ell_+^2)
 (\rho^2-\ell_-^2)}\,,
\end{aligned}
\end{equation}
By performing the change of coordinates
\begin{equation}
\label{moving frame coordinate change}
\begin{aligned}
\alpha' &= \hat\alpha'+\frac{a_2 b_1\nu}{a_1-b_1\nu}\,t\,,\\
\beta' &= \hat\beta'+\frac{a_1 b_2}{a_1-b_1\nu}\,t\,,\\
x' &= \hat x'+\frac{a_1b_3+a_3b_1\nu}{a_1-b_1\nu}\,t\,,
\end{aligned}
\end{equation}
we can remove the off-diagonal $(t,\hat\alpha')$,  $(t,\hat\beta')$,  $(t,\hat x')$
metric elements in the asymptotic region. In the new coordinates we obtain
\begin{equation}
\label{moving gen interval in rho coordinate}
ds_6^2 = \ell_s^2\,\left(\hat {\cal G}_{tt}\,dt^2{+}
{\cal G}_{\rho\rho}\,d\rho^2{+}2{\cal G}_{\rho\theta'}\,d\rho\,d\theta'{+}{\cal G}_{\theta'\theta'}d\theta^{\prime 2}
{+} {\cal G}_{ab}\,(d \hat X^a{+}{\cal R}^a\,dt)\,(d \hat X^b{+}{\cal R}^b\,dt)\right)\,,
\end{equation}
where we defined
\begin{equation}
\label{hat hat X def}
\hat X^a = \left(\hat \alpha',\;\hat \beta',\; \hat x'\right)\,.
\end{equation}
The rest of the components of the metric are given by
\begin{equation}
\begin{aligned}
\frac{1}{k}\,{\cal G} _ {\rho\rho} &= \frac{\rho^2}{(\rho^2-\rho_+^2)(\rho^2-\rho_-^2)}\,,\\
\frac{1}{k}\,{\cal G} _ {\rho\theta} &=\frac{1}{k}\,{\cal G} _ {\theta\rho}=
\frac{\rho a_2b_2\sin(2\theta')}{(\rho^2-\rho_+^2)(\rho^2-\rho_-^2)}\,,\\
\frac{1}{k}\,{\cal G}_{\theta'\theta'} &= 1-\frac{a_2^2b_2^2\sin(2\theta')^2}
{(\rho^2-\rho_+^2)(\rho^2-\rho_-^2)}\,,\\
\frac{1}{k}\,{\cal G}_{\hat\alpha'\hat\alpha'} &=
\frac{1}{k}\, {\cal G}_{\hat\beta'\hat\beta'} = 1-\frac{2a_2b_2\cos(2\theta')}{\rho^2+2a_3b_3}\,,\\
\frac{1}{k}\,{\cal G}_{\hat\alpha'\hat\beta'} &= \frac{1}{k}\,{\cal G}_{\hat\beta'\hat\alpha'} =
\cos(2\theta')-\frac{a_2b_2(3+\cos(4\theta'))}{2(\rho^2+2a_3b_3)}\,,\\
\frac{1}{k}\,{\cal G}_{\hat\alpha'\hat x'} &= \frac{1}{k}\,{\cal G}_{\hat x'\hat\alpha'} =
-\frac{a_2b_3+a_3b_2\cos(2\theta')}{\rho^2+2a_3b_3}\,,\\
\frac{1}{k}\,{\cal G}_{\hat\beta'\hat x'} &=\frac{1}{k}\, {\cal G}_{\hat x'\hat\beta'} =
-\frac{a_3b_2+a_2b_3\cos(2\theta')}{\rho^2+2a_3b_3}\,,\\
\frac{1}{k}\,{\cal G}_{\hat x'\hat x'} &=\frac{\rho^2}{\rho^2+2a_3b_3}\,.\\
\end{aligned}
\end{equation}
Notice that dependence on the arbitrary gauge parameter $\nu$
completely disappeared.
In the asymptotic region $\rho\rightarrow\infty$
we obtain the linear dilaton background times $\mathbb{S}^3\times \mathbb{S}^1$
\begin{equation}
ds_6^2 =k\, \ell_s^2\,\left(-dt^2+\frac{d\rho^2}{\rho^2} +ds_{\mathbb{S}^3}^2 +d\hat x^{\prime 2}\right)\,,
\end{equation}
where the $SU(2)\sim \mathbb{S}^3$ metric is given by\footnote{By performing the change of
coordinates $\hat\alpha' = (\phi_2-\phi_1) / 2$, $\hat\beta' = (\phi_2+\phi_1) / 2$, we can bring this
metric to the Hopf form
\begin{equation}
\label{Hopf metric}
ds_{\mathbb{S}^3}^2 = d\theta^{\prime 2 } + \sin^2(\theta')\,d\phi_1^2 + \cos^2(\theta') \,d\phi_2^2\,.
\end{equation}}
\begin{equation}
ds_{\mathbb{S}^3}^2 = d\hat\alpha^{\prime 2}+d\hat\beta^{\prime 2}+d\theta^{\prime 2}
+2\cos(2\theta')d\hat\alpha'd\hat\beta'\,.
\end{equation}

By picking a new radial coordinate $\chi$ we can 
eliminate the ${\cal G}_{\chi\theta}$ off-diagonal term in the metric tensor. For instance, let us choose
\begin{equation}
\chi = \rho^2 - \rho_+^2 + a_1b_1\,.
\end{equation}
The new radial coordinate in the domain of parameters (\ref{positive a1b1})
is then defined on the interval $\chi \in [\chi_+,+\infty)$, where
\begin{equation}
\chi_\pm = \pm a_1 b_1\,,
\end{equation}
and $\chi_+$ represents the outer black hole horizon.
The interval in the new coordinates is given by
\begin{equation}
\begin{aligned}
\label{interval in chi coordinates}
\frac{1}{k \ell_s^2}\,ds_6^2 &= -\frac{(\chi-\chi_+)(\chi-\chi_-)}{\ell^2}\,dt^2
+\frac{d\chi^2}{4(\chi-\chi_+)(\chi-\chi_-)} +d\theta^{\prime 2}\\
&+\frac{{\cal G}_{ab}}{k}\,(d \hat X^a+{\cal R}^a\,dt)\,(d \hat X^b+{\cal R}^b\,dt)\,,
\end{aligned}
\end{equation}
where we defined
\begin{equation}
\ell^2 = (\chi+1)^2-(1-a_1^2)(1-b_1^2)\,.
\end{equation}
In the new coordinates we can also express
\begin{equation}
\label{cal R components}
\begin{aligned}
{\cal R}^{\hat\alpha'} &=-\frac{a_2\,(\chi+b_1^2)}{\ell^2}\,,\\
{\cal R}^{\hat\beta'} &=\frac{b_2\,(\chi+a_1^2)}{\ell^2}\,,\\
{\cal R}^{\hat x'} &=\frac{a_1^2b_3-b_1^2a_3+(b_3-a_3)\,\chi}{\ell^2}\,,
\end{aligned}
\end{equation}
and\footnote{In Hopf coordinates (\ref{Hopf metric}) we have ${\cal G}_{\phi_1\phi_2} = 0$.}
\begin{equation}
\frac{{\cal G}_{ab}}{k}=\left(
\begin{array}{ccc}
 1-\frac{2 a_2 b_2 \cos (2 \theta')}{\ell_0^2}
  & \frac{\cos (2 \theta') (a_3 b_3+\chi +1)-a_2 b_2}{\ell_0^2}
 & -\frac{a_2 b_3+a_3 b_2 \cos (2 \theta')}{\ell_0^2} \\
 \frac{\cos (2 \theta') (a_3 b_3+\chi +1)-a_2 b_2}{\ell_0^2}
 & 1-\frac{2 a_2 b_2 \cos (2 \theta')}{\ell_0^2} 
 & -\frac{a_2 b_3 \cos (2 \theta')+a_3 b_2}{\ell_0^2} \\
 -\frac{a_2 b_3+a_3 b_2 \cos (2 \theta')}{\ell_0^2}
 & -\frac{a_2 b_3 \cos (2 \theta')+a_3 b_2}{\ell_0^2}
 & 1-\frac{2 a_3 b_3}{\ell_0^2} \\
\end{array}
\right)\,,
\end{equation}
where we denoted
\begin{equation}
\label{ell0 def}
\ell_0^2 = \chi+1+a_3b_3+a_2b_2\cos(2\theta')\,.
\end{equation}
The metric (\ref{interval in chi coordinates}) defines a rotating black hole 
with the event horizon at $\chi\equiv \chi_+$, $\theta' \equiv \textrm{const}$,
$d\hat X^a = -{\cal R}^a dt$, $a=\hat\alpha'$, $\hat\beta'$, $\hat x'$. The 
latter indicates that the angular velocity at the horizon is given by\footnote{Here the factor
of $\sqrt{k}$ originates from the fact that our time coordinate $t$ is normalized
so that the asymptotic behavior of the metric is $ds^2 = -k\,\ell_s^2\,dt^2+...$}
\begin{equation}
\label{Omega in terms of cal R}
\Omega ^ a = -\sqrt{k}\, {\cal R} ^ a |_{\chi = \chi_+}\,,\qquad a=\hat\alpha',\;\hat\beta',\; \hat x'\,.
\end{equation}
Using (\ref{cal R components}) we then obtain
\begin{equation}
\label{result for angular velocities}
\Omega^{\hat\alpha ' } = \frac{\sqrt{k}\, a_2b_1}{a_1+b_1}\,,\qquad
\Omega^{\hat\beta ' } = - \frac{\sqrt{k}\, a_1b_2}{a_1+b_1}\,,\qquad
\Omega^{\hat x ' } =\sqrt{k}\,  \frac{a_3b_1 - a_1b_3}{a_1+b_1}\,.
\end{equation}
These can be further split into the left- and right-moving components
\begin{equation}
\begin{aligned}
\label{result for angular velocities LR}
\Omega^{\hat\alpha ' }_L &= \frac{\sqrt{k}\, a_2b_1}{a_1+b_1}\,,\qquad \Omega^{\hat\alpha ' }_R = 0\,,\qquad
\Omega^{\hat\beta ' } _L=0\,,\qquad \Omega^{\hat\beta ' }_R = - \frac{\sqrt{k}\, a_1b_2}{a_1+b_1}\,,\\
\Omega^{\hat x ' } _L&= \frac{\sqrt{k}\, a_3b_1 }{a_1+b_1}\,,\qquad
\Omega^{\hat x ' } _R= - \frac{ \sqrt{k}\, a_1b_3}{a_1+b_1}\,,
\end{aligned}
\end{equation}
so that
\begin{equation}
\label{result for angular velocities tot}
\Omega^{\hat\alpha ' } = \Omega^{\hat\alpha ' }_L+\Omega^{\hat\alpha ' }_R\,,\qquad
\Omega^{\hat\beta ' } = \Omega^{\hat\beta ' } _L+\Omega^{\hat\beta ' } _R\,,\qquad
\Omega^{\hat x ' } =\Omega^{\hat x ' } _L+\Omega^{\hat x ' } _R\,.
\end{equation}

The full metric can also be written as (recall that $\ell_0$ is given by (\ref{ell0 def}))\footnote{Setting $a_1=\cos\psi$, $a_3=\sin\psi$, $b_1=1$,
$a_2=b_2=b_3=0$,  changing the radial coordinate to $\chi = 2\,e^{2r}\cos\psi-1$, and performing Kaluza-Klein
reduction of the $x$ coordinate we recover the metric of two-dimensional charged black hole
in string frame, $ds^2 = k\ell_s^2
\left(-f(r)dt^2+dr^2/f(r)+ds_{\mathbb{S}^3}^2\right)$, where $f(r) = 1-e^{-2r}\sec\psi+\frac{1}{4}e^{-4r}\tan^2\psi$.
The dilaton is then given by $\Phi = \Phi_0 - \frac{1}{2}\log(2\cos\psi) - r$, in agreement with \cite{Giveon:2003ge}.}
\begin{equation}
\label{full metric result}
\resizebox{.9\hsize}{!}{$\frac{{\cal G}_{\mu\nu}}{k} = \left(
\begin{array}{cccccc}
-1+ \frac{2}{\ell_0^2}
 & 0 & \frac{b_2 \cos (2 \theta')-a_2}{\ell_0^2}
 & 0 & \frac{b_2-a_2 \cos (2 \theta')}{\ell_0^2} & \frac{b_3-a_3}{\ell_0^2} \\
 0 & \frac{1}{4 (\chi ^2- a_1^2 b_1^2)} & 0 & 0 & 0 & 0 \\
 \frac{b_2 \cos (2 \theta')-a_2}{\ell_0^2}
 & 0 & 1-\frac{2 a_2 b_2 \cos (2 \theta')}{\ell_0^2}
 & 0 & \frac{\cos (2 \theta') (a_3 b_3+\chi +1)-a_2 b_2}{\ell_0^2} & -\frac{a_2 b_3+a_3 b_2 \cos (2 \theta')}{\ell_0^2} \\
 0 & 0 & 0 & 1 & 0 & 0 \\
 \frac{b_2-a_2 \cos (2 \theta')}{\ell_0^2} & 0 & \frac{\cos (2 \theta') (a_3 b_3+\chi +1)-a_2 b_2}{\ell_0^2}
 & 0 & 1-\frac{2 a_2 b_2 \cos (2 \theta')}{\ell_0^2}
 & -\frac{a_2 b_3 \cos (2 \theta')+a_3 b_2}{\ell_0^2} \\
 \frac{b_3-a_3}{l^2} & 0 & -\frac{a_2 b_3+a_3 b_2 \cos (2 \theta')}{\ell_0^2}
 & 0 & -\frac{a_2 b_3 \cos (2 \theta')+a_3 b_2}{\ell_0^2} & 1-\frac{2 a_3 b_3}{\ell_0^2} \\
\end{array}
\right)$}
\end{equation}
The dilaton is  given by
\begin{equation}
\label{dilaton result}
\Phi =  \Phi_0 - \log\ell_0\,.
\end{equation}
The Ricci scalar is given by
\begin{equation}
kR = 2\frac{3\chi^2{+}10(1{+}a_3b_3)(\chi{-}a_2b_2\cos(2\theta')){+}7(a_1^2(1{-}b_3^2){-}b_2^2(1{-}a_3^2)){-}3a_2^2b_2^2\cos(2\theta')^2}
{\ell_0^4}\,.
\end{equation}
Finally, for the independent components of the anti-symmetric $B$-field we obtain
\begin{equation}
\label{B field result 1}
\begin{aligned}
\frac{1}{k}\,{\cal B}_{t\hat\alpha'} &=\frac{a_2b_1\nu}{a_1-b_1\nu}+\frac{b_2\cos(2\theta')+a_2}{\ell_0^2}\,,\\
\frac{1}{k}\,{\cal B}_{t\hat\beta'} &= - \frac{a_1b_2}{a_1-b_1\nu}+\frac{a_2\cos(2\theta')+b_2}{\ell_0^2}\,,\\
\frac{1}{k}\,{\cal B}_{t\hat x'} &=\frac{a_3b_1\nu-a_1b_3}{a_1-b_1\nu}+\frac{a_3+b_3}{\ell_0^2}\,,\\
\frac{1}{k}\,{\cal B}_{\hat\alpha'\hat \beta'} &= -\frac{a_2b_2+\cos (2 \theta') (a_3 b_3+\chi +1)}{\ell_0^2}\,,\\
\frac{1}{k}\,{\cal B}_{\hat\alpha'\hat x'} &=
\frac{a_3 b_2\cos (2 \theta')-a_2b_3}{\ell_0^2}\,,\\
\frac{1}{k}\,{\cal B}_{\hat\beta'\hat x'} &=
\frac{a_3b_2-a_2 b_3\cos (2 \theta')}{\ell_0^2}\,.
\end{aligned}
\end{equation}
Notice that the unphysical $\nu$-dependent terms can be readily eliminated using the gauge transformation
$B_{\mu\nu}\rightarrow B_{\mu\nu}+\p_{[\mu}\Lambda_{\nu]}$, for the following choice of transformation parameters
\begin{equation}
\Lambda_{\hat\alpha'} = - k\,\frac{a_2b_1\nu}{a_1-b_1\nu}\,t\,,\quad
\Lambda_{\hat\beta'} =  k \frac{a_1b_2}{a_1-b_1\nu}\,t\,,\quad
\Lambda_{\hat x'} = - k\, \frac{a_3b_1\nu-a_1b_3}{a_1-b_1\nu}\,t\,,
\end{equation}
with the other components being trivial, $\Lambda_i = 0$, $i\neq \hat\alpha',\;\hat\beta',\;\hat x'$.\footnote{In general, adding
constants to the components of the $B$-field cannot affect supergravity equations of motion, that depend on the $B$-field
via the tensor $H=dB$.} Finally, anomaly-free conditions (\ref{positive a1b1}) need to be imposed on the derived expressions
(\ref{full metric result}), (\ref{dilaton result}), (\ref{B field result 1}) for the metric, dilaton, and the $B$-field.

\subsection{Conserved charges}

The black brane background  (\ref{full metric result}), (\ref{dilaton result}), (\ref{B field result 1})
possess symmetries associated with translations in time $t$, angular coordinates 
$\hat\alpha'$, $\hat\beta'$ of the $\mathbb{S}^3$ and compact coordinate $\hat x'$
of the $\mathbb{S}^1$. In this subsection, we are going to derive the corresponding
expressions for the conserved energy and angular momenta.

Angular momenta components are defined via the $1$-forms
\begin{equation}
k^{G(a)} = {\cal G}_{a\mu}\,dX^\mu\,,\qquad
k^{B(a)} = {\cal B}_{a\mu}\,dX^\mu\,,\qquad 
a = \hat\alpha',\;\hat\beta',\;\hat x'\,,
\end{equation}
by the Komar formula (see, e.g., \cite{Barnich:2001jy,Straumann:2013spu}) in string frame
\begin{equation}
\label{Komar}
{\cal J}_a^{(G,B)} = \frac{1}{\kappa_0^2}\,V_{\mathbb{T}^4}\,\ell_s^3
\,\int \star (dk^{G,B(a)})\,e^{-2\phi} \,, \qquad a = \hat\alpha',\;\hat\beta',\;\hat x'\,,
\end{equation}
as the linear combinations
\begin{equation}
{\cal J}_a^{(L,R)} =\frac{{\cal J}_a^{(G)}\pm {\cal J}_a^{(B)}  }{2}\,.
\end{equation}
In (\ref{Komar}) the star denotes the usual Hodge dual of a $p$-form,
\begin{align}
dk^{G,B(a)} &= k^{G,B(a)}_{bc} \,d\hat X^b\wedge  d\hat X^c\,,\; k^{G,B(a)}_{bc} =\p_{[b} k^{G,B(a)}_{c]} 
=\frac{1}{2}\left( \p_{b} ({\cal G},{\cal B})_{ca}
-\p_{c}  ({\cal G},{\cal B})_{ba}\right)\,,\\
\star (dk^{G,B(a)}) &= \frac{1}{4!}\sqrt{\det {\cal G}}\,\epsilon_{a_1a_2a_3a_4a_5a_6}\,{\cal G}^{a_5b_5}
{\cal G}^{a_6b_6}\,k^{(a)G,B}_{b_5b_6} dX^{a_1}\wedge dX^{a_2}\wedge dX^{a_3}\wedge dX^{a_4}\,,\notag
\end{align}
and $\epsilon_{\tau\chi\hat\alpha'\theta'\hat\beta'\hat x'} = 1$.
In (\ref{Komar}) the integral is performed over the asymptotic boundary $\mathbb{S}^3\times\mathbb{S}^1$,
which selects the corresponding component $(\hat\alpha'\theta'\hat\beta'\hat x ')$ of the $4$-form $\star (dk_a)$.
We then obtain
\begin{equation}
\label{J momenta result}
\begin{aligned}
{\cal J}_{\hat \alpha ' }^G &={\cal J}_{\hat \alpha ' }^B
= \frac{4\pi^2}{\kappa^2}\, k\,\ell_s^2 \,V_{\mathbb{T}^4}\,V_{\mathbb{S}^1}\, a_2\,,\\
{\cal J}_{\hat \beta ' } ^G&= -{\cal J}_{\hat \beta ' } ^B= 
 - \frac{4\pi^2}{\kappa^2}\, k\,\ell_s^2\,V_{\mathbb{T}^4}\,V_{\mathbb{S}^1}\,b_2\,,\\
{\cal J}_{\hat x ' } ^G&= \frac{4\pi^2}{\kappa^2}\, k\,\ell_s^2\,V_{\mathbb{T}^4}\,V_{\mathbb{S}^1}\,(a_3-b_3)\,,
\qquad {\cal J}_{\hat x ' } ^B&= \frac{4\pi^2}{\kappa^2}\, k\,\ell_s^2\,V_{\mathbb{T}^4}\,V_{\mathbb{S}^1}\,(a_3+b_3)\,.
\end{aligned}
\end{equation}
For the energy/mass we  can analogously use the Komar formula
\begin{equation}
{\cal E}= - \frac{1}{2\kappa_0^2}\,V_{\mathbb{T}^4}\,\ell_s^3
\,\int \star (dk^{G(t)})\,e^{-2\phi}\,,
\end{equation}
where $ k^{G(t)} = {\cal G}_{t\mu}\,dX^\mu$, which gives\footnote{
Here the factor of $\sqrt{k}$ is due to our choice of normalization of the time coordinate,
which is dual to the energy ${\cal E}$, according to which $ds^2 = -k\,\ell_s^2\,dt^2+\dots$
in the asymptotic region.}\; \footnote{Recall that relation between $\kappa$ and $\kappa_0$
is given by (\ref{kappa and kappa0}).}
\begin{equation}
\label{energy gen result}
{\cal E}= \frac{4\pi^2}{\kappa^2}\, k^\frac{3}{2}\,\ell_s^2 \,V_{\mathbb{T}^4}\,V_{\mathbb{S}^1}\,.
\end{equation}
In the next section, we will demonstrate that the derived expressions for energy and angular
momenta are consistent with the first law of thermodynamics.

\section{Thermodynamics}
\label{sec:thermodynamics}

In this section, we are going to study thermodynamics of the 
background (\ref{full metric result}), (\ref{dilaton result}), (\ref{B field result 1}).
This background describes a black brane, rotating
with non-vanishing angular velocities along two of the $\mathbb{S}^3$ 
as well as the $\mathbb{S}^1$ directions. We are going to determine
the temperature and derive the associated
Bekenstein-Hawking entropy. Using expressions for the energy,
angular velocities, and angular momenta calculated in the previous section,
we will derive the free energy, and demonstrate explicitly the first law of thermodynamics.
Additionally, by explicit calculation of the regularized on-shell action, we will demonstrate
that the free energy vanishes, in agreement with the general
expectation for an asymptotically linear dilaton space-time.

The near-horizon limit of the metric (\ref{interval in chi coordinates}) can be taken by expanding
\begin{equation}
\chi = \chi_+\,(1+r^2)\,.
\end{equation}
around $r=0$.
This gives
\begin{equation}
ds_6^2 {=}\ell_s^2\left(k\left( \frac{1}{2}
\left(dr^2{-}\frac{4a_1^2b_1^2}{(a_1+b_1)^2} r^2dt^2\right) {+} d\theta^{\prime 2}\right)
{+} {\cal G}_{ab}\,(d \hat X^a{+}{\cal R}^a\,dt)\,(d \hat X^b{+}{\cal R}^b\,dt)\right)\,.
\end{equation}
The null interval at the horizon is then defined by
\begin{equation}
\label{horizon conditions}
\chi = \chi_+\,,\qquad d\theta' = 0\,,\qquad d\hat X^a = -{\cal R}^adt\,,
\end{equation}
while regularity at the horizon requires that Euclidean time has the period expressed in terms of the black hole temperature
as $\tau\sim \tau + 1/T$. Recall that our Lorentzian time coordinate $t$ is normalized
so that the asymptotic metric at the boundary is determined by $ds^2 = -k\,\ell_s^2\,dt^2+\dots$.
Then the temperature is given by
\begin{equation}
\label{general T result}
T = \frac{1}{\pi\ell_s\sqrt{k}}\,\frac{a_1b_1}{a_1+b_1}\,.
\end{equation}
The temperature (\ref{general T result}) takes values in the range $0\leq T\leq T_H$,
 where $T_H =\frac{1}{2\pi\sqrt{k}\,\ell_s}$ is the Hagedorn temperature of two-dimensional black hole.
The zero-temperature limit, $T=0$, is achieved for the choice of parameters $a_1b_1 = 0$, in which
case we obtain $\chi_+=\chi_-$. In the limit $a_1\rightarrow 1$, $b_1\rightarrow 1$ we get $T\rightarrow T_H$, which
corresponds to sending the outer horizon to asymptotic linear dilaton region of space-time.

At the outer horizon we obtain for the Einstein frame metric
\begin{equation}
\tilde V_{\mathbb{T}^4}\,\sqrt{\det ||\tilde {\cal G}_{ij}||} = e^{2(\Phi_0-\Phi)}\,V_{\mathbb{T}^4}\,
\sqrt{\det ||{\cal G}_{ij}||}\,\Bigg|_{\chi = \chi_+}
= k^2\ell_s^4\,(a_1+b_1)\,V_{\mathbb{T}^4}\,\sin(2\theta')\,,
\end{equation}
where $i,j=\hat\alpha',\theta',\hat\beta',\hat x'$,  and ${\cal G}_{ij}$ is  projection of the metric on the
associated sub-space, and we substituted expression (\ref{dilaton result}) for the dilaton.
The horizon area, obtained by integrating
over $\theta'\in \left[0,\frac{\pi}{2}\right]$, $\alpha'\in \left[0,\pi\right]$, $\beta'\in \left[0,2\pi\right]$
is therefore given by
\begin{equation}
\label{area of hoirzon}
{\cal A} = 2\pi^2\, k^2\ell_s^3\,(a_1+b_1)\,V_{\mathbb{S}^1}\,V_{\mathbb{T}^4}\,,
\end{equation}
where we also denoted dimensionful length of the $U(1)_x$ circle (in  string frame) as
\begin{equation}
V_{\mathbb{S}^1} = 2\pi \ell_sR_x\,,
\end{equation}
following from the periodicity condition for the dimensionless coordinate  $x\sim x+2\pi R_x$.
The Bekenstein-Hawking entropy is therefore given by\footnote{
One can parametrize
the asymmetric gauging of the $U(1)$ sub-group as (see \cite{Chakraborty:2020yka} and references therein)
\begin{equation}
\label{rotating string parameters}
a_1 = \cos(\chi-\psi)\,,\quad  b_1 = \cos\chi\,,\quad  a_2=b_2 = 0\,,
\end{equation}
\textit{i.e.}, decoupling the $SU(2)$ sub-group.
Taking the same limit in (\ref{BH entropy result}), we can compare it with the entropy of rotating black
string following from the expressions for the metric ${\cal G}_{xx}$ and dilaton $\Phi-\Phi_0$
given by eq. (3.36) in  \cite{Chakraborty:2020yka}. It suffices to point out that the horizon area
is given by
\begin{equation}
{\cal A} = e^{2(\Phi_0-\Phi)}\,2\pi^2 \,k^2\ell_s^3\,V_{\mathbb{T}^4} \,V_{\mathbb{S}^1}\,\sqrt{{\cal G}_{xx}}
=\sqrt{(1+\cos\psi)(1+\cos(2\chi-\psi))}\,,
\end{equation}
that after a simple manipulation one can see to coincide with (\ref{area of hoirzon})
for the parameters (\ref{rotating string parameters}).
}
\begin{equation}
\label{BH entropy result}
{\cal S} = \frac{{\cal A}}{4G_{N}}= \frac{2\pi{\cal A}}{\kappa^2} = \frac{4\pi^2}{\kappa^2}\,
\frac{1}{T}\,k^\frac{3}{2}\ell_s^2\,V_{\mathbb{S}^1}\,V_{\mathbb{T}^4}\,a_1b_1\,.
\end{equation}

Now let us proceed to calculation of the Euclidean on-shell action
\begin{equation}
S = S_0+S_{\textrm{GHY}}+S_{\textrm{c.t.}}\,,
\end{equation}
where $S_0$ is obtained by plugging in the background solution into
the bulk action (\ref{bulk gravity action}),  $S_{\textrm{GHY}}$ is the Gibbons-Hawking-York (GHY) boundary term,
and $S_{\textrm{c.t.}}$ is the counter-term action.
Let us start with the former. Using (\ref{g eom 2}) we can re-write (\ref{bulk gravity action})
on shell as\footnote{In the second line we have applied $\sqrt{\det {\cal G}}\,\nabla_\mu A^\mu
=\partial_\mu \left(\sqrt{\det {\cal G}}\,A^\mu\right)$ to $A^\mu=\p^\mu(e^{-2\Phi})$.}
\begin{equation}
\begin{aligned}
S_0 &= -\frac{2}{\kappa_0^2} \,\ell_s^4\,V_{\mathbb{T}^4}\,\int d^{6}x\, \sqrt{\det{\cal G}}\,e^{-2\Phi}\,
\left(\nabla^\mu\partial_\mu\Phi - 2\,\p_\mu\Phi \p^\mu\Phi\right)\\
&= -\frac{2}{\kappa_0^2} \,\ell_s^4\,V_{\mathbb{T}^4}\int d^{6}x\, \p_\mu \left(\sqrt{\det {\cal G}}\,e^{-2\Phi}\,\p^\mu\Phi\right)\,.
\end{aligned}
\end{equation}
Using the Stokes's theorem, we obtain
\begin{equation}
\label{S0 onshell prelim}
S_0 = -\frac{2}{\kappa_0^2} \,\ell_s^4\,V_{\mathbb{T}^4}\,\int d^5x\, \sqrt{\det \gamma}\,e^{-2\Phi}\,n_\mu\,\p^\mu\Phi\,,
\end{equation}
where
\begin{equation}
\label{def of boundary metric gamma}
\gamma_{ij} = {\cal G}_{ij}\,, \qquad i,j=t,\hat \alpha',\theta',\hat \beta',\hat x'\,,
\end{equation}
is the metric on the boundary $\chi = \Lambda$,\footnote{The corresponding contribution
from the lower limit of integration $\chi = \chi_+$ vanishes.} and $n_\mu$, given by (\ref{def of normal n}),
is the outward-pointing unit vector normal to the boundary.
We can then re-write (\ref{S0 onshell prelim}) as
\begin{equation}
\label{F0 prelim}
S _ 0 = - \frac{2}{\kappa_0^2} \,\ell_s^4\,V_{\mathbb{T}^4}\,\int _0^{1/(\sqrt{k}\,\ell_s\,T)}d\tau\int d^4x\,\sqrt{\det {\cal G}}
\,e^{-2\Phi}\,{\cal G}^{\chi\chi}\,\p_\chi\Phi\Bigg|_{\chi = \Lambda}\,,
\end{equation}
where, as before, $\det {\cal G}$ is determinant of the six-dimensional metric.
Using (\ref{full metric result}), (\ref{dilaton result}), we obtain for any $\chi$,
\begin{equation}
 \sqrt{\det{\cal G}}\,e^{-2\Phi} = \frac{k^3}{2}\,\sin(2\theta')\,e^{-2\Phi_0} \,.
\end{equation}
Combining this with
\begin{equation}
{\cal G}^{\chi\chi}\,\p_\chi\Phi|_{\chi = \Lambda} = -2\Lambda+2(1+a_2b_2\cos(2\theta')+a_3b_3)+{\cal O}
\left(\frac{1}{\Lambda}\right)\,,
\end{equation}
and integrating over $\theta'\in [0,\pi/2]$, etc., we  obtain
\begin{equation}
\label{F0 divergent}
S_0  = \frac{4\pi^2}{\kappa^2}\,k^\frac{3}{2}\ell_s^2\,\frac{1}{T}\,V_{\mathbb{S}^1}\,V_{\mathbb{T}^4}\,
\left(\Lambda - 1 - a_3b_3\right)\,.
\end{equation}
We can remove the divergence with the following counter-term,
\begin{equation}
\label{Sct}
S_{\textrm{c.t.}} = -\frac{2}{\kappa_0^2}\,\ell_s^4\,V_{\mathbb{T}^4}\,
\int _{\chi = \Lambda}d^5x\,\sqrt{\det\gamma}\,e^{-2\Phi}
=- \frac{4\pi^2}{\kappa^2}\,k^\frac{3}{2}\ell_s^2\,\frac{1}{T}\,V_{\mathbb{S}^1}\,V_{\mathbb{T}^4}\,\Lambda
\end{equation}
without affecting the finite part.

The GHY term at the boundary $\chi=\Lambda$ in  string frame is given by (\ref{GHY in string frame}),
\begin{equation}
\label{GHY gen def}
S_{\textrm{GHY}} = \frac{1}{\kappa_0^2}\,\ell_s^4\,V_{\mathbb{T}^4}\,
\int _{\chi = \Lambda}d^5x\, \sqrt{\det \gamma}\,e^{-2\Phi}\, K\,.
\end{equation}
In (\ref{GHY gen def}) we used the induced metric (\ref{def of boundary metric gamma}) on the boundary
$\chi = \Lambda$.
At the same time, $K = \nabla_\mu  n^\mu$
is trace of the extrinsic curvature tensor, 
 defined by the normalized outward-pointing unit vector
(\ref{def of normal n}) normal to the surface $\chi = \Lambda$. 
We then evaluate
\begin{equation}
\sqrt{\det \gamma}\, K =2k^2
\,\left(\chi -\frac{(\chi-\chi_+)(\chi-\chi_-)}
{\ell_0^2}\right)\,\sin(2\theta')\,,
\end{equation}
that finally gives
\begin{equation}
S_{\textrm{GHY}} = \frac{4\pi^2}{\kappa^2}\,k^\frac{3}{2}\ell_s^2\,\frac{1}{T}\,V_{\mathbb{T}^4}\,V_{\mathbb{S}^1}
\,(1+a_3b_3)\,.\label{GHY asymptotic}
\end{equation}

The free energy
\begin{equation}
{\cal F} = T\,(S_0+S_{\textrm{GHY}}+S_{\textrm{c.t.}})
\end{equation}
can be calculated by combing
 (\ref{F0 divergent}), (\ref{Sct}), (\ref{GHY asymptotic}), giving
\begin{equation}
\label{vanishing grand potential}
 {\cal F}= 0\,,
\end{equation}
consistently with the Hagedorn growth of density of states, and
in agreement with analogous result for the two-dimensional charged black hole \cite{Kutasov:1991pv,Kutasov:2000jp,Giveon:2005jv}.

On the other hand, the free energy is given by 
\begin{equation}
\label{F formula}
{\cal F} ={\cal E} -  T\,{\cal S} 
 -\Omega\cdot{\cal J}\,,
\end{equation}
where we introduced short-hand notation
\begin{equation}
\Omega\cdot{\cal J} = 
 \Omega^{\hat\alpha'}_L\,{\cal J}_{\hat\alpha'}^L
+\Omega^{\hat\alpha'}_R\,{\cal J}_{\hat\alpha'}^R
+ \Omega^{\hat\beta'}_L\,{\cal J}_{\hat\beta'}^L
+ \Omega^{\hat\beta'}_R\,{\cal J}_{\hat\beta'}^R
+ \Omega^{\hat x'}_L\,{\cal J}_{\hat x'}^L
+ \Omega^{\hat x'}_R\,{\cal J}_{\hat x'}^R\,,
\end{equation}
where components of the angular velocity have been derived in (\ref{result for angular velocities LR}),
and angular momenta are given by (\ref{J momenta result}).
Using these expressions and taking into account the anomaly-cancellation constraints, we derive
\begin{equation} 
\label{OmegaJ gen result}
\Omega\cdot{\cal J} = 
 \frac{4\pi^2}{\kappa^2}\, k^\frac{3}{2}\,\ell_s^2 \,V_{\mathbb{T}^4}\,V_{\mathbb{S}^1}\, 
 \left(1-a_1b_1\right)\,.
\end{equation}
Combining expressions
for the entropy (\ref{BH entropy result}), angular velocities and momenta (\ref{OmegaJ gen result}),
and mass (\ref{energy gen result})
in the expression for the free energy (\ref{F formula}) we can explicitly reproduce that the free energy vanishes, 
in agreement with (\ref{vanishing grand potential}).

We can also verify that the first law of thermodynamics is satisfied,
\begin{equation}
d{\cal E} -  T\,d{\cal S} - \Omega^{\hat\alpha'}_L\,d{\cal J}_{\hat\alpha'}^L
- \Omega^{\hat\beta'}_R\,d{\cal J}_{\hat\beta'}^R
- \Omega^{\hat x'}_L\,d{\cal J}_{\hat x'}^L
- \Omega^{\hat x'}_R\,d{\cal J}_{\hat x'}^R = 0\,.
\end{equation}
This identity can be demonstrated for arbitrary $da_{1,2}$, $db_{1,2}$ by 
substituting
$\Omega^{\hat\alpha'}_R =  \Omega^{\hat\beta'}_L = 0$,
using $d{\cal E} = 0$, and taking into account that anomaly cancellation conditions imply
\begin{equation}
da_3 = -\frac{a_1da_1+a_2da_2}{a_3}\,,\qquad
db_3 = -\frac{b_1db_1+b_2db_2}{b_3}\,.
\end{equation}

\section{Two-charge background}
\label{sec:two charge}

In section~\ref{sec:background}, we derived
the metric, dilaton, and $B$-field ten-dimensional background configuration
(\ref{full metric result}),  (\ref{dilaton result}), (\ref{B field result 1})
for string theory on the coset space (\ref{big coset}).
This background possesses an event horizon, that rotates
with the angular velocities given by (\ref{result for angular velocities})
in the directions of $\hat\alpha'$, $\hat\beta'$, $\hat x'$. It asymptotes
to a static linear dilaton two-dimensional space-time times $\mathbb{S}^3\times\mathbb{S}^1\times\mathbb{T}^4$.

In this section, we are going to construct the nine-dimensional background obtained
from (\ref{full metric result}),  (\ref{dilaton result}), (\ref{B field result 1}) by
compactifying the $U(1)_x$ coordinate $\hat x'$. To be precise, due to
the non-vanishing angular velocity $\Omega^{\hat x'}$
at the horizon, we first need to perform a change of coordinates
$\hat x'\rightarrow \tilde x'$, defined by\footnote{An extra factor of $1/\sqrt{k}$
is due to our conventions of separating an explicit common factor of $k$
in expression for the interval $ds^2$.}
\begin{equation}
\hat x' = \tilde x' + \Omega^{\hat x'}\,t /\sqrt{k}\,,
\end{equation}
such that 
the new coordinate $\tilde x'$ is not rotating at the horizon. By performing
the KK reduction of $\tilde x'$, from the ten-dimensional background ${\cal G}_{MN}$, ${\cal B}_{MN}$, $\Phi$,
we can derive the nine-dimensional background 
\begin{equation}
\label{9d background}
\begin{aligned}
\mathrm{G}_{mn} &= {\cal G}_{mn} - \frac{{\cal G}_{\tilde x' m}\,{\cal G}_{\tilde x' n}}{{\cal G}_{\tilde x'\tilde x'}}\,,\qquad
&&\mathrm{B}_{mn} = {\cal B}_{mn}\,,\qquad \Phi_9 = \Phi - \frac{1}{4}\log({\cal G}_{\tilde x'\tilde x'})\,,\\
\mathrm{A}_{m} &=   \frac{{\cal G}_{\tilde x' m}}{{\cal G}_{\tilde x'\tilde x'}}\,,\qquad
&&\mathrm{B}_{m} = \frac{1}{k}\,{\cal B}_{\tilde x'm}\,,
\end{aligned}
\end{equation}
where we chose indices $m$, $n$ to label coordinates of the nine-dimensional non-compact background.
This background features two $U(1)$ gauge fields $\mathrm{A}_{m}$, $\mathrm{B}_{m}$,
that are sourced by the corresponding charges $Q_A$, $Q_B$ at the chemical
potentials $\mu_A$, $\mu_B$ that we will calculate below.

We can write the five-dimensional coset sub-space metric of the nine-dimensional
background in the form analogous to (\ref{interval in chi coordinates}),
\begin{equation}
\begin{aligned}
\label{5d interval in chi coordinates}
\frac{1}{k \ell_s^2}\,ds_5^2 &= -\frac{(\chi-\chi_+)(\chi-\chi_-)}{\ell^2}\,dt^2
+\frac{d\chi^2}{4(\chi-\chi_+)(\chi-\chi_-)} +d\theta^{\prime 2}\\
&+\frac{\mathrm{G}_{rs}}{k}\,(d \tilde X^r+\tilde{\cal R}^r\,dt)\,(d \tilde X^s+\tilde{\cal R}^s\,dt)\,,
\end{aligned}
\end{equation}
where indices $r,s$ take values $\hat\alpha',\hat\beta'$,
the two-dimensional metric $\mathrm{G}_{rs}$ is given by
\begin{equation}
\begin{aligned}
\label{G alpha beta 2d}
\mathrm{G}_{\hat\alpha'\hat\alpha'} &=
\frac{\ell^2+b_2^2(1-a_1^2)\sin(2\theta')^2}{\ell_0^2(\ell_0^2-2a_3b_3)}\,,\\
\mathrm{G}_{\hat\alpha'\hat\beta'} &=
\frac{\ell^2\cos(2\theta')
-a_2b_2(\chi+1)\sin(2\theta')^2}{\ell_0^2(\ell_0^2-2a_3b_3)}\,,\\
\mathrm{G}_{\hat\beta'\hat\beta'} &=
\frac{\ell^2+a_2^2(1-b_1^2)\sin(2\theta')^2}{\ell_0^2(\ell_0^2-2a_3b_3)}\,,
\end{aligned}
\end{equation}
and we have defined
\begin{equation}
\tilde{\cal R}^{\hat\alpha'} = - \frac{a_2(\chi+b_1^2)}{\ell^2}\,,\qquad
\tilde{\cal R}^{\hat\beta'} =  \frac{b_2(\chi+a_1^2)}{\ell^2}\,.
\end{equation}
In the asymptotic region $\chi\rightarrow \infty$ we have a static background, $\tilde{\cal R}^{r} = {\cal O}(1/\chi)$,
while the metric $\mathrm{G}_{rs}$ approaches 
\begin{equation}
\mathrm{G}_{rs} = 
\left(
\begin{array}{cc}
 1 & \cos (2\theta') \\
 \cos (2 \theta') & 1 \\
\end{array}
\right)+{\cal O}\left(\frac{1}{\chi}\right)\,.
\end{equation}
The nine-dimensional
background then asymptotes
to a static linear dilaton two-dimensional space-time times $\mathbb{S}^3\times\mathbb{T}^4$.

We can calculate angular velocities at the horizon according to $\Omega^r = -\sqrt{k}\,\tilde{\cal R}^r|_{\chi = \chi_+}$,
$r=\hat\alpha',\hat\beta'$, analogously to (\ref{Omega in terms of cal R}), yielding
\begin{equation}
\label{result for angular velocities r}
\Omega^{\hat\alpha ' } = \frac{\sqrt{k}\, a_2b_1}{a_1+b_1}\,,\qquad
\Omega^{\hat\beta ' } = - \frac{\sqrt{k}\, a_1b_2}{a_1+b_1}\,.
\end{equation}
in agreement with (\ref{result for angular velocities}).
We can determine the corresponding angular momenta ${\cal J}_r^{(G,B)}$, $r=\hat\alpha',\hat\beta'$
using (\ref{Komar}) applied to the nine-dimensional background, rendering
\begin{equation}
\begin{aligned}
{\cal J}_{\hat \alpha ' }^G &={\cal J}_{\hat \alpha ' }^B
= \frac{4\pi^2}{\kappa^2}\, k\,\ell_s^2 \,V_{\mathbb{T}^4}\,a_2\,,\\
{\cal J}_{\hat \beta ' } ^G&= -{\cal J}_{\hat \beta ' } ^B= 
 - \frac{4\pi^2}{\kappa^2}\, k\,\ell_s^2\,V_{\mathbb{T}^4}\,b_2\,,
\end{aligned}
\end{equation}
in agreement with (\ref{J momenta result}).

One can see that at the horizon the time-like components of the gauge fields in co-rotating
reference frame vanish,
$\tilde{\mathrm{A}}_{t}|_{\chi = \chi_+}=0$, $\tilde{\mathrm{B}}_{t} |_{\chi = \chi_+} = 0$, up to a gauge transformation.
Recall that since the horizon is rotating in the directions of $\hat\alpha'$, $\hat\beta'$,
we actually need to calculate gauge fields
in the co-rotating coordinates 
\begin{equation}
\hat \alpha' = \tilde \alpha' + \Omega^{\hat \alpha'}\,t /\sqrt{k}\,,\qquad
\hat \beta' = \tilde \beta' + \Omega^{\hat \beta'}\,t /\sqrt{k}\,,
\end{equation}
resulting in the transformation laws
\begin{equation}
\tilde{\mathrm{A}}_t = \mathrm{A}_t+\Omega^{\hat \alpha'}\,A_{\hat\alpha'}+\Omega^{\hat \beta'}\,A_{\hat\beta'}\,,\qquad
\tilde{\mathrm{B}}_t = \mathrm{B}_t+\Omega^{\hat \alpha'}\,B_{\hat\alpha'}+\Omega^{\hat \beta'}\,B_{\hat\beta'}\,.
\end{equation}
This gives
\begin{equation}
\begin{aligned}
\tilde{\mathrm{A}}_t &= \frac{(a_3b_1-a_1b_3)(\chi-\chi_+)}{(a_1+b_1)(\ell_0^2-2a_3b_3)}\,,\\
\tilde{\mathrm{B}}_t &= \frac{(a_3b_1+a_1b_3)(\chi-\chi_+)}{(a_1+b_1)\ell_0^2}\,,
\end{aligned}
\end{equation}
where we have also performed a constant shift
\begin{equation}
\tilde{\mathrm{B}}_t \rightarrow \tilde{\mathrm{B}}_t  + \frac{a_3b_1+a_1b_3}{a_1+b_1}\,.
\end{equation}
The remaining non-trivial components of the gauge fields are given by
\begin{equation}
\begin{aligned}
\tilde{\mathrm{A}}_{\hat\alpha'} &= -\frac{a_2b_3+a_3b_2\cos(2\theta')}{\ell_0^2-2a_3b_3}\,,\qquad
\tilde{\mathrm{A}}_{\hat\beta'} = -\frac{a_3b_2+a_2b_3\cos(2\theta')}{\ell_0^2-2a_3b_3}\,,\\
\tilde{\mathrm{B}}_{\hat\alpha'} &= \frac{a_2b_3-a_3b_2\cos(2\theta')}{\ell_0^2}\,,\qquad
\tilde{\mathrm{B}}_{\hat\beta'} = \frac{-a_3b_2+a_2b_3\cos(2\theta')}{\ell_0^2}\,.
\end{aligned}
\end{equation}

From asymptotic behavior of the gauge potentials we can read off the chemical potential
\begin{equation}
\mu_A =\sqrt{k}\,   \frac{a_3b_1-a_1b_3}{a_1+b_1}\,,\qquad
\mu_B =  \sqrt{k}\, \frac{a_3b_1+a_1b_3}{a_1+b_1}\,,
\end{equation}
that we can subsequently split into the left-moving and right-moving components as
\begin{equation}
\label{mu L R}
\mu_L = \frac{\mu_A +\mu_B}{2} =\sqrt{k}\,  \frac{a_3b_1}{a_1+b_1}\,,\qquad
\mu_R = \frac{\mu_A -\mu_B}{2} = - \sqrt{k}\,  \frac{a_1b_3}{a_1+b_1}\,.
\end{equation}
Expressions for the chemical potential (\ref{mu L R})
agree with their angular velocities counterparts $\Omega^{\hat x'}_{L,R}$
before compactification, given by (\ref{result for angular velocities LR}),
\begin{equation}
\mu_{L,R} = \Omega^{\hat x'}_{L,R}\,.
\end{equation}
The corresponding charges $Q_{L,R}$ can be found analogously to \cite{Giveon:2005mi,McGuigan:1991qp}.
We first calculate
\begin{equation}
Q_{G,B} = - \frac{1}{\kappa_0^2}\,\frac{2\pi^2\ell_s^2}{\sqrt{k}}\,V_{\mathbb{T}^4}\,V_{\mathbb{S}^1}
\lim_{\chi\rightarrow\infty} \sqrt{-\det\mathrm{G}}\, e^{-2\Phi_9}\, {\cal G}_{\hat x'\hat x'}\,
F_{G,B}^{\chi t}\,,
\end{equation}
where the field strength tensors are defined as $F_{G}^{rs} = \p^r\mathrm{A}^s-\p^s\mathrm{A}^r$,
 $F_{B}^{rs} = \p^r\mathrm{B}^s-\p^s\mathrm{B}^r$, obtaining
 \begin{equation}
 Q_G =  \frac{4\pi^2}{\kappa^2}\, k\,\ell_s^2\,V_{\mathbb{T}^4}\,V_{\mathbb{S}^1}\,(a_3-b_3)\,,\qquad
 Q_B =  \frac{4\pi^2}{\kappa^2}\, k\,\ell_s^2\,V_{\mathbb{T}^4}\,V_{\mathbb{S}^1}\,(a_3+b_3)\,.
 \end{equation}
 Notice that these expressions match the angular momenta ${\cal J}_{\hat x ' } ^{G,B}$ of the ten-dimensional
 background (\ref{J momenta result}).
 The corresponding left- and right-moving 
 charges can then be found according to 
\begin{equation}
\label{Q L R}
\begin{aligned}
Q_L &= \frac{Q_G +Q_B}{2} = \frac{4\pi^2}{\kappa^2}\, k\,\ell_s^2\,V_{\mathbb{T}^4}\,V_{\mathbb{S}^1}\,a_3\,,\\
Q_R &= \frac{Q_G -Q_B}{2} = -  \frac{4\pi^2}{\kappa^2}\, k\,\ell_s^2\,V_{\mathbb{T}^4}\,V_{\mathbb{S}^1}\,b_3\,.
\end{aligned}
\end{equation}

The free energy of the nine-dimensional background is given by expression analogous to 
its ten-dimensional counterpart (\ref{F formula}), except that now we have rotations
only in $\hat \alpha'$, $\hat\beta'$ directions, as well as two $U(1)$ charges,
\begin{equation}
\label{F formula 9d}
{\cal F} ={\cal E} -  T\,{\cal S} 
 -\Omega_L^{\hat\alpha'}\,{\cal J}^L_{\hat\alpha'}
 -\Omega_R^{\hat\beta'}\,{\cal J}^R_{\hat\beta'}
 -\mu_L\,Q_L-\mu_R\,Q_R\,.
\end{equation}
Here we can calculate
\begin{equation}
\Omega_L^{\hat\alpha'}\,{\cal J}^L_{\hat\alpha'}
 +\Omega_R^{\hat\beta'}\,{\cal J}^R_{\hat\beta'}
 +\mu_L\,Q_L+\mu_R\,Q_R =  \frac{4\pi^2}{\kappa^2}\, k^\frac{3}{2}\,\ell_s^2 \,V_{\mathbb{T}^4}\,V_{\mathbb{S}^1}\,\, 
 \left(1-a_1b_1\right)\,,
\end{equation}
in agreement with (\ref{OmegaJ gen result}).
The values for temperature $T$, entropy ${\cal S}$,  energy ${\cal E}$, and grand potential ${\cal F}$
of the nine-dimensional background are the same as for the ten-dimensional
background, and are given by (\ref{general T result}), (\ref{BH entropy result}),
(\ref{vanishing grand potential}), (\ref{energy gen result}).
Finally, one can explicitly verify that the first law of thermodynamics is satisfied, 
\begin{equation}
d{\cal E} -  T\,d{\cal S} - \Omega^{\hat\alpha'}_L\,d{\cal J}_{\hat\alpha'}^L
- \Omega^{\hat\beta'}_R\,d{\cal J}_{\hat\beta'}^R
- \mu_L\,dQ_L- \mu_R\,dQ_R = 0\,.
\end{equation}

\section{Spectrum}
\label{sec:BRST}

In this section, we will discuss construction of spectrum of type-II superstring theory
on the coset space (\ref{big coset}). As we will review below, physical states of superstring theory
 satisfy the super-Virasoro constraints, ensuring superconformal
invariance of quantum theory on the world-sheet.
Additionally, physical states of the coset sigma-model  satisfy gauge-invariant
conditions, that can be formulated via the BRST
formalism \cite{Karabali:1989dk}.\footnote{See \cite{Goykhman:2013oja,Chakraborty:2020yka} for a recent discussion.} We will derive the corresponding constraints in section~\ref{sec: coset brst current}.

It is important to emphasize that
our conventions in this section are to use the coset BRST formalism
 to impose the $U(1)_L\times U(1)_R$
gauge symmetry in the spectrum of quantum theory. At the same time, we we will use the covariant
quantization approach (without employing the superconformal BRST formalism) to ensure superconformal invariance of
the spectrum in terms of the (super-)Virasoro constraints.

We are mostly interested in the bosonic excitations of the NS-NS sector,
in particular, the lowest tachyon state, and the massless states composing the gravity multiplet
of the type-II supergravity.
Therefore the distinction between type-IIA and type-IIB superstrings,
originating from the choice of space-time chirality of the fermionic 
states in the R-NS and NS-R sectors, will not play a role in most of the subsequent discussion.
We will also not consider bosonic excitations coming from the R-R sector,
leaving these topics for future work.

\subsection{Coset BRST constraints}
\label{sec: coset brst current}

In this subsection, we are going to derive the coset BRST constraints,
that have to be satisfied by the physical states on the coset (\ref{big coset}).
These constraints impose the $U(1)_L\times U(1)_R$ symmetry
on the states of the physical spectrum of the coset theory.
(We stress once again, that in this paper the superconformal BRST formalism is not used
to impose superconformal symmetry on the physical states.)

To lighten up the discussion, we will begin by performing  derivation of the
BRST constraints for the six-dimensional coset subspace
of the target space-time (\ref{big coset}), working in the purely bosonic sector. We will subsequently
formulate the supersymmetric version of
the obtained constraints.
Partition function of the gauged WZW model (\ref{gWZW result})
is given by
\begin{equation}
Z = \int [dG][dA][d\tilde A]\,e^{-S_{\textrm{gWZW}}[G,A,\tilde A]}\,.
\end{equation}
Using (\ref{gWZW compact}), (\ref{A tilde A in terms of u and v}), we can re-write it as
\begin{equation}
\label{Z det p}
Z = \int [dG][du][d v]\,\det\,\p\,\det\,\bar\p\,e^{-S_{\textrm{WZW}}'[UGV]+S_{\textrm{P}}[w]}\,,
\end{equation}
where
\begin{equation}
S_{\textrm{P}}[w]=\frac{1}{2\pi}\int d^2z\,\p w\bar\p w
\end{equation}
is the Polyakov action for the field $w = u - v$.
Introducing anti-commuting ghost fields $b$, $c$, $\tilde b$, $\tilde c$,
one can express the functional determinants in (\ref{Z det p}) as
\begin{equation}
\det\,\p\,\det\,\bar\p = \int [Db][Dc][D\tilde b][D\tilde c]
\,e^{-\frac{1}{2\pi}\int d^2z\,(b\bar\p c+\tilde b\p \tilde c)}\,.
\end{equation}
Ghost fields satisfy the following non-trivial OPEs
\begin{equation}
\label{c b ope}
c(z)b(w) \simeq \frac{1}{z-w}\,,\qquad
\tilde c(\bar z)\tilde b(\bar w) \simeq \frac{1}{\bar z-\bar w}\,.
\end{equation}
The ghosts $b$ and $c$ are conformal primaries of dimensions $1$
and $0$ respectively \cite{Karabali:1988au,Dijkgraaf:1991ba},\footnote{As a side note, in the BRST
quantization approach to superstring theory
one employs the superconformal ghosts comprised of the fermions $\mathtt{b}$, $\mathtt{c}$, and
bosons $\beta$, $\gamma$. Dimensions of these pairs of ghosts are $(2,1)$ and $(3/2,-1/2)$,
respectively. In particular, this is reflected in contributions of $\mathtt{b}$, $\mathtt{c}$ to the stress-energy
tensor, $T_{\mathtt{b}\mathtt{c}}=-2\,\mathtt{b}\,\p\,\mathtt{c}+\mathtt{c}\,\p\,\mathtt{b}$, which is manifestly different from
its coset BRST ghosts counterpart (\ref{b c contribution to T}). We do not perform superconformal
 BRST quantization of superstring theory,
and therefore the superconformal ghosts $\mathtt{b}$, $\mathtt{c}$, $\beta$, $\gamma$
do not appear in our derivation of the spectrum.} with the following expansions in integer modes,
\begin{equation}
b(z) = \sum_n \frac{b_n}{z^{n+1}}\,,\quad
c(z) = \sum_n \frac{c_n}{z^{n}}\,,\quad
\tilde b(\bar z) = \sum_n \frac{\tilde b_n}{\bar z^{n+1}}\,,\quad
\tilde c(\bar z) = \sum_n \frac{\tilde c_n}{\bar z^{n}}\,.
\end{equation}
Due to the OPEs (\ref{c b ope}), these modes satisfy the anti-commutation relations
\begin{equation}
\{c_n,\,b_m\}=\delta_{mn}\,,\qquad \{\tilde c_n,\,\tilde b_m\}=\delta_{mn}\,.
\end{equation}
Conformal dimensions of the ghosts can also be reproduced
from their  the stress-energy tensor,
\begin{equation}
\label{b c contribution to T}
T_{bc}(z)=b(z)\,\p c(z)\,,
\end{equation}
where the product in the r.h.s. is normally-ordered.

Finally, let us change integration variables $u\rightarrow u+w$,
$G\rightarrow U^{-1}GV^{-1}$ and divide the partition function (\ref{Z det p}) by the
constant factor of $\int [Dv]$, rendering 
\begin{equation}
Z = \int [dG][dw][Db][Dc][D\tilde b][D\tilde c]\,e^{-S_{\textrm{q}}}\,,
\end{equation}
where the full action of the model is given by
\begin{equation}
\label{S q def}
S_{\textrm{q}} = S_{\textrm{WZW}}'[G]
-S_{\textrm{P}}[w]+\frac{1}{2\pi}\int d^2z\,(b\bar\p c+\tilde b\p \tilde c)\,.
\end{equation}
Notice that from the action (\ref{S q def}) it follows that the two-point correlation
function of the field $w$ has the `wrong' sign
\begin{equation}
\label{ww correlator}
\langle \p w(z_1)\p w(z_2)\rangle = \frac{1}{2}\,\frac{1}{(z_1-z_2)^2}\,.
\end{equation}
This does not pose a problem, since the physical state (coset BRST) conditions
will eliminate the corresponding negative norm states from the spectrum.
These conditions are derived as follows.
Consider the holomorphic BRST transformation defined
by the fermionic parameter $\eta$, with the non-trivial variations given by \cite{Bastianelli:1990ey}
\begin{equation}
\label{hol BRST transform}
\delta G = i \eta\,c\,T_LG\,,\qquad \delta w = i \eta \,c\,,\qquad
\delta b = i \eta\,(J + 2 \p w)\,,
\end{equation}
where the gauge current $J$ is given by (\ref{general currents}).
Using (\ref{WZWvarFinal}) we can derive the corresponding
variation of the action (\ref{S q def}) as 
\begin{equation}
\begin{aligned}
\delta S_{\textrm{q}} &= - \frac{1}{2\pi}\,\int d^2z\,
\left(k\,\textrm{Tr}\,\left(P\,\p G G^{-1}\bar\p (\delta G G^{-1})\right)
+2\p w\,\bar\p \delta w -\delta b\bar\p c\right)\\
&= - \frac{i}{2\pi}\,\int d^2z\,\bar\p\eta\, c\,\left( J+2\p w\right)\,.
\end{aligned}
\end{equation}
Analogous calculation can be done for the anti-holomorphic BRST transformation,
\begin{equation}
\label{antihol BRST transform}
\delta G = - i \eta\,\tilde c\,GT_R\,,\qquad \delta w = i \eta \,\tilde c\,,\qquad
\delta b = i \eta\,(-\tilde J + 2\bar \p w)\,,
\end{equation}
giving variation of the action
\begin{equation}
\delta S_{\textrm{q}} =  - \frac{i}{2\pi}\,\int d^2z\,\p\eta\, \tilde c\,\left(-\tilde J+2\bar\p w\right)\,.
\end{equation}
For a global parameter $\eta$ the transformations (\ref{hol BRST transform}), (\ref{antihol BRST transform})
are therefore a symmetry of the action (\ref{S q def}). Choosing a local parameter $\eta$
we obtain the coset BRST currents $J^{\textrm{BRST}}=c\,J^{\textrm{(0)}}$, $\tilde J^{\textrm{BRST}} =\tilde c\,\tilde J^{\textrm{(0)}}$, where we denoted
the corresponding null currents as\footnote{Here we have taken the defiinitions (\ref{jx and jw defs}) into account.}
\begin{equation}
\label{BRST currents b}
J^{(0)} =  i\,J + 2j^ w\,,\qquad
\tilde J^{(0)} = -i\,\tilde  J + 2\tilde j^ w\,.
\end{equation}

Let us now switch to discussion of the full supersymmetric model. To begin with,
recall that the $U(1)_L\times U(1)_R$ currents $J$, $\tilde J$ of the purely bosonic
sector, given by (\ref{general currents}), are to be replaced with their supersymmetric theory counterparts
$\mathtt{J}$, $\tilde{\mathtt{J}}$, given by
(\ref{J tJ in terms of constituents}). These total bosonic gauge currents incorporate 
contributions from the fermions in adjoint representation of the $sl(2,\mathbb{R})$
and $su(2)$ algebras. The corresponding null currents in the supersymmetric coset theory are then given by
\begin{equation}
\label{BRST currents}
\mathtt{J}^{(0)} =  i\,\mathtt{J} + 2j^ w\,,\qquad
\tilde{\mathtt{J}}^{(0)} = -i\,\tilde {\mathtt{J}} + 2\tilde j^ w\,.
\end{equation}
Indeed, combining (\ref{JJ cor}), (\ref{ww correlator}) we recover the BRST null conditions
\begin{equation}
\label{J brst null}
\langle \mathtt{J}^{(0)} (z)\mathtt{J}^{(0)} (w)\rangle =0\,,\qquad
\langle \tilde{\mathtt{J}}^{(0)} (\bar z)\tilde{\mathtt{J}}^{(0)} (\bar w)\rangle =0\,.
\end{equation}
The associated BRST charges
\begin{equation}
Q_{\textrm{BRST}} = \frac{1}{2\pi i}\oint dz\,\mathtt{J}^{\textrm{BRST}}\,,\qquad
\tilde Q_{\textrm{BRST}} = \frac{1}{2\pi i}\oint dz\,\tilde{\mathtt{J}}^{\textrm{BRST}}\,,
\end{equation}
 of the currents
 $\mathtt{J}^{\textrm{BRST}}=c\,\mathtt{J}^{\textrm{(0)}}$, $\tilde{\mathtt{J}}^{\textrm{BRST}} =
 \tilde c\,\tilde{\mathtt{J}}^{\textrm{(0)}}$
are therefore nilpotent, $Q_{\textrm{BRST}} ^2 =0$,  $\tilde Q_{\textrm{BRST}}^2 = 0$.
Physical states in the spectrum of the coset model (\ref{S q def})
are to be closed w.r.t. these charges,
\begin{equation}
\label{BRST U1 conditions}
Q_{\textrm{BRST}}  |\textrm{phys}\rangle = 0\,,\qquad 
\tilde Q_{\textrm{BRST}}  |\textrm{phys}\rangle = 0\,,
\end{equation}
and are defined up to the BRST-exact states.
Expanding the null currents (\ref{BRST currents}),
\begin{equation}
\label{BRST currents expansion}
\mathtt{J}^{(0)}(z) = \sum_{n}\,\frac{\mathtt{J}^{(0)}_n}{z^{n+1}}\,,\qquad
\tilde{\mathtt{J}}^{(0)}(\bar z) = \sum_{n}\,\frac{\tilde{\mathtt{J}}^{(0)}_n}{\bar z^{n+1}}\,,
\end{equation}
the physical state conditions are expressed as
\begin{equation}
\label{BRST constraints}
\mathtt{J}^{(0)} _n  |\textrm{phys}\rangle = 0\,,\qquad
\tilde{\mathtt{J}}^{(0)} _n  |\textrm{phys}\rangle = 0\,,\qquad n\geq 0\,,
\end{equation}
while the BRST exact states are obtained as $\mathtt{J}^{(0)} _{-1}  |0\rangle$,
$\tilde{\mathtt{J}}^{(0)} _{-1}  |0\rangle$, where $|0\rangle$ is the physical vacuum.
Using  (\ref{J brst null}) we also obtain the commutation relations
\begin{equation}
[\mathtt{J}^{(0)}_m,\,\mathtt{J}^{(0)}_n] =0\,,\qquad
[\tilde{\mathtt{J}}^{(0)}_m,\,\tilde{\mathtt{J}}^{(0)}_n] =0\,.
\end{equation}

The  (anti-)holomorphic superpartners of the gauge currents $\mathtt{J}$, $\tilde{\mathtt{J}}$
are given by the fermions (\ref{psi tpsi in terms of constituents}).
At the same time, the superpartners of the bosonic currents $j^ w$, $\tilde j^ w$
are given by the fermions $\psi^w$, $\tilde\psi^w$, such that\footnote{
Notice that the sign of these correlation functions is opposite to the sign
of the two-point functions of physical fermions (\ref{psi x ope}). This is analogous
to having opposite signs of physical and auxiliary bosonic currents (\ref{jx jx and others opes}).
Unphysical $w$ polarization is eliminated from the spectrum by the BRST constraints.}
\begin{equation}
\label{psi w ope}
\psi^w (z) \psi^w(w) = - \frac{1}{2}\,\frac{1}{z - w}\,,\qquad
\tilde \psi^w (z) \tilde \psi^w(w) = - \frac{1}{2}\,\frac{1}{\bar z -\bar w}\,.
\end{equation}
We can use these to write down
the fermionic  null currents superpartners \cite{Figueroa-OFarrill:1995vqf}\footnote{The fermionic sector possesses its own ghosts, which are bosonic counterparts
of $b$ and $c$. We skip detailed discussion of this formalism.}
\begin{equation}
\label{psi BRST currents}
\psi^{(0)} = i \,\psi + 2\, \psi^w\,,\qquad 
\tilde \psi^{(0)} = - i \,\tilde \psi + 2\, \tilde \psi^w\,,
\end{equation}
that  satisfy the null conditions
\begin{equation}
\label{psi brst null}
\langle \psi^{(0)} (z)\psi^{(0)} (w)\rangle =0\,,\qquad
\langle \tilde{\psi}^{(0)} (\bar z)\tilde{\psi}^{(0)} (\bar w)\rangle =0\,.
\end{equation}
Physical states in the spectrum of NS states of the supersymmetric coset theory should satisfy the constraints
\begin{equation}
\psi^{(0)} _r |\textrm{phys}\rangle = 0\,,\qquad
\tilde \psi^{(0)} _r |\textrm{phys}\rangle = 0\,,\qquad
r=\frac{1}{2},\,\frac{3}{2},\dots\,,
\end{equation}
where we defined the NS sector mode expansion amplitudes  as
\begin{equation}
\label{psi brst mode expansion}
\psi^{(0)}  = \sum_{r\in\mathbb{Z}+\frac{1}{2}}\frac{\psi^{(0)} _r}{z^{r+\frac{1}{2}}}\,,\qquad
\tilde \psi^{(0)}  = \sum_{r\in\mathbb{Z}+\frac{1}{2}}\frac{\tilde\psi^{(0)} _r}{\bar z^{r+\frac{1}{2}}}\,,
\end{equation}
where $r$ is half-integer in the NS sector. 
Using  (\ref{psi brst null}) we obtain
\begin{equation}
\{\psi^{(0)}_r,\,\psi^{(0)}_s\} =0\,,\qquad
\{\tilde \psi^{(0)}_r,\,\tilde \psi^{(0)}_s\} =0\,.
\end{equation}

\subsection{Virasoro constraints}
\label{sec:virasoro constraints}

In section~\ref{sec: coset brst current}, we used the coset BRST formalism to
derive the $U(1)_L\times U(1)_R$ 
null physical state constraints $\mathtt{J}^{(0)}_n|\textrm{phys}\rangle = 0$, $n\geq 0$,
$\psi^{(0)}_r|\textrm{phys}\rangle = 0$, $r=1/2,\,3/2,...$, that are satisfied by the states
in the spectrum of the coset theory (\ref{big coset}). 
At the same time, the physical states are defined up to the exact states, $\mathtt{J}^{(0)}_{-n}\,|0\rangle$ and
 $\psi^{(0)}_{-r}\,|0\rangle$
where $|0\rangle$ is the physical vacuum. 

Besides these constraints, the physical states of superstring theory have to satisfy the (super-)Virasoro constraints.
These constraints reflect superconformal invariance of the quantum world-sheet theory.
The superconformal transformations are generated by the stress energy tensor $T(z)$,
and the super-current $G(z)$, and their anti-holomorphic counterparts $\tilde T(\bar z)$,
$\tilde G(\bar z)$. These currents form (anti-)holomorphic affine superconformal
algebra, that we briefly review in appendix~\ref{app:effective wzw action}.

Components of the stress-energy tensor of the coset
world-sheet theory  (\ref{big coset})
are given by
\begin{equation}
\begin{aligned}
&T=\frac{1}{k}j^aj^a +\frac{1}{k}j'^aj'^a+j^xj^x - j^w j^w +T _f +T_{\mathbb{T}^4} +T_{\textrm{ghosts}},\\
&\tilde{T}=\frac{1}{k}\tilde{j}^a\tilde{j}^a +\frac{1}{k}\tilde{j}'^a\tilde{j}'^a +\tilde j^x\tilde j^x - \tilde j^w \tilde j^w
+\tilde T _f +\tilde{T}_{\mathbb{T}^4} +\tilde T_{\textrm{ghosts}}\,,
\end{aligned}
\end{equation}
where $T_{\mathbb{T}^4}$, $\tilde{T}_{\mathbb{T}^4}$ are the components of the world-sheet stress-energy
tensor with target space 
$\mathbb{T}^4$ parametrized by coordinates $z^i$, $i=1,2,3,4$,
\begin{equation}
T_{\mathbb{T}^4}  = \sum_{i=1}^4 j^i\,j^i\,,\qquad
\tilde T_{\mathbb{T}^4}  = \sum_{i=1}^4 \tilde j^i\,\tilde j^i\,.
\end{equation}
We have also denoted contributions from the fermionic sector as
\begin{equation}
T_f= - \frac{1}{k}\,\psi^a\p \psi^a - \frac{1}{k}\,\psi^{\prime a}\p \psi^{\prime a}
-\psi^i\p\psi^i+\psi^w\p\psi^w\,,
\end{equation}
and similarly for the anti-holomorphic sector,
while contributions from coset BRST ghosts, $T_{\textrm{ghosts}}$
comes from the ghosts sector, that includes $T_{bc}$ given by (\ref{b c contribution to T}),
and its supersymmetric counterpart.

The (anti-)holomorphic supercurrents $G(z)$, $\tilde G(\bar z)$
are fermionic operators. We are mostly interested in the NS sector,
and therefore the fermions are expanded in half-integer modes.
Specifically, expanding the stress-energy tensor and the supercurrent, we obtain
\begin{equation}
\begin{aligned}
T(z) &=\sum _{n\in \mathbb{Z}}\frac{L_n}{z^{n+2}}\,,\qquad
&&\tilde T(\bar z) = \sum _{n\in \mathbb{Z}}\frac{\tilde L_n}{\bar z^{n+2}}\,,\\
G(z) &= \sum_{r\in \mathbb{Z} + 1/2} \frac{G_r}{z^{r+\frac{3}{2}}}\,,
\qquad
\qquad &&\tilde G(\bar z) = \sum_{r\in \mathbb{Z} + 1/2} \frac{\tilde G_r}{\bar z^{r+\frac{3}{2}}}\,.
\end{aligned}
\end{equation}
The physical state conditions in terms of the corresponding
mode expansion amplitudes are given by\footnote{
In terms of the vertex operator ${\cal V}$ corresponding
to the state $|\textrm{phys}\rangle$, 
one calculates, \textit{e.g,} $L_n \cdot {\cal V}(0) \simeq \frac{1}{2\pi i} \oint z^n \, T(z)\,{\cal V}(0)$
and sets it to zero.}
\begin{equation}
\label{super Virasoro}
\begin{aligned}
\left(L_n -\frac{1}{2}\,\delta_{n, 0}\right)|\textrm{phys}\rangle &= 0\,,\qquad
\left(\tilde L_n -\frac{1}{2}\,\delta_{n, 0}\right)|\textrm{phys}\rangle = 0\,,\qquad n=0,1,2,\dots\\
G_r|\textrm{phys}\rangle &= 0\,,\qquad
\tilde G_r|\textrm{phys}\rangle = 0\,,\qquad r=\frac{1}{2},\frac{3}{2},\frac{5}{2},\dots\,.\\
\end{aligned}
\end{equation}
Additionally, the physical states are defined up to exact states $L_{-n}\,|0\rangle = 0$,
 $G_{-r}\,|0\rangle = 0$, and similarly for the anti-holomorphic sector.

\subsection{Ground state vertex operator}
\label{sec:ground state}

We now proceed to construction of the primary vertex operator 
of the  (\ref{big coset}) coset model. This
operator describes the ground state of the model, that we write as
\begin{equation}
\label{ground state W VN}
V^{(0)}\tilde  V^{(0)}= W^{(0)}(z,\bar z)\,V^{(0)}_{\mathbb{T}^4}(z,\bar{z})\,,
\end{equation}
where $W^{(0)}(z,\bar z)$ is the ground-state vertex operator
of the six-dimensional coset sub-space of (\ref{big coset}),
while\footnote{Recall
that in our conventions all the coordinates are dimensionless. At the same time, we will define all the
momenta to be dimensionful.}
\begin{equation}
\label{T4 ground state}
V^{(0)}_{\mathbb{T}^4} = e^{i\ell_s\sum_{i=1}^4(p_L^iz_L^i+p_R^iz^i_R)}\,.
\end{equation}
is due to the four-dimensional flat sub-space $\mathbb{T}^4 = U(1)^4$.

The corresponding ground state $V^{(0)}\tilde  V^{(0)}|0\rangle$ should satisfy physical conditions, that include
the coset BRST constraints and the superconformal super-Virasoro constraints.
In this subsection we will focus on the former.

We start by expressing the coset ground state vertex operator  $W^{(0)}(z,\bar z)$ as a product of the 
$SL(2,\mathbb{R})$, $SU(2)$, $U(1)_x$, and $U(1)_w$\footnote{
The $U(1)_w$ is non-compact, and therefore $\kappa_L = \kappa_R = \kappa$.} primary operators,
\begin{equation}
\label{def of ground state v.o.}
W^{(0)}(z,\bar z) = 
V^{\hat\omega}_{j;m,\bar m}(z,\bar z)\,
V^{\prime\,\hat\omega',\hat{\bar\omega}'}_{j;m',\bar m'}(z,\bar z)\,
e^{i\ell_s(p_L x_L(z,\bar z)+p_Rx_R(z,\bar z))}\,
e^{i\ell_s\kappa w}\,,
\end{equation}
where $\hat\omega\in \mathbb{Z}$
is the $SL(2,\mathbb{R})$ spectral flow parameter and $\hat\omega',\;\hat{\bar\omega}'\in\mathbb{Z}$
are the $SU(2)$ left- and right-moving spectral flow parameters.
The left- and right-moving momenta on the compact $U(1)_x$ circle with radius $R_x$ are given by
\begin{equation}
p_{L,R} = \frac{n}{R_x} \pm \frac{\textrm{w}\,R_x}{\ell_s^2}\,,
\end{equation}
where $n$ and $\textrm{w}$ are the momentum and winding numbers respectively.

Recall that the $SL(2,\mathbb{R})$ and $SU(2)$ primary operators satisfy (\ref{jvopesf}), (\ref{sfvjpope})
\begin{equation}
\label{j3 and V OPE}
\begin{aligned}
\J_3(z) V^{\hat\omega}_{j;m,\bar m}(0) &= \frac{m+\frac{k}{2}\hat\omega}{z}\,V^{\hat\omega}_{j;m,\bar m}\,,\;
&&\tilde \J_3(\bar z) V^{\hat\omega}_{j;m,\bar m}(0)
= \frac{ \bar m+\frac{k}{2}\hat\omega}{\bar z}\,V^{\hat\omega}_{j;m,\bar m}\,,\\
\J_3'(z) V_{j';m',\bar m'}^{\prime\,\hat\omega',\hat{\bar\omega}'}(0) &= \frac{m'+\frac{k}{2}\hat\omega'}{z}\,V_{j';m',\bar m'}
^{\prime\,\hat\omega',\hat{\bar\omega}'}\,,\;
&&\tilde \J_3'(\bar z) V_{j';m',\bar m'}^{\prime\,\hat\omega',\hat{\bar\omega}'}(0) =
 \frac{\bar m'+\frac{k}{2}\hat{\bar\omega}'}{\bar z}\,V_{j';m',\bar m'}^{\prime\,\hat\omega',\hat{\bar\omega}'}\,,
\end{aligned}
\end{equation}
while for the $U(1)_x$, $U(1)_w$ primary operators one finds
\begin{equation}
\label{jxw and V OPE}
\begin{aligned}
j^x(z)\, e^{i\ell_s(p_L x_L(0)+p_Rx_R(0))} &= \frac{\ell_s p_L}{2z}\,e^{i\ell_s(p_L x_L+p_Rx_R)}\,,\quad
j^w(z) \, e^{i\ell_s\kappa w(0)} = - \frac{\ell_s\kappa}{2z}\,e^{i\ell_s\kappa w}\\
\tilde j^x(\bar z)\, e^{i\ell_s(p_R x_L(0)+p_Rx_R(0))} &= \frac{\ell_sp_R}{2\bar z}\,e^{i\ell_s(p_L x_L+p_Rx_R)}\,,\quad
\tilde j^w(\bar z)\, e^{i\ell_s\kappa w(0)} = -\frac{\ell_s\kappa}{2\bar z}\,e^{i\ell_s\kappa w}\,.
\end{aligned}
\end{equation}
Using (\ref{J tJ in terms of constituents}), (\ref{BRST currents}) we then obtain
\begin{equation}
\begin{aligned}
\mathtt{J}^{(0)}(z)W^{(0)}(0) &= 2\left(\frac{1}{\sqrt{k}}
\left[a_1\,\left(m+\frac{k}{2}\hat\omega\right)+a_2\,\left(m'+\frac{k}{2}\hat\omega'\right)\right]+\frac{\ell_s p_La_3}{2}
-\frac{\ell_s\kappa}{2}\right)\,\frac{W^{(0)}}{z}\,,\\
\tilde{\mathtt{J}}^{(0)}(\bar z)W^{(0)}(0) &= 2\left(-\frac{1}{\sqrt{k}}
\left[b_1\,\left(\bar m+\frac{k}{2}\hat\omega\right)+b_2\,\left(\bar m'+\frac{k}{2}\hat{\bar\omega}'\right)\right]
-\frac{\ell_sp_Rb_3}{2}-\frac{\ell_s\kappa}{2}\right)\,\frac{W^{(0)}}{\bar z}\,,
\end{aligned}
\end{equation}
which is consistent with $m,\bar m\in \mathbb{R}$, $m',\bar m'\in \mathbb{R}$.
The BRST constraints (\ref{BRST constraints}) therefore take the form
\begin{equation}
\label{BRST constraints prelim res}
\begin{aligned}
\frac{1}{\sqrt{k}}
\left[a_1\,\left(m+\frac{k}{2}\hat\omega\right)+a_2\,\left(m'+\frac{k}{2}\hat\omega'\right)\right]+\frac{\ell_s p_La_3}{2}
-\frac{\ell_s\kappa}{2} &= 0\,,\\
\frac{1}{\sqrt{k}}
\left[b_1\,\left(\bar m+\frac{k}{2}\hat\omega\right)+b_2\,\left(\bar m'+\frac{k}{2}\hat{\bar\omega}'\right)\right]
+\frac{\ell_sp_Rb_3}{2}+\frac{\ell_s\kappa}{2} &=0\,.
\end{aligned}
\end{equation}
Combining the constraints (\ref{BRST constraints prelim res}) we obtain
\begin{equation}
\label{BRST constr res}
\begin{aligned}
&\frac{1}{\sqrt{k}}
\left[a_1\,\left(m+\frac{k}{2}\hat\omega\right)+a_2\,\left(m'+\frac{k}{2}\hat\omega'\right)
+b_1\,\left(\bar m+\frac{k}{2}\hat\omega\right)+b_2\,\left(\bar m'+\frac{k}{2}\hat{\bar\omega}'\right)\right]\\
&+\frac{\ell_s p_La_3}{2}
+\frac{\ell_sp_Rb_3}{2} = 0\,.
\end{aligned}
\end{equation}

\subsection{Asymptotic behavior of  vertex operators}
\label{sec:asymptotic vertex operator}

In this subsection, we are going to explore asymptotic behavior of
vertex operators describing string excitation modes. We begin with the ground state vertex operator
(\ref{def of ground state v.o.}). Specifically, we are interested in the plane-wave behavior
in the asymptotic region, that allows one to read off energy of the string excitation mode.
In fact, the same energy characterizes any string excitation obtained by acting
with the creation operators on the given ground state vertex operator.

In the $SL(2,\mathbb{R})$ parametrization given by (\ref{groups parametrizations}),
the asymptotic region is found by taking the limit $\theta\rightarrow \infty$.
The $SL(2,\mathbb{R})$ ground state vertex operator in this limit behaves as \cite{Elitzur:2002rt}
\begin{equation}
\label{sl2r V asymptotic behavior}
V^{\hat{\omega}}_{j;m,\bar{m}}(\theta\to\infty)\simeq
e^{iy\left[ \left(m+\frac{k\hat{\omega}}{2}\right)\nu+\left(\bar{m}+\frac{k\hat{\omega}}{2}\right)\right]}\left[e^{2\theta j}+(-1)^{-(2j+1)}R(j;m,\bar{m}) e^{-2\theta(j+1)} \right]\,,
\end{equation}
where (this coincides with the coefficient in (\ref{VV2ptfn}))
\begin{eqnarray}
R(j;m,\bar{m})=\frac{\Gamma\left(1-\frac{2j+1}{k}\right)\Gamma(j+1-m)\Gamma(j+1+\bar{m})
\Gamma(-2j-1)}{\Gamma\left(1+\frac{2j+1}{k}\right)\Gamma(-j-m)\Gamma(-j+\bar{m})\Gamma(2j+1)}~.
\end{eqnarray}
In the parafermionic decomposition one represents locally $SU(2)\sim \frac{SU(2)}{U(1)}\times U(1)_{\textrm{y}}$,
that allows one to re-write the $SU(2)$ ground state vertex operator as
\begin{eqnarray}
\label{su2 V parafermionic decomposition}
{V'}^{\hat{\omega}',\hat{\bar{\omega}}'}_{j';m',\bar{m}'}=e^{i\left[\left(m'+\frac{k\hat{\omega}'}{2}\right) 
y_{\textrm{su}}+\left(\bar{m}+\frac{k\hat{\bar{\omega}}'}{2}\right) \bar{y}_{\textrm{su}}\right]} \Psi_{j';m,\bar{m}'},
\end{eqnarray}
where $\Psi_{j';m,\bar{m}'}$ is the $SU(2) / U(1)$ vertex operator.

Combining expressions (\ref{sl2r V asymptotic behavior}), (\ref{su2 V parafermionic decomposition})
for the $SL(2,\mathbb{R})$ and $SU(2)$ ground state vertex operators in the
expression (\ref{def of ground state v.o.}) for the ground vertex operator of the coset sub-space of (\ref{big coset})
and focusing on the plane-wave phase in the asymptotic region, we obtain
\begin{equation}
W^{(0)}\simeq e^{i\left[\left( \left(m+\frac{k\hat{\omega}}{2}\right)\nu+\left(\bar{m}+\frac{k\hat{\omega}}{2}\right)\right)y
+\left(m'+\frac{k\hat{\omega}'}{2}\right) 
y_{\textrm{su}}+\left(\bar{m}+\frac{k\hat{\bar{\omega}}'}{2}\right) \bar{y}_{\textrm{su}}
+\ell_s(p_L x_L+p_Rx_R+
\kappa w)\right]}\,.
\end{equation}
Here we can substitute expressions for $m$, $\bar m$, obtained by solving the 
BRST constraints (\ref{BRST constraints prelim res}), which after some re-arrangement gives
\begin{equation}
\label{moving frame asymptotic W}
W^{(0)}\simeq e^{i\left[\sqrt{k}\,\kappa t
+\left(m'+\frac{k\hat\omega'}{2}\right)\hat y_{\textrm{su}}
+\left(\bar m'+\frac{k\hat{\bar{\omega}}'}{2}\right)\hat{\bar{y}}_{\textrm{su}}
+\ell_s(p_L \hat x_L+p_R\hat x_R
+\kappa w)\right]}\,,
\end{equation}
where we defined the time coordinate $t$
\begin{equation}
\label{asymptotic V t choice}
y=\frac{2a_1b_1t}{a_1-b_1\nu}\,,
\end{equation}
as well as the moving frame coordinates
\begin{equation}
\label{asymptotic V x y choice}
\begin{aligned}
\hat x_L &= x_L - \frac{a_3b_1\nu\,\sqrt{k}\,t}{a_1-b_1\nu}\,,\qquad
\hat x_R = x_R - \frac{a_1b_3\,\sqrt{k}\,t}{a_1-b_1\nu}\,,\\
\hat y_{\textrm{su}} &= y_{\textrm{su}}  - \frac{2a_2b_1\nu\,t}{a_1-b_1\nu}\,,\qquad
\hat{\bar y}_{\textrm{su}} = \bar y_{\textrm{su}}  - \frac{2a_1b_2\,t}{a_1-b_1\nu}\,.
\end{aligned}
\end{equation}
Notice that the time coordinate change (\ref{asymptotic V t choice})
and the moving frame coordinates change (\ref{asymptotic V x y choice})
agree with the corresponding expressions (\ref{y to t change}),
(\ref{moving frame coordinate change}) obtained
by analyzing the asymptotic region in derivation of the coset background geometry.\footnote{
The relative factor of $\sqrt{k}$ in the transformation laws for $x = x_{L} + x_R$ is due to
the relation between $x$ and $x'$ given by
(\ref{def of x prime}).}

Defining time as $t\rightarrow \frac{t}{\sqrt{k}}-\ell_s w$, we conclude from 
the asymptotic plane-wave expression
(\ref{moving frame asymptotic W}) for the vertex operator $W^{(0)}$ that the frequency of the
excited string state defined by this
vertex operator is given by $\kappa$.
Notice that this is true regardless of whether we consider a spectrally flowed sector.

\subsection{Mass-shell condition and spectrum}

In  subsection~\ref{sec:ground state},
we constructed ground state vertex operator (\ref{ground state W VN})
of the coset model (\ref{big coset}), and derived coset BRST physical state
constraints (\ref{BRST constraints prelim res}) that such an operator
needs to satisfy.
In subsection~\ref{sec:virasoro constraints}
we wrote down the (super-) Virasoro constraints that need to be satisfied
by the (NS sector) states of superstring theory on the coset  (\ref{big coset}).
Only a finite number of non-trivial constraints, depending on the string excitation
number, need to be imposed for each state, with the rest of infinitely many
constraints satisfied automatically. However, every state needs to satisfy
the lowest Virasoro $L_0$, $\tilde L_0$ constraints, known as the mass-shell condition. 
In this subsection we will derive the mass-shell condition for  string states on the coset (\ref{big coset}).

Let us denote the (anti-)holomorphic superstring vertex operator at the level ($\bar N$) $N$ as
($\tilde V^{(\bar N)}$) $V^{(N)}$.
It is obtained by acting on the ground state vertex operator $V^{(0)}\tilde V^{(0)}$
(that corresponds to $N=\bar N=0$) with the product  ($\tilde \alpha _{\bar N}$) $\alpha_N$ of creation operators
such as ($\tilde\psi^M_{-r}$, $\tilde{\J}^L_{-n}$) $\psi^M_{-r}$, $\J^L_{-n}$, $r>0$, $n>0$,
\begin{equation}
V^{(N)}=\alpha_N\,V^{(0)}\,,\qquad
\tilde V^{(N)}=\tilde \alpha_{\bar N}\,\tilde V^{(0)}\,,
\end{equation}
such that\footnote{
See (\ref{L j psi (anti)commutators}) for commutation relations of the Virasoro amplitudes
with the super-Kac-Moody currents.}
\begin{equation}
\label{L0 alpha commutator}
[L_0, \, \alpha_N] = N\,\alpha_N\,,\qquad 
[\tilde L_0, \, \tilde\alpha_{\bar N}] = \bar N\,\tilde\alpha_{\bar N}\,.
\end{equation}

Using (\ref{L0 alpha commutator}) together with the OPE of the ground state vertex\
operator (\ref{ground state W VN})  with the stress-energy tensor, we derive
\begin{equation}
\begin{aligned}
L_0\cdot V^{(N)}&=\left(\Delta_{sl}+\Delta_{su}+\frac{\ell_s^2p_L^2}{4}
-\frac{\ell_s^2\kappa^2}{4}+N+\Delta_{\mathbb{T}^4} \right)V^{(N)},\\
\tilde L_0\cdot \tilde V^{(\bar N)}&=\left(\bar{\Delta}_{sl}
+\bar{\Delta}_{su}+\frac{\ell_s^2p_R^2}{4}
-\frac{\ell_s^2\kappa^2}{4}+\bar{N}+\bar \Delta_{\mathbb{T}^4}\right)\tilde V^{(\bar N)}\,,
\end{aligned}
\end{equation}
where $N,\bar {N}=1,2,\dots$ are the string excitation numbers, 
\begin{equation}
\begin{aligned}\label{delta}
\Delta_{sl}=-\frac{j(j+1)}{k}-m\hat{\omega}-\frac{k\hat{\omega}^2}{4}, && \bar{\Delta}_{sl}=
-\frac{j(j+1)}{k}-\bar{m}\hat{\omega}-\frac{k\hat{\omega}^2}{4},\\
\Delta_{su}=\frac{j'(j'+1)}{k}+m'\hat{\omega}'+\frac{k\hat{\omega}'^2}{4}, && \bar{\Delta}_{su}=
\frac{j'(j'+1)}{k}+\bar{m}'\hat{\bar{\omega}}'+\frac{k\hat{\bar{\omega}}'^2}{4}
\end{aligned}
\end{equation}
are scaling dimensions of spectrally-flowed ground
state vertex operators of the $SL(2,\mathbb{R})$ and $SU(2)$ WZW models
(see (\ref{deltaslsf}), (\ref{deltabarslsf}), (\ref{deltasfsu})), and
\begin{equation}
\Delta_{\mathbb{T}^4} = \frac{\ell_s^2}{2} \sum_{i=1}^4(p_{L}^i)^2\,,\qquad
\bar \Delta_{\mathbb{T}^4} = \frac{\ell_s^2}{2} \sum_{i=1}^4(p_{R}^i)^2
\end{equation}
are scaling dimensions of the ground state vertex operator (\ref{T4 ground state})
on $\mathbb{T}^4$.
The mass-shell condition, given by the $L_0$, $\tilde L_0$ Virasoro constraints of (\ref{super Virasoro}), then reads
\begin{equation}
\label{onshell}
\begin{aligned}
\frac{\ell_s^2\kappa^2}{4}&=\Delta_{sl}+\Delta_{su}+\frac{\ell_s^2p_L^2}{4}+\Delta_{\mathbb{T}^4}+N-\frac{1}{2}\,,\\
\frac{\ell_s^2\kappa^2}{4}&=\bar{\Delta}_{sl}+\bar{\Delta}_{su}+\frac{\ell_s^2p_R^2}{4}+\bar\Delta_{\mathbb{T}^4}
+\bar{N}-\frac{1}{2}\,.
\end{aligned}
\end{equation}

In section~\ref{sec:asymptotic vertex operator} we  established
that the energy of string state, defined by the plane-wave behavior of the
vertex operator in the asymptotic region,\footnote{The plane-wave calculation is true for any excitation
numbers $N=\bar N$, and is determined only by asymptotic behavior of the ground state vertex
operator $W^{(0)}$.} is given by $\kappa$.
The $L_0$, $\tilde L_0$ Virasoro constraints (\ref{onshell}) then have the physical interpretation
of the mass-shell conditions, with the terms in the r.h.s. of these expressions given by the sum of the momenta
(on $SL(2,\mathbb{R})$ and $SU(2)$ the role of momenta are played by the
representation numbers $j$, $m$, $j'$, $m'$) and oscillator numbers $N-1/2$, $\bar N-1/2$. Substituting \eqref{delta} in \eqref{onshell} and  imposing the BRST constraints \eqref{BRST constraints prelim res} to eliminate $m,\bar{m}$, one obtains\footnote{Recall that the parameters $a_{1,2,3}$, $b_{1,2,3}$ need to satisfy the anomaly-free conditions
(\ref{time-like gauging}).}
\begin{equation}
\begin{aligned}\label{kappa}
\left(\kappa-\frac{(a_1-b_1)\sqrt{k}}{2a_1b_1 \ell_s}\hat{\omega}\right)^2=&\left(\frac{n}{R_x}\right)^2+\left(\frac{\mathrm{w}R_x}{\ell_s^2}\right)^2+\frac{2}{\ell_s^2}\Bigg( -2 \frac{j(j+1)-j'(j'+1)}{k}\\
&+\frac{k\hat{\omega}^2}{8}\left[\left(\frac{1}{a_1}-\frac{1}{b_1}\right)^2+4\right]+(m'\hat{\omega}'+\bar{m}'\hat{\bar{\omega}}')+\frac{k}{4}(\hat{\omega}'^2+\hat{\bar{\omega}}'^2)\\
&+\frac{a_2\hat{\omega}}{a_1}\left(m'+\frac{k\hat{\omega}'}{2}\right)+\frac{b_2\hat{\omega}}{b_1}\left(\bar{m}'+\frac{k\hat{\bar{\omega}}'}{2}\right)\\
&+\frac{\sqrt{k}\ell_s\hat{\omega}}{2a_1b_1}\left((a_3b_1+a_1b_3)\frac{n}{R_x}+(a_3b_1-a_1b_3)\frac{\mathrm{w}R_x}{\ell_s^2}\right)+\\
&+\Delta_{\mathbb{T}^4}+\bar{\Delta}_{\mathbb{T}^4}+\left(N-\frac{1}{2}\right)+\left(\bar{N}-\frac{1}{2}\right)\Bigg).
\end{aligned}
\end{equation}

For the ground state vertex operator (\ref{ground state W VN}) in the NS-NS sector we 
have  $N=\bar N = 0$.\footnote{Also setting spectral flow parameters and the $U(1)_x$ winding number to zero, $\hat{\omega}=\hat{\omega}'=\hat{\bar{\omega}}'=\mathrm{w}=0$.} The state is then tachyonic, $\kappa^2 < 0$, at zero momenta.
Such a state is removed from the physical spectrum by the GSO projection.
The first excited states of the NS sector, with ($\bar N = \frac{1}{2}$) $ N = \frac{1}{2}$
defined by the (anti-)holomorphic vertex operators ($\tilde V^{(\frac{1}{2})}$)
$V^{(\frac{1}{2})}$ are massless, and represent the lowest-lying
state in the physical spectrum of string excitation. Irreducible representations of tensor product of these states
form the NS-NS sector of the supergravity multiplet, and will be studied below in section~\ref{sec:massless states}.

\subsection{Massless states}
\label{sec:massless states}

Massless states of the type-II superstring theory form the type-II supergravity multiplet.
Half of its bosonic degrees of freedom are contained in the NS-NS sector,
and form a  gravity multiplet, consisting of the graviton, dilaton, and $B$-field.
In section~\ref{sec:coset construction}, we have derived  background values of these fields,
(\ref{full metric result}), (\ref{dilaton result}), (\ref{B field result 1}),
for superstring theory on the coset (\ref{big coset}), exactly in $k$. 

In this subsection, we are interested in  excitations of the massless NS-NS fields on top of the background
(\ref{full metric result}), (\ref{dilaton result}), (\ref{B field result 1}).
Performing the KK reduction on the $U(1)_x$ circle, one can also describe
excitations on top of the nine-dimensional background (\ref{9d background}).
These excitations can be obtained
by first separately constructing (anti-)holomorphic massless vertex operators in the NS sectors
of the (right-) left-moving string states, and then taking the direct product of these vertex operarors.
Finally, the resulting closed string states can be grouped into irreducible representations of the target space-time
symmetry group, rendering the graviton, dilaton, and $B$-field excitations.

The massless holomorphic vertex operator $V^{(\frac{1}{2})}$ in the NS sector on the coset  (\ref{big coset}) is
obtained by acting with the creation operators $\psi^M_{-1/2}$ on the tachyon ground state vertex operator
(\ref{ground state W VN}). For the most general polarizations $\xi_M$ in the target space-time we obtain
the state
\begin{align}
\label{V0 def}
V^{(\frac{1}{2})} &=\left[ \xi_+\,\psi^+_{-\frac{1}{2}} \,\J^-_0 + \xi_-\,\psi^-_{-\frac{1}{2}} \,\J^+_0
+\xi\,\psi^{(0)}_{-\frac{1}{2}}\right.\\
&+\left.  \xi_+'\,\psi^{\prime +}_{-\frac{1}{2}} \,\J^{\prime -}_0+  \xi_-'\,\psi^{\prime -}_{-\frac{1}{2}}\,\J^{\prime +}_0
+ \xi_3' \, \psi^{\prime 3}_{-\frac{1}{2}} + \xi _w\, \psi^w_{-\frac{1}{2}} 
+\xi_x\, \psi^x_{-\frac{1}{2}} + \sum_{i=1}^4
\xi_{i}\,\psi^{i}_{-\frac{1}{2}} \right]V^{(0)}\,,\notag
\end{align}
where we introduced linear combinations of the $sl(2,\mathbb{R})$ and $su(2)$ fermions
\begin{equation}
\psi^\pm = \psi^1 \pm i\psi^2\,,\qquad
\psi^{\prime\pm} = \psi^{\prime 1} \pm i\psi^{\prime 2}\,.
\end{equation}
In (\ref{V0 def}) we have chosen to replace the basis target space direction $\psi^3$
with the null spinor $\psi ^{(0)}$ given by (\ref{psi BRST currents}).
Physical state conditions, that include the coset BRST constraints
and super-Virasoro constraints, and remove the pure-gauge exact states, restrict the allowed polarizations $\xi_M$,
rendering eight independent polarizations.

Due the coset BRST constraints,
the state (\ref{V0 def}) needs to be closed w.r.t. the null spinor $\psi ^{(0)}$, given by (\ref{psi BRST currents}).
In terms of the amplitudes (\ref{psi brst mode expansion})
this condition imposes
\begin{equation}
\psi ^{(0)}_\frac{1}{2}\, V^{(0)} = 0\,.
\end{equation}
In terms of the basis fermionic operators, this can be written explicitly as
\begin{equation}
\label{brst psi12 general}
\left(\frac{a_1}{\sqrt{k}} \, \psi_\frac{1}{2}^3
+\frac{a_2}{\sqrt{k}}\, \psi_\frac{1}{2}^{\prime 3}
+a_3\,\psi_\frac{1}{2}^x +\psi^w_\frac{1}{2}
\right)\,V^{(0)} = 0\,.
\end{equation}
The BRST constraint (\ref{brst psi12 general}) can be solved as
\begin{equation}
\xi_w = a_3\,\xi_x+a_2\sqrt{k}\,\xi_3^\prime\,.
\end{equation}
The most general massless holomorphic BRST closed state is then given by
\begin{align}
\label{V0 brst closed}
V^{(\frac{1}{2})} &=\left[ \xi_+\,\psi^+_{-\frac{1}{2}} \,\J^-_0+ \xi_-\,\psi^-_{-\frac{1}{2}} \,\J^+_0
+\xi\,\psi^{(0)}_{-\frac{1}{2}} + \sum_{i=1}^4
\xi_{i}\,\psi^{i}_{-\frac{1}{2}}\right.\\
&+\left.  \xi_+'\,\psi^{\prime +}_{-\frac{1}{2}} \,\J^{\prime -}_0+  \xi_-'\,\psi^{\prime -}_{-\frac{1}{2}}\,\J^{\prime +}_0
+ \xi_3' \, (\psi^{\prime 3}_{-\frac{1}{2}}+a_2\sqrt{k}\,\psi^w_{-\frac{1}{2}} )
+\xi_x\, (\psi^x_{-\frac{1}{2}} +a_3\,\psi^w_{-\frac{1}{2}}) \right]V^{(0)}\,.\notag
\end{align}
Here the polarization $\xi$ is un-physical, and represents a pure-gauge exact state on the coset.
After imposing the coset BRST physical
state conditions, the vertex operator (\ref{V0 brst closed}) then possesses ten independent polarizations.

Next, let us discuss the super-Virasoro constraints (\ref{super Virasoro}). For the massless state, $V^{(\frac{1}{2})} $,
the only non-trivial constraints are given by
\begin{equation}
\left(L_0-\frac{1}{2}\right)\, V^{(\frac{1}{2})}  = 0\,,\qquad G_\frac{1}{2}\, V^{(\frac{1}{2})} =0\,,
\end{equation}
and its counterparts in the anti-holomorphic sector.
The $L_0$ constraint is simply (\ref{onshell}) with $N=\bar N=1/2$,
giving the mass-shell condition for the massless state.
The super-Virasoro $G_{1/2}$ constraint restricts allowed physical polarizations.
First, using (\ref{j3 and V OPE}), (\ref{jxw and V OPE}), (\ref{G psi anticom}), (\ref{BRST constraints prelim res})
we obtain
\begin{equation}
\{G_\frac{1}{2},\, \psi^{(0)}_{-\frac{1}{2}}\} \, V^{(0)}= 0\,,
\end{equation}
and therefore the BRST-exact state $\psi^{(0)}_{-\frac{1}{2}}\,V^{(0)}$ in (\ref{V0 brst closed})
automatically satisfies the super-Virasoro $G_{1/2}$ constraint.
Using (\ref{jvopesf}), (\ref{jvopesu}) we obtain
\begin{equation}
\label{jpm acting on V}
\begin{aligned}
\J^\mp _0\,\J^\pm_0\, V_{j;m,\bar m} &= (m\mp j)(m\pm 1 \pm j)\, V_{j;m,\bar m}\,,\\
\J^{\prime \mp} _0\,\J^{\prime \pm}_0\, V^\prime_{j';m',\bar m'} &=
 (j'(j'+1)-m'(m'\pm 1))\, V^\prime _{j';m',\bar m'}\,.
\end{aligned}
\end{equation}
Then using (\ref{G psi anticom}) we arrive at
\begin{equation}
\label{G12 V}
\begin{aligned}
G_\frac{1}{2}\,V^{(\frac{1}{2})}  {=} \left[\hat \xi_+ {+} \hat \xi_- {+}\hat \xi_+' {+} \hat \xi_-' 
{+}\xi_3'\,\left(m' {-}\frac{a_2\ell_s\sqrt{k}\,\kappa}{2}\right){+}\frac{\ell_s}{2}\,\left(
\xi_x\,(p_L{-}a_3\kappa) {+}\sum_{i=1}^4 \xi_{i}\,p_L^i\right) \right]\,V^{(\frac{1}{2})} 
\end{aligned}
\end{equation}
where we denoted
\begin{equation}
\begin{aligned}
\hat\xi_+ &= (m+j)(m-1-j)\,\xi_+\,,\qquad &&\hat\xi_- = (m-j)(m+1+j)\,\xi_-\,,\\
\hat\xi_+' &= (j'(j'+1) - m'(m'-1))\,\xi_+'\,,\qquad &&\hat\xi_-' = (j'(j'+1) - m'(m'+1))\,\xi_-'\,.
\end{aligned}
\end{equation}
Taking into account the BRST constraints (\ref{BRST constraints prelim res}), that we
solve for $\kappa$, the Virasoro constraint following from (\ref{G12 V}) then reads
\begin{equation}
\begin{aligned}
\hat \xi_+ + \hat \xi_- +\hat \xi_+' + \hat \xi_-' +m'\,\xi_3'
&+\frac{\ell_s}{2}\,\left(\xi_x\,p_L+\sum_{i=1}^4 \xi_{i}\,p_L^i\right)\\
&=\left(\xi_3'\,a_2\,\sqrt{k}+\xi_x\,a_3\right)\,
\left(\frac{\ell_s\,p_L\,a_3}{2}+\frac{m\,a_1+m'\,a_2}{\sqrt{k}}\right)\,.
\end{aligned}
\end{equation}
We can solve this for $\hat \xi_\pm$ simply as
\begin{equation}
\label{solution for xipm}
\begin{aligned}
\hat \xi_\pm &= 
\pm\,\hat\xi -\frac{1}{2}\,\left(\hat \xi_+' + \hat \xi_-' +m'\,\xi_3'
+\frac{\ell_s}{2}\,\left(\xi_x\,p_L+\sum_{i=1}^4 \xi_{i}\,p_L^i\right)\right.\\
&+\left.\left(\xi_3'\,a_2\,\sqrt{k}+\xi_x\,a_3\right)\,
\left(\frac{\ell_s\,p_L\,a_3}{2}+\frac{m\,a_1+m'\,a_2}{\sqrt{k}}\right)\right)\,.
\end{aligned}
\end{equation}
The most general massless holomorphic physical vertex operator is then given by
(\ref{V0 brst closed}) with polarizations satisfying (\ref{solution for xipm}).
It is also defined up to 
the Virasoro exact massless null state, given by
\begin{equation}
G_{-\frac{1}{2}}\,V^{(0)} = \left[\frac{2}{k}\,\left(\psi^a_{-\frac{1}{2}}\,\J^a_0
+\psi^{\prime a}_{-\frac{1}{2}}\,\J^{\prime a}_0\right)+\psi^x_{-\frac{1}{2}}\,j^x_0
-\psi^w_{-\frac{1}{2}}\,j^w_0+\sum_{i=1}^4\psi^i_{-\frac{1}{2}}\,j^i_0\right]\,V^{(0)} \,,
\end{equation}
therefore rendering a massless string state with eight independent physical polarizations.\\

In preparation for   discussion of holography
in section~\ref{sec:holography}, we are now going to consider (anti-)holomorphic
massless vertex operators polarized along the direction 
of propagation of excitations, as well as along the $U(1)_x$ circle.
Without loss of generality, we can define the former to be 
the coordinate $z^1$ on the torus $\mathbb{T}^4$, $p^1_L = p^1_R = p$, 
setting the rest of the momenta on the torus to zero, $p^i_{L,R} = 0$, $i=2,3,4$.
We are also going to perform the KK reduction on the $U(1)_x$ circle,
setting the corresponding momenta to zero, $p_L=p_R = 0$. (Anti-)holomorphic vertex operators polarized along the $U(1)_x$
circle will then be found as constituents of the gauge field vertex operators of type-II superstring theory
in the nine-dimensional space-time.

In other words, we will be looking for the physical massless vertex operators
of the form $\Psi^1 _{-\frac{1}{2}}\,V^{(0)}$ and $\Psi^x _{-\frac{1}{2}}\,V^{(0)}$,
with $\Psi^1 _{-\frac{1}{2}} = \psi^1_{-\frac{1}{2}}+\dots$ and $\Psi^x _{-\frac{1}{2}} = \psi^x_{-\frac{1}{2}}+\dots$,
where we will fill in the missing terms accordingly with the expression for the physical vertex operator
(\ref{V0 brst closed}), (\ref{solution for xipm}).
Specifically, in the former case we set $\xi_1 = 1$, $\xi_x = 0$, while
in the latter case we set $\xi_x = 1$, $\xi_1 = 0$. We also set $\xi_{2,3,4} = 0$, $\xi_\pm ' =0$, $\xi_3'=0$, $\hat\xi = 0$
everywhere. For convenience, we will denote
\begin{equation}
\Psi^i = \psi^i\,,\qquad \tilde\Psi^i = \tilde\psi^i\,, \qquad i=2,3,4\,.
\end{equation}
In general, physical massless states can then be defined as
\begin{equation}
V^{(\frac{1}{2})} = \xi_m\,\Psi^m_{-\frac{1}{2}}\cdot V^{(0)}\,,\qquad
\tilde V^{(\frac{1}{2})} = \tilde\xi_m\,\tilde \Psi^n_{-\frac{1}{2}}\cdot V^{(0)}\,.
\end{equation}
With such a convention, physical massless states are constructed using
creation operators $\Psi^m_{-\frac{1}{2}}$, $\tilde\Psi^m_{-\frac{1}{2}}$,
that automatically take into account the coset BRST and super-Virasoro constraints.
Using (\ref{V0 brst closed}), (\ref{solution for xipm}), we then obtain
\begin{align}
\Psi^1_{-\frac{1}{2}} &=\psi^1_{-\frac{1}{2}}-\frac{\ell_s\, p }{4(m+j)(m-1-j)}\,\J^-_0\psi^+_{-\frac{1}{2}}
-\frac{\ell_s\, p }{4(m-j)(m+1+j)}\,\J^+_0\psi^-_{-\frac{1}{2}}\,,\\
\Psi^x_{-\frac{1}{2}} &=\psi^x_{-\frac{1}{2}}{-}\frac{(m\,a_1+m'\,a_2)\,a_3}{2\sqrt{k}\,(m{+}j)(m{-}1{-}j)}\,\J^-_0\psi^+_{-\frac{1}{2}}
{-}\frac{(m\,a_1+m'\,a_2)\,a_3}{2\sqrt{k}\,(m{-}j)(m{+}1{+}j)}\,\J^+_0\psi^-_{-\frac{1}{2}}\,,
\end{align}
and similarly for the anti-holomorphic components.

Using (\ref{jpm acting on V}), (\ref{psi r anticom}), we obtain
the two-point correlation functions
\begin{align}
\notag
\langle \Psi^1_{\frac{1}{2}} \Psi^1_{-\frac{1}{2}}\rangle_0&=
\left(\frac{1}{2} + \frac{k\,\ell_s^2\,p^2}{16}\,
F_{j;m}\right)\,\langle V^{(0)}V^{(0)}\rangle\,,\\
\langle \Psi^x_{\frac{1}{2}} \Psi^x_{-\frac{1}{2}}\rangle_0&=
\left(\frac{1}{2} + \frac{(m\,a_1+m'\,a_2)^2\,a_3^2}{4}\,
F_{j;m}\right)\,\langle V^{(0)}V^{(0)}\rangle\,,\label{holomorphic correlators}\\
\langle \Psi^x_{\frac{1}{2}} \Psi^1_{-\frac{1}{2}}\rangle_0&=
\langle \Psi^1_{\frac{1}{2}} \Psi^x_{-\frac{1}{2}}\rangle_0=
\frac{(ma_1+m'a_2)a_3\ell_sp\sqrt{k}}{8}\,
F_{j;m}\,\langle V^{(0)}V^{(0)}\rangle\,,\notag
\end{align}
where we have denoted
\begin{equation}
\label{Fjm}
F_{j;m} = 
\frac{1}{(m-j)(m+1+j)}+\frac{1}{(m+j)(m-1-j)}\,,
\end{equation}
and used a short-hand notation $\langle \cdots \rangle_0$
for averages in the state $V^{(0)}\tilde V^{(0)} |0\rangle$.
Analogous correlation functions can be written down in the anti-holomorphic sector.
The tachyon two-point function $\langle V^{(0)}V^{(0)}\rangle$
is given by a product of the two-point functions
of its constituent vertex operators (\ref{ground state W VN}), (\ref{def of ground state v.o.}).
For the full state, including both the anti-holomorphic and the holomorphic sectors, we obtain
\begin{equation}
\langle V^{(0)}V^{(0)}\rangle 
\langle \tilde V^{(0)}\tilde V^{(0)}\rangle = \langle V_{j;m,\bar m} V_{j;m,\bar m}\rangle
\langle V_{j';m',\bar m'}^\prime V_{j';m',\bar m'}^\prime\rangle
\times \textrm{flat\;space\;correlators}\,,
\end{equation}
where one should substitute (\ref{VV2ptfn}), (\ref{su2ptfn})
for the two-point functions of the $SL(2,\mathbb{R})$
and $SU(2)$ ground state vertex operators.
Notice that zero of (\ref{VV2ptfn}) at $m = - j$
cancels out the simple pole of (\ref{Fjm}) at $m = -j$.
At the same time, $F_{j;m}$ has a simple pole at $m = j$,
which is inherited by all the correlation functions (\ref{holomorphic correlators}).
Similarly, the anti-holomorphic sector has a simple pole at $\bar m = - j$,
while exhibiting regular behavior at $\bar m = j$.

\subsection{Spectrum of single-trace $T\bar{T}$-deformed symmetric product CFT}
\label{sec:spectrum of single-trace TTbar}

We will end this section with a discussion of the spectrum of the long strings
in the coset model (\ref{big coset}),
and its relation to the spectrum obtained by  a deformation of CFT$_2$ by
the $T\bar{T}$ operator. As it turns out, the spectrum of a single long string in the sector with the unit winding number,  $\mathrm{w}=1$, corresponds to the spectrum of the
double-trace $T\bar{T}$ deformed CFT \cite{Chakraborty:2020yka}. For a detailed review of the single-trace
$T\bar{T}$ deformation of string theory in $AdS_3$, and its relation to the long strings, we refer the reader to \cite{Chakraborty:2019mdf}.

Let us begin by reviewing certain aspects of the theory
on a single long string in $AdS_3$. Via the AdS/CFT correspondence,
string theory in $AdS_3\times \mathcal{N}_7$ is dual to a CFT$_2$
living on the boundary of $AdS_3$.\footnote{In section~\ref{sec:holography}
we will discuss holographic description of a $4+1$ dimensional boundary field theory,
with the spatial coordinates spanned by the $\mathbb{T}^4$.}
Here $\mathcal{N}_7$ stands for a seven-dimensional compact space,
{\it{e.g.}},  $\mathcal{N}_7=\mathbb{S}^3\times\mathbb{T}^4$. In the presence of
pure NS-NS flux in the bulk, the space-time theory has an $SL(2,\mathbb{C})$-invariant vacuum.
This is either an NS vacuum, that corresponds to global $AdS_3$ in the bulk, or 
an R vacuum, that corresponds to a massless BTZ (denoted by $BTZ_{M=J=0}$).
The NS sector contains states that belong to the discrete as well as 
the continuous series representation of $SL(2,\mathbb{R})$.  The continuous series representation starts above a gap of order $k/2$. The R sector, on the other hand, contains a continuum of long strings above a gap of order $1/k$. In the rest of the discussion of
 the spectrum of long strings in this section, we will consider only the long string states in the R sector. 

On the $AdS_3\times \mathcal{N}_7$ with pure NS-NS flux,  the theory on a single long string is given by
the product of the ${\cal N}=2$ Liouville theory $\mathbb{R}_\phi$ and the compact SCFT on
$\mathcal{N}_7$ \cite{Seiberg:1999xz}
 \begin{equation}
 \mathcal{M}_{6k}^{(L)}=\mathbb{R}_\phi\times \mathcal{N}_7\,,
 \end{equation}
where $c_{\mathcal{M}}=6k$ is the central charge of $\mathcal{M}_{6k}^{(L)}$  and the theory on $\mathbb{R}_\phi$ has a dilaton linear in the radial direction of $AdS_3$ denoted by $\phi$ with the slope
\begin{equation}
Q^{(L)}=(k-1)\sqrt{\frac{2}{k}}\,.
\end{equation}
The effective coupling on the long strings is given by
\begin{equation}
g^{(L)}\simeq e^{Q^{(L)}\phi}
\end{equation}
implying that for $k>1$, the long strings become strongly coupled as they move towards the boundary. But there is a wide range of positions along the radial direction where they are weakly coupled. The effective theory of $N$ such long strings is described by the symmetric product \cite{Argurio:2000tb,Chakraborty:2019mdf}
\begin{equation}\label{symprodls}
\frac{(\mathcal{M}_{6k}^{(L)})^N}{S_N}.
\end{equation} 

Let us now proceed to discussion of long strings on the coset (\ref{big coset}).
Recall that the value of
$\kappa$ is equal to the energy of a string excitation mode
propagating in the coset background. The long strings in that background belong to the continuous representation of $SL(2,\mathbb{R})$, for which we define, due to (\ref{C}),
\begin{equation}
j=-\frac{1}{2}+is\,,\qquad s\in \mathbb{R}\,.
\end{equation}
For  a non-vanishing winding $\mathrm{w}$ around the $U(1)_x$-circle,
the energy $E$ of such a long string  in the spectrally un-flowed sector ({\it{i.e.}}, $\hat{\omega}=\hat{\omega}'=\hat{\bar{\omega}}'=0$), above the BPS configuration is given by
\begin{equation}\label{eabps}
E=\kappa-\frac{\mathrm{w}R_x}{\ell_s^2}\,,
\end{equation}
where due to (\ref{kappa}) we have
\begin{equation}
\label{unspectkappa}
\kappa^2=\left(\frac{n}{R_x}\right)^2+\left(\frac{\mathrm{w}R_x}{\ell_s^2}\right)^2+\frac{2}{\ell_s^2}\left( -2\, \frac{j(j+1)-j'(j'+1)}{k}
+\Delta_{\mathbb{T}^4}+\bar{\Delta}_{\mathbb{T}^4}+N+\bar{N}-1\right).
\end{equation}

Due to the dispersion relation of the long strings in the winding $\mathrm{w}$ sector in $BTZ_{M=J=0}\times \mathbb{S}^3\times \mathbb{T}^4$ that is given by \cite{Parsons:2009si}
\begin{equation}\label{lse}
\begin{aligned}
E_{L}&=\frac{1}{\mathrm{w}}\left[-\frac{j(j+1)}{k}+\frac{j'(j'+1)}{k}+\Delta_{\mathbb{T}^4}+N-\frac{1}{2} \right],\\
E_{R}&=\frac{1}{\mathrm{w}}\left[-\frac{j(j+1)}{k}+\frac{j'(j'+1)}{k}+\bar{\Delta}_{\mathbb{T}^4}+\bar{N}-\frac{1}{2} \right]\,,
\end{aligned}
\end{equation}
expression (\ref{unspectkappa}) can be written as 
\begin{equation}
\label{unspectkappa1}
\kappa^2=\left(\frac{n}{R_x}\right)^2+\left(\frac{\mathrm{w}R_x}{\ell_s^2}\right)^2+\frac{2\,\mathrm{w}}{\ell_s^2}\left(
E_L+E_R\right).
\end{equation}

Expression \eqref{lse} also gives the spectrum of the $\mathbb{Z}_\mathrm{w}$ twisted sector of the symmetric product CFT $\mathcal{M}^N/S_N$ where the block CFT $\mathcal{M}$ is the CFT with central charge \cite{Argurio:2000tb}
\begin{eqnarray}\label{cm}
c_\mathcal{M}=6k 
\end{eqnarray}
living on a single long string with winding one. This is not surprising because the effective theory on the long strings are well described by a symmetric product \eqref{symprodls} at weak coupling \cite{Argurio:2000tb}. Spectrum of operators of dimension $h_{\mathrm{w}}$ in the $\mathbb{Z}_\mathrm{w}$ twisted sector of the symmetric product CFT in the Ramond sector above the Ramond vacuum is given by \cite{Klemm:1990df,Argurio:2000tb} 
\begin{eqnarray}\label{spectsymo}
E_L=h_\mathrm{w}-\frac{k\mathrm{w}}{4}~, \ \ \ \ E_{R}=\bar{h}_\mathrm{w}-\frac{k\mathrm{w}}{4}~.
\end{eqnarray}
For every operator of dimension $h_1$ in the untwisted sector there exists an operator in the $\mathbb{Z}_\mathrm{w}$ twisted sector whose dimension is related to $h_1$ as \cite{Klemm:1990df}
\begin{eqnarray}\label{homega}
h_\mathrm{w}=\frac{h_1}{\mathrm{w}}+\frac{k}{4}\left(\mathrm{w}-\frac{1}{\mathrm{w}}\right)\,,\qquad
\bar h_\mathrm{w}=\frac{\bar h_1}{\mathrm{w}}+\frac{k}{4}\left(\mathrm{w}-\frac{1}{\mathrm{w}}\right).
\end{eqnarray}
Next we solve \eqref{eabps} for $\kappa$ in terms of $E$,
and we also substitute  (\ref{homega}) into  (\ref{spectsymo}).
Plugging the result into \eqref{unspectkappa1} one obtains \cite{Giveon:2017nie,Chakraborty:2018vja,Chakraborty:2019mdf,Chakraborty:2020yka}
\begin{eqnarray}\label{spect1}
\left(E+\frac{\mathrm{w} R_x}{\ell_s^2}\right)^2-\left(\frac{\mathrm{w} R_x}{\ell_s^2}\right)^2=\frac{2}{\ell_s^2}\left(h_1+\bar{h}_1-\frac{k}{2}\right)+\left(\frac{n }{R_x}\right)^2,
\end{eqnarray}
and 
\begin{eqnarray}\label{hhbardiff}
\bar{h}_1-h_1=n.
\end{eqnarray}
In the untwisted sector corresponding to $\mathrm{w}=1$, \eqref{spect1} takes the form \cite{Smirnov:2016lqw,Cavaglia:2016oda}
\begin{eqnarray}\label{ttbarspect}
E=-\frac{R_x}{\ell_s^2}+\sqrt{\left(\frac{R_x}{\ell_s^2}\right)^2+\frac{2}{\ell_s^2}\left(h_1+\bar{h}_1-\frac{c_{\mathcal{M}}}{2}\right)+\left(\frac{h_1-\bar{h}_1}{R_x}\right)^2}~,
\end{eqnarray}
which is the spectrum of a CFT$_2$ on a cylinder of radius $R_x$ deformed by an irrelevant operator $\delta\mathcal{L}=-t \,T\bar{T}$ with $t=\ell_s^2>0$.  For $\mathrm{w}>1$, \eqref{spect1} is the spectrum of the $\mathbb{Z}_\mathrm{w}$ twisted sector of single trace $T\bar{T}$ deformation of a symmetric product CFT \cite{Hashimoto:2019hqo}.

\section{Holography}
\label{sec:holography}

In this section, we are going to derive various two-point correlation
functions of the field theory holographically dual to the bulk string theory
on the coset (\ref{big coset}).
We are going to explore different regimes of the holographic correspondence,
and calculate various correlation functions in the NS-NS sector of string theory in the bulk. 

In subsection~\ref{sec:collective excitations}, we will consider the bulk configuration, derived in section~\ref{sec:two charge},
obtained by compactification of the $U(1)_x$
circle of the coset (\ref{big coset}).
This background is characterized by a rotating black brane geometry with an event horizon,
finite temperature, and two independent non-trivial $U(1)$ gauge field profiles.
We will study massless  excitations modes
described by the metric, $B$-field, dilaton, and gauge field, on top of this  bulk configuration.
The corresponding holographic
dual $4+1$ dimensional field theory in the asymptotic region is a finite-temperature
system, with two conserved $U(1)$ charges having non-vanishing vacuum expectations.
We are interested in gapless collective excitations
in this system, that we determine via holographic world-sheet calculation of two-point functions
of the stress-energy tensor and charge currents.

In subsection~\ref{sec:2d correlation functions}, we compactify the $\mathbb{T}^4$
sub-space of the coset (\ref{big coset}), which allows us to study the $1+1$ dimensional boundary field
theory. The resulting expressions that we obtain
for for the two-point correlation functions in momentum space interpolate between
the well-known CFT$_2$ expressions in the IR, calculated within the AdS$_3$/CFT$_2$
correspondence, and non-local expressions for LST in the UV.

\subsection{Collective excitation modes}
\label{sec:collective excitations}

In section~\ref{sec:massless states}, we constructed the (anti-)holomorphic vertex
operators ($\tilde V^{(\frac{1}{2})}$) $V^{(\frac{1}{2})}$
describing massless NS (right-) left-moving string states. The  
NS-NS states of closed type-II superstring are obtained by taking
a direct product of the (anti-)holomorphic states. The corresponding
vertex operator  $V^{(\frac{1}{2})}\otimes \tilde V^{(\frac{1}{2})}$
can then be split into irreducible representations of the target space symmetry group,
rendering the graviton, $B$-field, and dilaton vertex operators, and forming
the NS-NS bosonic degrees of freedom of the type-II supergravity multiplet.

These operators describe excitation on top of the metric, $B$-field, and dilaton ten-dimensional background configuration
 (\ref{full metric result}), (\ref{B field result 1}),  (\ref{dilaton result}) 
that we derived in section~\ref{sec:background}, as well as the nine-dimensional
two-charge background (\ref{9d background}) derived in section~\ref{sec:two charge}.
An excitation mode is characterized by the energy $\kappa$, and momentum $p$.
Without loss of generality, we can align the momentum with
the direction $z^1$ on the torus $\mathbb{T}^4$, $p^1_L = p^1_R = p$, 
setting the rest of the momenta to zero, $p^i_{L,R} = 0$, $i=2,3,4$.

In this section, we are going to discuss holographic interpretation
of string excitations in the bulk in terms of the dual $4+1$ dimensional field theory on the boundary.\footnote{
This is to be contrasted with discussion in section~\ref{sec:spectrum of single-trace TTbar},
where we compactify $\mathbb{T}^4$ and study  $1+1$ dimensional field theory on the boundary.}
Recall that the field theory dual of a charged black brane describes
matter at finite density.\footnote{See, \textit{e.g.}, \cite{Goykhman:2012vy,Goykhman:2014enw} and references therein.} In particular, we are interested in
the corresponding  collective excitation modes, that we can compare with the 
hydrodynamic theory.
These appear as poles of two-point correlation functions of the stress-energy tensor.
In general, these correlation functions also include conserved charge current, and form correlation matrices.
Correspondingly,
the bulk holographic dual of the stress-energy tensor is given by the fluctuating metric field,
which is coupled to fluctuations of the gauge fields, $B$-field, and dilaton.\footnote{See \cite{Parnachev:2005hh,Goykhman:2013oja} for analogous calculation in coset theories involving
$SL(2,\mathbb{R})/U(1)$ and $SL(2,\mathbb{R}) \times U(1)/U(1)$.}

For the space-like polarizations $m, n=x,\,i $, where $i=1,2,3,4$, along the 
circle $U(1)_x$ and the torus sub-space $\mathbb{T}^4$,
we write down the NS-NS type-II supergravity vertex operators
\begin{equation}
\begin{aligned}
g^{mn} &= \left(\Psi^m_{-\frac{1}{2}}\,\tilde \Psi^n_{-\frac{1}{2}}
+\Psi^n_{-\frac{1}{2}}\,\tilde \Psi^m_{-\frac{1}{2}}\right)\,V^{(0)}\,\tilde V^{(0)}\,,\\
b^{mn} &= \left(\Psi^m_{-\frac{1}{2}}\,\tilde \Psi^n_{-\frac{1}{2}}
-\Psi^n_{-\frac{1}{2}}\,\tilde \Psi^m_{-\frac{1}{2}}\right)\,V^{(0)}\,\tilde V^{(0)}\,,\\
\varphi &= \Psi^m_{-\frac{1}{2}}\,\tilde \Psi^m _{-\frac{1}{2}}\,V^{(0)}\,\tilde V^{(0)}\,.
\end{aligned}
\end{equation}
Performing the Kaluza-Klein reduction on the $U(1)_x$ circle, we set $p_L=p_R = 0$,
and derive the gauge fields vertex operators
\begin{equation}
\begin{aligned}
a^{m} &= \left(\Psi^x_{-\frac{1}{2}}\,\tilde \Psi^m_{-\frac{1}{2}}+\Psi^m_{-\frac{1}{2}}\,\tilde \Psi^x_{-\frac{1}{2}}
\right)\,V^{(0)}\,\tilde V^{(0)}\,,\\
b^{m} &= \left(\Psi^x_{-\frac{1}{2}}\,\tilde \Psi^m_{-\frac{1}{2}}
-\Psi^m_{-\frac{1}{2}}\,\tilde \Psi^x_{-\frac{1}{2}}\right)\,V^{(0)}\,\tilde V^{(0)}\,.
\end{aligned}
\end{equation}

These vertex operators can be further split into three groups,
defined by the spin rotation properties in the sub-space $\alpha,\,\beta=2,3,4$
transverse to the direction $z^1$ of propagation of excitations:
\begin{equation}
\begin{aligned}
\textrm{spin-0\;\;sound\;channel:}\qquad &g^{11},\;a^1,\; b^1\,,\\
\textrm{spin-1\;\;shear\;channel:}\qquad &g^{1\alpha},\;b^{1\alpha},\;a^\alpha,\; b^\alpha\,,\\
\textrm{spin-2\;\;scalar\;channel:}\qquad &g^{\alpha\beta},\;b^{\alpha\beta},\;\varphi\,.
\end{aligned}
\end{equation}
Notice that due to rotation symmetry, operators from different groups do not couple to each other.

Furthermore, the $B$-field excitations and the $b$ gauge field mix
with the graviton excitations and the $a$ gauge field. These can be decoupled from each other
into the left-moving and right-moving sectors
as follows:
\begin{equation}
\begin{aligned}
\textrm{spin-0\;\;sound\;channel:}\qquad &\left\{ \begin{aligned} 
 W^1 &= \frac{1}{2}\,(a^1+b^1) ,\;g^{11}\\
 U^1 &= \frac{1}{2}\,(a^1-b^1) 
\end{aligned} \right.\\
\textrm{spin-1\;\;shear\;channel:}\qquad &\left\{ \begin{aligned} 
 W^\alpha &= \frac{1}{2}\,(a^\alpha+b^\alpha) ,
 \;S^{1\alpha}=\frac{1}{2}(g^{1\alpha}+b^{1\alpha})\\
 U^\alpha &= \frac{1}{2}\,(a^\alpha-b^\alpha) ,
 \; R^{1\alpha}=\frac{1}{2}(g^{1\alpha}-b^{1\alpha})
\end{aligned} \right.\\
\textrm{spin-2\;\;scalar\;channel:}\qquad &\left\{ \begin{aligned} 
S^{\alpha\beta}&=\frac{1}{2}(g^{\alpha\beta}+b^{\alpha\beta})\\
R^{\alpha\beta}&=\frac{1}{2}(g^{\alpha\beta}-b^{\alpha\beta})
\end{aligned} \right.
\end{aligned}
\end{equation}
Here we have (recall that $m,\,n=1,2,3,4$, and $\alpha,\,\beta=2,3,4$)
\begin{align}
W^m&=  \Psi^x_{-\frac{1}{2}}\,\tilde\Psi^m_{-\frac{1}{2}}\,V^{(0)}\,\tilde V^{(0)}\,,\qquad
U^m= \Psi^m_{-\frac{1}{2}}\, \tilde\Psi^x_{-\frac{1}{2}}\,V^{(0)}\,\tilde V^{(0)}\,,\\
S^{mn} &=\Psi^m_{-\frac{1}{2}}\,\tilde \Psi^n_{-\frac{1}{2}}\,V^{(0)}\,\tilde V^{(0)}\,,\qquad
R^{mn} =\Psi^n_{-\frac{1}{2}}\,\tilde \Psi^m_{-\frac{1}{2}}\,V^{(0)}\,\tilde V^{(0)}\,.
\end{align}
Non-trivial two-point correlation functions in the sound channel are then given by
\begin{equation}
\begin{aligned}
\langle W^1W^1\rangle_0 &= \langle \tilde\Psi^1_{\frac{1}{2}}\tilde\Psi^1_{-\frac{1}{2}}
 \rangle _0\,\langle \Psi^x_{\frac{1}{2}}\Psi^x _{-\frac{1}{2}}\rangle _0\,,\\
\langle g^{11}W^1\rangle_0 &= \langle \Psi^1_{\frac{1}{2}}
\Psi^x_{-\frac{1}{2}} \rangle _0\,\langle \tilde\Psi^1_{\frac{1}{2}}\tilde\Psi^1_{-\frac{1}{2}} \rangle _0\,,\\
\langle g^{11}g^{11}\rangle_0 &= \langle \Psi^1_{\frac{1}{2}}
\Psi^1_{-\frac{1}{2}} \rangle _0\,\langle \tilde\Psi^1_{\frac{1}{2}}\tilde\Psi^1_{-\frac{1}{2}} \rangle _0\,,
\end{aligned}
\end{equation}
and
\begin{equation}
\langle U^1U^1\rangle_0 = \langle \tilde\Psi^x_{\frac{1}{2}}\tilde\Psi^x_{-\frac{1}{2}} \rangle _0
\,\langle \Psi^1_{\frac{1}{2}}\Psi^1_{-\frac{1}{2}} \rangle _0\,.
\end{equation}
Similarly, the shear channel non-vanishing correlation functions are  given by
\begin{equation}
\begin{aligned}
\langle W^\alpha W^\beta \rangle_0 &= \langle \tilde\Psi^\alpha_{\frac{1}{2}}\tilde\Psi^\beta_{-\frac{1}{2}}
 \rangle _0\,\langle \Psi^x_{\frac{1}{2}}\Psi^x _{-\frac{1}{2}}\rangle _0\,,\\
 \langle S^{1\alpha} S^{1\beta} \rangle_0 &= \langle \Psi^1_{\frac{1}{2}}\Psi^1_{-\frac{1}{2}}
 \rangle _0\,\langle \tilde\Psi^\alpha_{\frac{1}{2}}\tilde\Psi^\beta _{-\frac{1}{2}}\rangle _0\,,\\
 \langle W^\alpha S^{1\beta} \rangle_0 &= \langle \tilde\Psi^\alpha_{\frac{1}{2}}\tilde\Psi^\beta_{-\frac{1}{2}}
 \rangle _0\,\langle \Psi^x_{\frac{1}{2}}\Psi^1 _{-\frac{1}{2}}\rangle _0\,,
\end{aligned}
\end{equation}
and
\begin{equation}
\begin{aligned}
\langle U^\alpha U^\beta \rangle_0 &= \langle \Psi^\alpha_{\frac{1}{2}}\Psi^\beta_{-\frac{1}{2}}
 \rangle _0\,\langle \tilde\Psi^x_{\frac{1}{2}}\tilde\Psi^x _{-\frac{1}{2}}\rangle _0\,,\\
 \langle R^{1\alpha} R^{1\beta} \rangle_0 &= \langle \tilde\Psi^1_{\frac{1}{2}}\tilde\Psi^1_{-\frac{1}{2}}
 \rangle _0\,\langle \Psi^\alpha_{\frac{1}{2}}\Psi^\beta _{-\frac{1}{2}}\rangle _0\,,\\
 \langle U^\alpha R^{1\beta} \rangle_0 &= \langle \Psi^\alpha_{\frac{1}{2}}\Psi^\beta_{-\frac{1}{2}}
 \rangle _0\,\langle \tilde\Psi^x_{\frac{1}{2}}\tilde\Psi^1 _{-\frac{1}{2}}\rangle _0\,.
\end{aligned}
\end{equation}
In these expressions we have $\langle \Psi^\alpha_{\frac{1}{2}}\Psi^\beta_{-\frac{1}{2}}
\rangle _0=\frac{1}{2}\,\delta^{\alpha\beta}$, $\langle \tilde\Psi^\alpha_{\frac{1}{2}}\tilde\Psi^\beta_{-\frac{1}{2}}
\rangle _0=\frac{1}{2}\,\delta^{\alpha\beta}$, while the rest of the two-point
functions are given by (\ref{holomorphic correlators}) and its anti-holomorphic counterparts.
These correlation functions were shown in section~\ref{sec:massless states}
to possess simple poles at $j = m$ and $j = -\bar m$. 

Let us locate the gapless shear diffusion mode. At zero propagation velocity, for the mass $M$, pressure $P$,
volume $V$, and viscosity $\eta$, we have (for the dimensionless $\omega$, $p$, measured in units of string length)
\begin{equation}
\omega = -\frac{i\eta}{(M+P)/V}\,p^2\,.
\end{equation}
Defining entropy density $s = {\cal S} / V$, taking into account
that due to (\ref{vanishing grand potential}) the pressure vanishes, $P = {\cal F} = 0$,
 and using expressions for the mass $M = {\cal E}$
given by (\ref{energy gen result}), and entropy ${\cal S}$ given by (\ref{BH entropy result}), we obtain
\begin{equation}
\label{gen hydro eta over s}
\frac{\eta}{s} = \frac{{\cal E}}{{\cal S}}\,\frac{\omega}{-ip^2} = \frac{1}{\pi\sqrt{k}\,(a_1+b_1)}\,\frac{\omega}{-i\,\ell_s\,p^2}\,.
\end{equation}
At the same time, the mass-shell condition (\ref{onshell}) at $N=\bar N=1/2$ gives
\begin{equation}
\label{diff mass shell}
\frac{\ell_s^2\,\kappa^2}{4} = \frac{j(j+1)-j'(j'+1)}{k}+\frac{\ell_s^2\,p^2}{4}
\end{equation}
while the BRST constraints (\ref{BRST constraints prelim res}) impose
\begin{equation}
\label{diff brst constraint}
\frac{\ell_s\,\kappa}{2} = \frac{m\,a_1+m'\,a_2}{\sqrt{k}}
 = - \frac{\bar m\,b_1+\bar m'\,b_2}{\sqrt{k}}\,.
\end{equation}
Using results of section~\ref{sec:asymptotic vertex operator} and performing Wick rotation, we obtain $\kappa = - i \omega$.
From (\ref{diff brst constraint}) we can therefore express
\begin{equation}
m = - \frac{1}{a_1}\,\left( i\,\frac{\ell_s\,\sqrt{k}\,\omega}{2} + m'\,a_2 \right)\,,\qquad
\bar m = \frac{1}{b_1}\,\left(i\,\frac{\ell_s\,\sqrt{k}\,\omega}{2} + \bar m'\,b_2 \right)\,,
\end{equation}
while from (\ref{diff mass shell}) we derive
\begin{equation}
j  = \frac{1}{2}\,\left(-1 + \sqrt{1+4j'(j'+1)-k\,\ell_s^2\,(p^2+\omega^2)}\right)\,.
\end{equation}

As discussed above, excitation modes are located at $j = m$ and $j = -\bar m$,
coming from the holomorphic and the anti-holomorphic sector respectively.
At $j'=0$, $m'=0$, these give rise to the modes
\begin{equation}
\omega_L = -i\frac{\sqrt{k} \, a_1}{2}\,\ell_s\,p^2\,,\qquad
\omega_R = -i\frac{\sqrt{k} \, b_1}{2}\,\ell_s\,p^2\,,
\end{equation}
Using this in (\ref{gen hydro eta over s})
we obtain 
\begin{equation}
\frac{\eta_L}{s} = \frac{a_1}{2(a_1+b_1)\pi}\,, \qquad \frac{\eta_R}{s} = \frac{b_1}{2(a_1+b_1)\pi}\,.
 \end{equation}
 Analogous systems of two decoupled liquids has been found in \cite{Goykhman:2013oja}.

 \subsection{Correlation functions on two-dimensional boundary}
 \label{sec:2d correlation functions}

In this subsection, we will compute correlation functions of operators in momentum space of the space-time theory dual to string theory on the coset (\ref{big coset})
in the pure NS-NS sector. Unlike the previous section, we will work with the setup where
the sub-space $\mathbb{T}^4$ has been compactified. The dual boundary theory is therefore two-dimensional.

Since the space-time theory is not a local field theory, \footnote{The space-time theory is non-local (LST) in the sense that
it asymptotes to a linear dilaton background and the UV behavior of the boundary theory is not described by a fixed point.} it is not immediately clear if there exist local operators on the boundary theory. To get around this issue, we restrict ourselves to correlation functions of operators in momentum space. One can, of course, Fourier-transform the momentum space correlation function to position space but the interpretation of such position space two-point correlation function is not clear.

 To start with, let us review the computation of two-point correlation function of local operators in CFT$_2$ dual to superstring theory in $AdS_3\times \mathbb{S}^3\times \mathbb{T}^4$. A large class of scalar observables of string theory in  $AdS_3\times \mathbb{S}^3\times \mathbb{T}^4$ in the $(-1,-1)$ picture is given by the expression \cite{Kutasov:1999xu}
\begin{equation}\label{localo}
\mathcal{O}(x,\bar{x})=\int_{\Sigma} d^2z\, e^{-\varphi-\bar{\varphi}}\, \Phi_{h}(z,\bar{z};
x,\bar{x})\,V_{su}(z,\bar{z})\, V_{\mathbb{T}^4}(z,\bar{z})\,,
\end{equation}
where $x,\bar{x}$ are coordinates of the space-time theory, $z,\bar{z}$ are coordinates of the world-sheet
Riemann surface $\Sigma$, $\varphi,\bar{\varphi}$ are the world-sheet superconformal ghosts, 
$\Phi_{h}$ is the usual non-normalizable vertex operator of the world-sheet theory on $AdS_3$
with world-sheet dimensions  $(\Delta_h,\bar{\Delta}_h)$ where
\begin{eqnarray}
\Delta_h=\bar{\Delta}_h=-\frac{h(h-1)}{k}\,,
\end{eqnarray}
$V_{su}$ is the $\mathcal{N}=1$ superconformal primary of the CFT on $\mathbb{S}^3$ with dimensions $(\Delta_{su},\bar{\Delta}_{su})$, and $V_{\mathbb{T}^4}$ is the $\mathcal{N}=1$  superconformal primary of the CFT on $\mathbb{T}^4$ with dimensions $(\Delta_{\mathbb{T}^4},\bar{\Delta}_{\mathbb{T}^4})$.  The local operator $\mathcal{O}(x,\bar{x})$,
given by \eqref{localo}, of the space-time theory dual to string theory in $AdS_3$ has dimensions $(h,h)$.  For simplicity let us assume $\Delta_{su}=\bar{\Delta}_{su}$ and $\Delta_{\mathbb{T}^4}=\bar{\Delta}_{\mathbb{T}^4}$.

The onshell condition of the operator \eqref{localo} reads
\begin{eqnarray}
-\frac{h(h-1)}{k}+\Delta_{su}+\Delta_{\mathbb{T}^4}=\frac{1}{2}\,.
\end{eqnarray}
The vertex operators $V_{su}$ and $V_{\mathbb{T}^4}$ are normalized so that 
\begin{eqnarray}\label{v2pt}
\langle V_{su}(z_1)V_{su}(z_2) \rangle=\frac{1}{|z_1-z_2|^{4\Delta_{su}}}\,, \ \ \ 
 \langle V_{\mathbb{T}^4}(z_1)V_{\mathbb{T}^4}(z_2) \rangle=\frac{1}{|z_1-z_2|^{4\Delta_{\mathbb{T}^4}}}\,,
\end{eqnarray}
while the world-sheet operators $\Phi_h(z;x)$ are normalized so that 
\begin{eqnarray}\label{phi2pt}
\langle \Phi_{h}(z_1;x_1)  \Phi_{h'}(z_2;x_2)  \rangle=\delta(h-h')\frac{B(h)}{|z_1-z_2|^{4\Delta_h}|x_1-x_2|^{4h}}~,
\end{eqnarray}
where the coefficient $B(h)$ is given in \eqref{bh}. Taking the Fourier transform in $x,\bar{x}$ one recovers the correlation function in momentum space $p,\bar{p}$,
\begin{eqnarray}\label{phi2ptmom}
\langle \Phi_{h}(z_1;p)  \Phi_{h'}(z_2;-p)  \rangle= \delta(h-h')\pi B(h)\gamma{(1-2h)}\left(\frac{p^2}{4}\right)^{2h-1}\frac{1}{|z_1-z_2|^{4\Delta_h}}~,
\end{eqnarray}
where $\gamma(x)$ is defined in \eqref{bh} and $p^2=p\bar{p}$. 
Thus the two point function of the operator $\mathcal{O}(p)$ of the space-time theory takes the form 
\begin{eqnarray}\label{oo2ptmom}
\langle \mathcal{O}(p,\bar{p})\mathcal{O}(-p,-\bar{p})\rangle=\pi (2h-1)B(h)\gamma(1-2h)\left(\frac{p^2}{4}\right)^{2h-1}\,.
\end{eqnarray}
Fourier transforming \eqref{oo2ptmom} back to position space, one obtains
\begin{eqnarray}\label{ox1ox2}
\langle \mathcal{O}(x_1)\mathcal{O}(x_2)\rangle \simeq  \frac{1}{|x_1-x_2|^{4h}}\,.
\end{eqnarray}

Next let us focus on computation of two-point function of operators in momentum space of the boundary theory dual to string theory in the background \eqref{big coset}. As stated earlier, the boundary field theory is not conformal because the bulk space-time does not asymptote to $AdS_3$. One can write the $(-1,-1)$ picture  operators in momentum space
in the coset construction  as
\begin{equation}
\label{calO def}
\mathcal{O}(p)=\int d^2z \, e^{-\varphi-\bar{\varphi}}\,
\Phi_{h_p}(z;p)\,V_{su}(z)\, e^{il_sp_x x} \, e^{i\ell_s\kappa w}\,V_{\mathbb{T}^4}(z)~.
\end{equation}
The onshell condition is given by
\begin{equation}\label{hponshell}
-\frac{h_p(h_p-1)}{k}+\Delta_{su}+\frac{\ell_s^2(p_x^2-\kappa^2)}{4}+\Delta_{\mathbb{T}^4}=\frac{1}{2}\,,
\end{equation}
 where $(h_p,h_p)$\footnote{One can interpret $(h_p,h_p)$ as the dimension of the operator $\mathcal{O}(p)$ in the space-time theory dual to string theory in the coset \eqref{big coset} (see \cite{Guica:2021pzy,Guica:2021fkv} on related discussion on the momentum dependent dimension of operators in non-local CFTs).} is the space-time dimension of $\Phi_{h_p}$,
 and $h=\lim_{|p|\to0}h_p$.
 (Additionally, the BRST constraints (\ref{BRST constraints prelim res}) need to be satisfied.)
 Then from \eqref{hponshell} we derive
\begin{equation}\label{hhp}
 2h_p-1=\sqrt{(2h-1)^2+\ell_s^2kp^2}\,,
\end{equation}
where the momentum of the space-time theory is given by
\begin{equation}
p^2=p_x^2-\kappa^2\,.
\end{equation}
Notice that, as before, the $\Phi_{h_p}$, $V_{su}$ and $V_{\mathbb{T}^4}$ are vertex operators of the CFT on $AdS_3$, $SU(2)$, and $\mathbb{T}^4$ with the two-point functions
given by \eqref{phi2ptmom} (with $h$ replaced by $h_p$),   and \eqref{v2pt}. Thus, for string theory on (\ref{big coset}),
the two-point function of the dual space-time theory is given by
\begin{eqnarray}\label{doo2ptmom}
\langle \mathcal{O}(p,\bar{p})\mathcal{O}(-p,-\bar{p})\rangle=\pi (2h_p-1)B(h)\gamma(1-2h_p)\left(\frac{p^2}{4}\right)^{2h_p-1}.
\end{eqnarray}
This matches exactly with the two-point function computed in similar coset backgrounds that asymptotes to a linear-dilaton background \cite{Chakraborty:2020yka,Asrat:2017tzd}. 
As expected, in the deep IR, ${\it{i.e.}}, |\ell_s p|\ll1$, one recovers the CFT$_2$ correlation function \eqref{oo2ptmom}. In the deep UV, ${\it{i.e.}}, |\ell_s p|\gg1$, the two-point function \eqref{doo2ptmom}, however,  behaves as  $|p|^{2\sqrt{k}\ell_s|p|}$ which is a highly non-local behavior. 

The normalization of the two-point function of primary operators in a CFT \eqref{oo2ptmom} can be changed by rescaling the operators by an arbitrary analytic function of their dimensions. Physics remains invariant due to such a rescaling of the operators. In the coset theory, however, it seems  that the two-point function depends on the normalization of the operators due the the momentum dependence of $h_p$. At this point, it is unclear how to fix the normalization of the operators of the space-time theory.

Let us end this section by giving a brief sketch of the general structure of the three-point
correlation function of the operator (\ref{calO def}) on the
 boundary theory in momentum space, in the un-spectrally flowed sector.
 Analogously to  calculation of the two-point function, we determine
\begin{equation}
\langle \mathcal{O}_1(p_1)\mathcal{O}_2(p_2)\mathcal{O}_3(p_3)\rangle=\tilde{C}(h_{p_1},h_{p_2},h_{p_3})C_{sl}(h_{p_1},h_{p_2},h_{p_3})
C_{su}(j_i',m_i',\bar m_i')
\end{equation}
where OPE coefficients due to $SL(2,\mathbb{R})$ and $SU(2)$ factors are given by
(\ref{Csl}), (\ref{Csu}), and
\begin{equation}
\tilde{C}(h_{1},h_{2},h_{3})=(2\pi)^2\delta^2(p_1+p_2+p_3)\int d^2x_1d^2x_2\frac{e^{-ip_1.x_1}e^{-ip_3.x_3}}{|x_{12}|^{2(h_{1}+h_{2}-h_{3})}|x_2|^{2(h_{2}+h_{3}-h_{1})}|x_1|^{2(h_{3}+h_{1}-h_{2})}}\,,
\end{equation}
where
 \begin{equation}
  2h_{p_i}-1=\sqrt{(2h_{i}-1)^2+\ell_s^2kp_{i}^2}\,.
 \end{equation}

\section{Discussion}
\label{sec:discussion}

In this paper, we have investigated type-II superstring theory on the coset
space (\ref{big coset}).
It is described by the superconformal gWZW model at level $k$,
such that $R=\sqrt{k}\,\ell_s$ is the characteristic length scale of the
target space-time.
We have derived an exact (in $k$) ten-dimensional target space-time background configuration
of the NS-NS fields, including the metric, $B$-field, and dilaton.
This background interpolates between local $AdS_3\times\mathbb{S}^3\times\mathbb{T}^4$
in the IR and linear dilaton times $\mathbb{S}^3\times\mathbb{S}^1\times\mathbb{T}^4$ in the UV.
It possesses an event horizon, with non-vanishing angular
velocities along two of the $\mathbb{S}^3$ directions, as well as along the $\mathbb{S}^1$ circle.
We have calculated the corresponding temperature, entropy, energy, conserved angular momenta,
and free energy. Whenever applicable, our results reproduce the known expressions for thermodynamics
of (charged) black brane backgrounds obtained in the near-horizon limit of the NS5-F1 systems,
and described by coset models that can be found as particular cases of (\ref{big coset}).

We have also constructed a nine-dimensional background obtained by performing the KK reduction
of the ten-dimensional background on the $U(1)_x$ circle $\mathbb{S}^1$.
Such a background is particularly relevant in the context of holographic description
of strongly-coupled systems at finite density. This is due to the fact that KK reduction on $\mathbb{S}^1$
generates non-trivial profiles of gauge fields, that furnish holographic dual of non-vanishing
densities of conserved $U(1)$ charges of the boundary field theory. In our case, the role
of the boundary theory is played by LST found in the asymptotic linear-dilaton region of the bulk
space-time, and characterized by non-vanishing vacuum expectations
of two independent $U(1)$ charges.

Due to exact solvability of the corresponding gWZW model, the bulk theory enjoys
an exact in $k$ world-sheet control, going beyond the effective supergravity approximation $R/\ell_s\gg 1$.
In particular, it allows one to access the regime of dual field theory characterized by a finite number
of degrees of freedom, determined by $k$. We take advantage of this power of exact world-sheet
theory to calculate two-point correlation functions of massless NS-NS vertex operators
describing the graviton, $B$-field, and dilaton excitations on top of the constructed background.
Holographically, these give us correlation matrices
of the stress-energy tensor and two $U(1)$ currents of the dual field theory.
These correlation matrices manifest a rich structure of collective density excitations,
that include gapless diffusion modes. We derive the corresponding dispersion
relations and comment on their hydrodynamic interpretation.

We also derived the two- and three-point correlation functions of the scalars in the $1+1$-dimensional boundary theory in momentum space. The expressions for the correlation functions  take the form of CFT$_2$ correlation functions with the dimensions $h$ being replaced by $h_p$ \eqref{hhp}.
(In fact, such a prescription seem to hold for higher-order correlation functions as well.)
At low energies we recover the CFT$_2$ correlators, but at high energies the correlators seem to develop highly non-local behavior. This is in harmony with the fact the the short-distance physics is governed by LST rather than a local fixed point. However, the correlation functions at an arbitrary point on the RG flow seem to depend on the normalization of the operators at the IR fixed point. In this paper we do not provide a recipe to fix the normalization ambiguity in studying holography in non-AdS backgrounds. It would be nice to see if the Ward identities of the theory at a generic point on the RG flow can impose constraints in the normalization. It would also be interesting to perform an LSZ reduction on the correlators and keep only contributions that give rise to physical onshell particles. This may shed some light on the normalization of the  operators of the boundary theory.

One of the questions left unanswered in this paper concerns microscopic interpretation
of the obtained background in terms of extended string theory objects. Particularly, we expect
that the background configuration that we constructed in this paper can be found in the near-horizon
limit of a certain system involving separated rotating NS5-brane in the double-scaling limit \cite{Giveon:1999px,Giveon:1999tq} and fundamental F1-strings with non-trivial
momentum quantum numbers. We believe this question can be resolved in a manner analogous
to \cite{Chakraborty:2020yka} that recently provided an embedding of the $(SL(2,\mathbb{R})\times U(1))/U(1)$
target-space background as a near-horizon limit of a certain NS5-F1 system.
We would like to have an exact map between the gauge-invariant operators of the boundary theory and the world-sheet theory  and identify the operators in the chiral ring algebra of the boundary theory.

The supergravity background obtained in this paper is characterized by asymptotically
linear dilaton geometry. In order to describe this background as a near-horizon
limit of NS5-F1 system, it would be helpful to embed it into asymptotically
flat space-time.
We expect that such an embedding 
can be done in the spirit of \cite{Bena:2011dd} which considered D1-D5 system.
We leave this for future work.

Our work paves the avenue for various future research. In section \ref{sec:2d correlation functions} we computed the correlation function of scalar operators in the momentum space. It would be an interesting to check if the same correlation functions are recoverable from the supergravity analysis. 
As stated earlier, the boundary field theory is a certain vacua of LST that flows to a CFT$_2$ in the IR. At a generic point on the RG flow, the boundary field theory is not local. It has been shown before \cite{Chakraborty:2018kpr,Chakraborty:2020udr} that in similar non-local setups, the Ryu-Takayanagi curve develops non-local features below a certain scale (non-locality scale proportional to $\ell_s$). It would be interesting to understand the non-local features of boundary theory through the lenses of holographic entanglement entropy. In the same spirit it is  worth computing the holographic complexity where we expect to observe exotic divergences reflecting the non-local features of LST.  It would also be nice to set up a computation of  holographic Wilson-loop to understand the potential energy between a probe quark-anti-quark pair in the boundary theory \cite{Chakraborty:2018aji}.  From the analytic structure of the quark-anti-quark potential energy, one may expect to understand the various phases of the theory.

\section*{Acknowledgements} 

The authors would like to thank
M. Guica, N. Kovensky
for helpful discussions, and S.~Massai for comments on the draft.
The work of SC is supported in part by the ERC Starting Grant 679278 Emergent-BH.
The work of MG is supported by DOE grant DE-SC0011842.

\appendix

\section{Conventions and review}
\label{app:conventions}

In this appendix, we review some well-known facts about
the WZW models that are relevant for this paper.
We parametrize the two-dimensional Euclidean string world-sheet
by complex coordinates $z=\sigma^1+i\sigma^2$, $\bar z=\sigma^1-i\sigma^2$. 
The world-sheet metric in the $\sigma^{1,2}$ coordinate frame
is given by $h_{ab} = \delta_{ab}$, $a,b=1,2$, while
in the complex coordinates $z,\bar z$ frame we have
 $h_{z\bar z} = h_{\bar z  z} =  1/2$, $h^{z\bar z} = h^{\bar z z} =  2$.
Consider the field $g(z,\bar z)$, taking values on the group manifold 
$\mathbb{G}$. The WZW action for this field, at the level $k=1$, is given by
the expression
\begin{equation}
\label{WZWgen}
\mathtt{S}[g] = S_\sigma [g] + S_{\textrm{WZ}}[g]\,,
\end{equation}
where the sigma-model action is given by
\begin{equation}
\label{sigmamodeldef}
S_\sigma[g] = {1\over 4\pi} \int d^2z\,\Tr \left(g^{-1}\p g g^{-1}\bar\p g\right)\,,
\end{equation}
while the Wess-Zumino (WZ) term is defined by
\begin{equation}
\label{WZdef}
S_{\textrm{WZ}}[g] = {i\over 24\pi} \int _{{\cal B}}\,\Tr\left( g^{-1}d g
\wedge g^{-1}d g\wedge g^{-1}d g\right)\,,
\end{equation}
and the trace $\Tr$ is taken in the given representation of $\mathbb{G}$.
We define the measure of integration as $d^2z\equiv d\sigma^1 d\sigma^2$.
The integral in (\ref{WZdef}) is taken over a three-dimensional compact space ${\cal B}$
with the boundary given by the string world-sheet. 

By taking a variation $\delta g$ of the field $g$ we derive from (\ref{sigmamodeldef})
\begin{equation}
\label{Ssigmavar}
\delta S_{\sigma} = -{1\over 4\pi}\int d^2z\,
\Tr\left(g^{-1}\delta g\left(\partial (g^{-1}\bar\partial g\right)
+\bar\partial \left(g^{-1}\partial g\right)\right)\,.
\end{equation}
At the same time, from the WZ action (\ref{WZdef}) we derive
\begin{equation}
\label{SWZvar}
\delta S_{\textrm{WZ}} = {i\over 24\pi} \int _{{\cal B}}\,3\,\Tr\left(
(-g^{-1}\delta g g^{-1} dg + g^{-1} d\delta g)\wedge g^{-1}d g\wedge
g^{-1}d g \right)\,,
\end{equation}
Performing cyclical permutation under the trace,
while taking into account anti-symmetric property of the wedge product,
we can re-arrange the first term in the r.h.s. of (\ref{SWZvar}), obtaining as a result
\begin{equation}
\label{SWZvarArr}
\delta S_{\textrm{WZ}} = {i\over 8\pi} \int _{{\cal B}}\,\Tr\left(
-g^{-1} dg\wedge g^{-1}\delta g g^{-1} dg\wedge g^{-1} dg + g^{-1} d\delta g\wedge g^{-1}d g\wedge
g^{-1}d g \right)\,.
\end{equation}
Using 
\begin{equation}
\label{ExacDg}
d(g^{-1}d g\wedge g^{-1}d g) = 0\,,
\end{equation}
we can subsequently re-write (\ref{SWZvarArr}) as
\begin{equation}
\label{SWZvarTotDer}
\delta S_{\textrm{WZ}} = {i\over 8\pi} \int _{{\cal B}}\,d\,\Tr\left(
g^{-1} \delta g \, g^{-1}d g\wedge
g^{-1}d g \right)={i\over 8\pi} \int \,\Tr\left(
g^{-1} \delta g\,  g^{-1}d g\wedge g^{-1}d g \right)\,.
\end{equation}
Substituting here $dg = \p g \,dz+\bar\p g\, d\bar z$, and using
\begin{equation}
\label{dzdbzwedge}
dz\wedge d\bar z = -2i\,d \sigma^1
\wedge d\sigma ^2\equiv -2i\, d^2\sigma\equiv -2i \, d^2z\,,
\end{equation}
we obtain
\begin{equation}
\label{SWZvarInt}
\delta S_{\textrm{WZ}} ={1\over 4\pi} \int d^2z\,\Tr\left(g^{-1}\delta g \,(
g^{-1} \p g g^{-1} \bar\p g-g^{-1} \bar\p g g^{-1} \p g) \right)\,,
\end{equation}
that we finally re-write as
\begin{equation}
\label{SWZvarFinal}
\delta S_{\textrm{WZ}} ={1\over 4\pi} \int  d^2z\,\Tr\left(g^{-1}\delta g \,(
\bar\p (g^{-1}\p g)-\p(g^{-1}\bar\p g) \right)\,.
\end{equation}
Combining (\ref{Ssigmavar}), (\ref{SWZvarFinal}), we obtain variation of the WZW action 
(\ref{WZWgen}) determined by the expression
\begin{equation}
\label{WZWvarFinal}
\delta \mathtt{S} =-{1\over 2\pi} \int  d^2z\,\Tr\left(g^{-1}\delta g \,\p(g^{-1}\bar\p g) \right)\,.
\end{equation}

In particular, for the variation determined by a single local parameter $\xi(z,\bar z)$
and the generators $T_{L,R}$ belonging to a $u(1)$ sub-algebra of the group algebra of $\mathbb{G}$,
\begin{equation}
\label{gvar}
\delta g(z,\bar z) = \xi(z,\bar z)\,\left(T_L g+gT_R\right)\,,
\end{equation}
we obtain
\begin{equation}
\label{SWZWxiGenVar}
\delta {\cal S} = -{1\over 2\pi}\int d^2z\,\xi\,\left(
\p\tilde {\cal J} + \bar\p {\cal J}\right)\,,
\end{equation}
where we have defined the (anti-)holomorphic $U(1)$ currents
at the level $k=1$ as
\begin{equation}
\label{JtildeGenDef}
{\cal J} = \Tr\left(T_L\,\p g g^{-1}\right)\,\qquad
\tilde {\cal J} = \Tr\left(T_R\, g^{-1}\bar\p g\right)\,.
\end{equation}
Using (\ref{gvar}) we derive the following gauge transformations of these currents,
\begin{equation}
\label{tildeJgaugetransform}
\delta {\cal J}
=\left(\Tr(T_L^2)+\Tr(T_LgT_Rg^{-1})\right)\,\p\xi\,,\qquad
\delta \tilde {\cal J}
=\left(\Tr(T_R^2)+\Tr(T_LgT_Rg^{-1})\right)\,\bar\p\xi\,.
\end{equation}

From expression (\ref{WZWvarFinal}) one can see that it is possible
to gauge the corresponding $U(1)$ sub-group of $\mathbb{G}$
parametrized by $\xi$. To that end, introduce auxiliary (non-dynamical) gauge fields
$A$, $\tilde A$, that transform as
\begin{equation}
\label{AtildeAgentransform}
\delta A = -\p\xi\,,\qquad \delta\tilde A = -\bar\p\xi
\end{equation}
under the infinitezimal gauge transformation, and define the gauged WZW action as
\begin{equation}
\label{gWZWgen}
{\cal S}_g = \mathtt{S}+{1\over 2\pi}\int d^2z\,
\left(A\tilde {\cal J} + \tilde A{\cal J} +{\cal M}'\,A\tilde A\right)\,,
\end{equation}
where ${\cal M}'$ is a gauge-invariant function of $g$ to be determined below.
Algebraic equations of motion for the gauge fields $A$, $\tilde A$
that follow from the action (\ref{gWZWgen}) are given by
\begin{equation}
\label{AAtildeeom}
{\cal J} +{\cal M}'\,A =0\,,\qquad
\tilde {\cal J} +{\cal M}'\,\tilde A =0\,.
\end{equation}
Demanding that the equations (\ref{AAtildeeom}) are invariant w.r.t. the gauge transformation,
and using (\ref{tildeJgaugetransform}), (\ref{AtildeAgentransform}),
we recover the anomaly cancellation
condition
$\Tr(T_L^2) = \Tr(T_R^2)$,
as well as the gauge-invariant object
${\cal M}' = \Tr(T_L^2)+ \Tr(T_R g^{-1}T_Lg)$.
It is convenient to use the following normalization of the $u(1)$ generators
\begin{equation}
\label{anomalyCancelGenNorm}
\Tr(T_L^2) = \Tr(T_R^2) = 2 \,,
\end{equation}
and define
\begin{equation}
\label{MdefNorm}
{\cal M}' = {\cal M} +2\,,\qquad {\cal M} = \Tr(T_R g^{-1}T_Lg)\,.
\end{equation}
Gauge transformations (\ref{tildeJgaugetransform}) can then be re-written as
\begin{equation}
\label{JtildeJgaugetr}
\delta {\cal J}
=(2+{\cal M})\,\p\xi\,,\qquad
\delta \tilde {\cal J}
=(2+{\cal M})\,\bar\p\xi\,.
\end{equation}
Since we have explicitly established that e.o.m. (\ref{AAtildeeom}) are
gauge-invariant, variation of non-WZW terms in the action (\ref{gWZWgen})
can be obtained either before
or after integrating out non-dynamical fields $A$, $\tilde A$. Let us integrate these
out, arriving at
\begin{equation}
\label{gWZWIntegrateoutAtildeA}
{\cal S}_g = \mathtt{S}-{1\over 2\pi}\int d^2z\,
{{\cal J} \tilde {\cal J}\over {\cal M} + 2}\,.
\end{equation}
Using (\ref{SWZWxiGenVar}), (\ref{JtildeJgaugetr}), we then observe that the action 
(\ref{gWZWIntegrateoutAtildeA}) is indeed gauge-invariant.\\

\subsection{WZW model on $SU(2)$}
\label{sec:su2wzw}

Consider the following parametrization of $SU(2)$ group element,\footnote{
Performing a constant shift of parameters $\alpha\rightarrow\alpha-\pi/4$, $\beta\rightarrow\beta+\pi/4$
in (\ref{SUtwoparam}) we obtain $g = e^{i\alpha\sigma_3}e^{i\theta\sigma_2}e^{i\beta\sigma_3}$
as another parametrization of $SU(2)$.}
\begin{equation}
\label{SUtwoparam}
g = e^{i\alpha\sigma_3}e^{i\theta\sigma_1}e^{i\beta\sigma_3}\,,
\end{equation}
where $\alpha\in \left[0,\pi\right]$, $\theta\in \left[0,\frac{\pi}{2}\right]$, $\beta\in \left[0,2\pi\right]$ are real-valued parameters,
 and the
Pauli matrices are given by
\begin{equation}
\label{Pauli}
\sigma_1=\left({0\atop 1}\;{1\atop 0}\right)\,,\quad
\quad\sigma_2=\left({0\atop i}\;{-i\atop 0}\right)\,,
\qquad\sigma_3=\left({1\atop 0}\;{0\atop -1}\right)\,.
\end{equation}
This can also be written as
\begin{equation}
g = \left({a+ib\atop ic+d}\;{ic-d\atop a-ib}\right)\,,
\end{equation}
where the $SU(2)\sim S^3$ manifold $a^2+b^2+c^2+d^2 = 1$ is parametrized by 
\begin{equation}
a = \cos\theta\,\cos \gamma_+\,,\quad b = \cos\theta\,\sin \gamma_+\,,\quad
c=\sin\theta\,\cos \gamma_-\,,\quad d=\sin\theta\,\sin\gamma_-\,,
\end{equation}
for $\gamma_\pm = \alpha\pm\beta$.
After a straightforward calculation we obtain
\begin{equation}
\label{Ssigmasutwo}
S_\sigma [g] = \frac{1}{ 2\pi}\int d^2z\,
G_{\mu\nu}\partial X^\mu\bar\partial X^\nu\,,
\end{equation}
where we denoted the target-space coordinate as $X^\mu = (\alpha,\theta,\beta)$, and defined
the corresponding background metric as
\begin{equation}
\label{su2 background metric}
G_{\mu\nu}=k\,
\left(
\begin{array}{ccc}
-1& 0 & -\cos(2\theta)\\
0 & -1 & 0 \\
-\cos(2\theta) & 0 & -1 \\
\end{array}
\right)
\end{equation}
For the representation (\ref{SUtwoparam}) we also obtain
\begin{equation}
\Tr\left( g^{-1}d g
\wedge g^{-1}d g\wedge g^{-1}d g\right) = 6d\,\left(\cos(2\theta)\,d\alpha\wedge d\beta\right)\,,
\end{equation}
and therefore the Wess-Zumino term induces the background $B$-field via
\begin{equation}
\label{Swzsutwo}
S_{\textrm{WZ}}[g]= \frac{1}{ 2\pi}\int d^2z\,
B_{\mu\nu}\partial X^\mu\bar\partial X^\nu\,,
\end{equation}
where we denoted 
\begin{equation}
\label{su2 background bfield}
B_{\mu\nu}=
k\,\left(
\begin{array}{ccc}
0& \;\;0 & \;\;\cos(2\theta)\\
0 & \;\;0 & 0 \\
-\cos(2\theta) & \;\;0 & 0 \\
\end{array}
\right)
\end{equation}

We conclude this section by writing down expression for the volume of the unit $3$-sphere.
Using negative of the metric (\ref{su2 background metric}), which is achieved
by selecting the level $k=-1$, we obtain
\begin{equation}
\label{3sphere vol}
V_{\mathbb{S}^3} = \int _0^{2\pi}d\alpha
\int _0^\pi d\beta\int _0^\frac{\pi}{2}d\theta\,\sqrt{-\det G}= \int _0^{2\pi}d\alpha
\int _0^\pi d\beta\int _0^\frac{\pi}{2}d\theta\,\sin(2\theta)=2\pi^2\,.
\end{equation}

\subsection{WZW model on $SL(2,\mathbb{R})$}
\label{sec:sl2rwzw}

Consider the following parametrization of $SL(2,\mathbb{R})$ group element,
\begin{equation}
\label{SLtwoparam}
g = e^{\alpha\sigma_3}e^{\theta\sigma_1}e^{\beta\sigma_3}\,,
\end{equation}
where $\alpha$, $\theta$, $\beta$ are real-valued parameters,
and Pauli matrices are as in (\ref{Pauli}).
After a straightforward calculation we obtain
\begin{equation}
\label{Ssigmasltwo}
S_\sigma [g] = \frac{1}{ 2\pi}\int d^2z\,
G_{\mu\nu}\partial X^\mu\bar\partial X^\nu\,,
\end{equation}
where we denoted the target-space coordinate as $X^\mu = (\alpha,\theta,\beta)$, and defined
the corresponding background metric as
\begin{equation}
\label{su2 background metric}
G_{\mu\nu}=
k\,\left(
\begin{array}{ccc}
1& \;\;0 & \;\;\cosh(2\theta)\\
0 & \;\;1 & 0 \\
\cosh(2\theta) & \;\;0 & 1 \\
\end{array}
\right)
\end{equation}
For the representation (\ref{SLtwoparam}) we also obtain
\begin{equation}
\Tr\left( g^{-1}d g
\wedge g^{-1}d g\wedge g^{-1}d g\right) = -6d\,\left(\cosh(2\theta)\,d\alpha\wedge d\beta\right)\,,
\end{equation}
and therefore the Wess-Zumino term induces the background $B$-field via
\begin{equation}
\label{Swzsutwo}
S_{\textrm{WZ}}[g]= \frac{1}{ 2\pi}\int d^2z\,
B_{\mu\nu}\partial X^\mu\bar\partial X^\nu\,,
\end{equation}
where we denoted 
\begin{equation}
\label{sl2r background bfield}
B_{\mu\nu}=
k\,\left(
\begin{array}{ccc}
0& \;\;0 & -\cosh(2\theta)\\
0 & \;\;0 & 0 \\
\cosh(2\theta) & \;\;0 & 0 \\
\end{array}
\right)
\end{equation}

\subsection{Effective action and superconformal WZW model}
\label{app:effective wzw action}

We have reviewed the classical (gauged) WZW models
describing bosonic string theory
with the target space given by the group manifold $\mathbb{G}$.
In particular, we considered examples of $\mathbb{G} = SL(2,\mathbb{R})$
and $\mathbb{G}=SU(2)$, and derived the corresponding classical target space background.
The latter is generally specified by the metric, dilaton, and $B$-field. 

The  WZW model is characterized by the level $k$,
such that the classical action of the WZW model is given by $S = k\,\mathtt{S}$,
where the $k=1$ action $\mathtt{S}$ is given by (\ref{WZWgen}).
Perturbative corrections to the classical theory,
due to the world-sheet path integral expansion,
therefore behave as $1/k$.
The semi-classical limit is then achieved in the limit of large $k$.

In terms of the target space theory, the semi-classical limit translates into the
small curvature or, equivalently, the large length scale limit. Indeed, from our discussion in sections~\ref{sec:su2wzw},
\ref{sec:sl2rwzw}, one can see that the target space metric scales as $k$,
that sets the characteristic length scale of the semi-classical geometry to be $\sqrt{k}\,\ell_s$.

Similar observations can be made when one considers gauged
WZW models. In particular, integrating out auxiliary gauge fields, via their equations of motion, in the classical WZW action
on a coset space renders a classical sigma-model. Classical background fields,
including the metric, dilaton and $B$-field, can be read off from the sigma-model action.\\

We now proceed to reviewing how  classical results get modified when $1/k$ corrections,
due to expansion of the world-sheet path integral,
are taken into account. It is illustrative to follow two different perspectives, that ultimately are shown
to render consistent results. The first perspective involves the (anti-) holomorphic symmetry argument
and the affine algebra relations for the corresponding currents, while the second one involves calculation 
of the effective action for the WZW model \cite{Bars:1992sr,Tseytlin:1992ri,Tseytlin:1993my}.

Recall that the   WZW model at level $k$ on the group manifold $\mathbb{G}$
enjoys the $\mathbb{G}_L\times \mathbb{G}_R$ Kac-Moody symmetry.
The  corresponding affine currents $j^a$, $\tilde j^a$ satisfy the OPE (similar OPE can be written
down for the anti-holomorphic current components $\tilde j^a$)
\begin{equation}
\label{bosonic wzw ope}
j^a(z)j^b(w) \simeq \frac{k}{2}\frac{k^{ab}}{(z-w)^2} + \frac{if^{ab}_{\;\;\;\;c}\, j^c(w)}{z-w}\,,
\end{equation}
 where the index $a$ takes values in the adjoint representation
of $\mathbb{G}$, $k_{ab}$ is the Cartan metric in the 
algebra space, and $f^{abc}$ are the algebra structure constants.
Indices are raised and lowered with the tensor $k_{ab}$.

The WZW model also enjoys conformal symmetry, with the generators  
given by the (anti-)holomorphic stress-energy tensor components $T$, $\tilde T$.
These can be determined in terms of the Kac-Moody currents via the Sugawara expression as
\begin{equation}
\label{T in bosonic WZW}
T = \frac{1}{k+c_G}\,j^a\,j^a\,,
\end{equation}
where the dual Coxeter number $c_G$ is defined via\footnote{
Recall that $c_{SL(2,\mathbb{R})} = - 2$ and $c_{SU(2)} =  2$. While the structure constants $f^{abc}=\epsilon^{abc}$,
given by the Levi-Civita symbol, 
are the same for both algebras, the difference between the values of the dual Coxeter numbers
 can be seen to originate from the $k_{ab} =\textrm{diag}\{1,\,1,\, -1\}$ Lorentzian metric  of the
$sl(2,\mathbb{R})$, versus the $k_{ab} =\textrm{diag}\{1,\,1,\, 1\}$ Euclidean metric  of the $su(2)$.}
\begin{equation}
f_{abc}f_{d}^{\;\; bc} = c_G\,\eta_{ad}\,.
\end{equation}
In quantum theory the product of currents in the r.h.s. of (\ref{T in bosonic WZW}) is normally-ordered. The overall
prefactor of the r.h.s. is determitned from the requirement that
$j^a$ is a primary field of dimension one,
\begin{equation}
T(z) j^a(w)\simeq \frac{ j^a(w)}{(z-w)^2} + \frac{\p j^a (w)}{z-w}\,,
\end{equation}
or, equivalently, that the stress-energy tensor satisfies the Virasoro OPE
\begin{equation}
\label{T T OPE}
T(z)T(w) \simeq \frac{c/2}{(z-w)^4} + \frac{2T(w)}{(z-w)^2} + \frac{\p T(w)}{z-w}\,.
\end{equation}

Expression  (\ref{T in bosonic WZW})  for the stress-energy tensor is exact in quantum world-sheet theory 
for all values of the level $k$.
All of the $1/k$ corrections of the world-sheet perturbation theory are accounted for in the difference 
between the semi-classical expression $T=\frac{1}{k}\, j^aj^a$, and the full quantum expression (\ref{T in bosonic WZW}).
The same conclusion can be arrived at from the
perspective of the effective WZW action in quantum  theory
\cite{Tseytlin:1992ri,Tseytlin:1993my}. A non-trivial functional determinant
appearing in the Legendre transform calculation of the effective action via the world-sheet path integral 
can be reformulated as $e^{-c_G\,\mathtt{S}[g]}$. The effective action
of the WZW model is then given by
\begin{equation}
\label{Gamma g bosonic WZW}
\Gamma[g] = (k+c_G) \,\mathtt{S}\,.
\end{equation}
This is simply the original classical WZW action with the shifted level, $k\rightarrow k+c_G$.
The level $k+c_G$ in the effective action results in the prefactor
$1/(k+c_G)$ in the expression for the stress-energy tensor, in agreement with  (\ref{T in bosonic WZW}),
manifesting consistency of the effective action and the symmetry
 perspectives \cite{Tseytlin:1992ri,Tseytlin:1993my}.\\

Let us now proceed to  discussion of supersymmetric WZW models.
Analogously to the case of bosonic WZW models, the symmetry and the effective action perspectives
can be considered \cite{Kazama:1988qp,Tseytlin:1992ri,Tseytlin:1993my}.

The (anti-) holomorphic Kac-Moody symmetry and the conformal symmetry of the bosonic WZW model
get upgraded to the super Kac-Moody algebra and the superconformal symmetry in the
supersymmetric WZW model. To start with, the bosonic degrees of freedom represented by the affine currents 
$j^a$, $\tilde j^a$ acquire superpartners, given by the Majorana-Weyl fermions $\psi^a$, $\tilde\psi^a$,
where index $a=1,\dots,\textrm{dim}(\mathbb{G})$ takes values in the adjoint representation
of the group $\mathbb{G}$.
The fermions are free, in particular, their OPEs with the bosonic currents $j^a$, $\tilde j^a$ are non-singular.
At the same time, we have
\begin{equation}
\label{psi psi OPE}
\psi^a(z)\psi^b(w) \simeq \frac{k}{2}\,\frac{k^{ab}}{z-w}\,,
\end{equation}
and similarly for the anti-holomorphic sector fermions $\tilde\psi^a$.

The Kac-Moody currents of supersymmetric WZW model then acquire contributions
both from the bosonic degrees of freedom $g\in \mathbb{G}$, via the purely bosonic currents
$j^a$, $\tilde j^a$, as well as from the fermions $\psi^a$, $\tilde \psi^a$, resulting in the total affine currents
\begin{equation}
\label{tt j def}
\mathtt{j}^a = j^a - \frac{i}{k} f^a_{\;\; bc} \,\psi^b\psi^c\,,
\end{equation}
where the product of fermions in the second term in the r.h.s. is normally-ordered.
The total bosonic currents $\mathtt{j}^a$ satisfy the Kac-Moody OPE at level $k$, 
\begin{equation}
\mathtt{j}^a(z)\mathtt{j}^b(w)
\simeq \frac{k}{2}\frac{k^{ab}}{(z-w)^2} + \frac{if^{ab}_{\;\;\;\;c}\, \mathtt{j}^c(w)}{z-w}\,.
\end{equation}
This implies that the purely bosonic Kac-Moody current components $j^a$ satisfy the
OPE  at  level $\mathtt{k} = k - c_G$,
\begin{equation}
\label{j j bos OPE}
j^a(z)j^b(w) \simeq \frac{k-c_G}{2}\frac{k^{ab}}{(z-w)^2} + \frac{if^{ab}_{\;\;\;\;c}\, j^c(w)}{z-w}\,.
\end{equation}
The currents $\J^a$, $\J^a$ are primary operators of dimension $1$,
\begin{equation}
\label{T j general}
T(z) \J^a(w)\simeq \frac{ \J^a(w)}{(z-w)^2} + \frac{\p \J^a (w)}{z-w}\,,
\end{equation}
While the purely bosonic currents $j^a$ are decoupled from the fermions $\psi^a$,
\begin{equation}
j^a(z)\psi^b(w)  = {\cal O}(z - w)\,,
\end{equation}
the total bosonic currents $\mathtt{j}^a$ satisfy non-trivial OPEs with the fermions $\psi^a$,
\begin{equation}
\label{psi j OPE}
\mathtt{j}^a(z)\psi^b(w)\simeq \psi^a(z)\mathtt{j}^b(w)\simeq\frac{if^{ab}_{\;\;\;\; c}\,j^c(w)}{z-w}\,.
\end{equation}
Combining everything together,  relations (\ref{psi psi OPE}), (\ref{j j bos OPE}), (\ref{psi j OPE})
define the super Kac-Moody affine algebra.

The stress-energy tensor in supersymmetric WZW model is given by 
\begin{equation}
\label{set susy wzw}
T(z) = \frac{1}{k}\,j^aj^a - \frac{1}{k}\,\psi^a\p\psi^a\,,
\end{equation}
where products of fields in  r.h.s. are normally-ordered.
A couple of comments are in order here. First of all, we have contributions
to (\ref{set susy wzw}) coming from the free fermionic sector, such that the field $\psi^a$
is a primary of dimension $1/2$,
\begin{equation}
\label{T psi general}
T(z)\psi^a(w)\simeq \frac{1}{2}\,\frac{\psi^a(w)}{(z-w)^2} + \frac{\p\psi^a(w)}{z-w}\,.
\end{equation}
At the same time, the first term in r.h.s.  of (\ref{set susy wzw}), coming from the bosonic
sector, is similar to the stress-energy tensor (\ref{T in bosonic WZW}) of the bosonic WZW model,
except that the overall prefactor is $1/k$. This is due to the bosonic currents $j^a$
in the supersymmetric theory
satisfying the OPE relation (\ref{j j bos OPE}) at the level $\mathtt{k} = k-c_G$; to be contrasted
with the OPE relation (\ref{bosonic wzw ope}) at level $k$ for such currents in the bosonic WZW model.
The rest of the calculation is the same as in the bosonic WZW model:
the classical prefactor $1/\mathtt{k}$ in expression for the stress-energy tensor in supersymmetric WZW model
changes as $1/\mathtt{k}\rightarrow 1/(\mathtt{k}+c_G) = 1/k$ in  quantum theory.

Conformal symmetry of the bosonic WZW model
gets upgraded to superconformal symmetry. The conformal symmetry generators
$T$, $\tilde T$ become supplemented with the supersymmetry generators $G$, $\tilde G$.
The latter are conformal primaries of dimension $3/2$,
\begin{equation}
\label{T G OPE}
T(z)G(w)\simeq G(z)T(w) \simeq \frac{3}{2}\,\frac{G(w)}{(z-w)^2} + \frac{\p G(w)}{z-w}\,.
\end{equation}
The other non-trivial OPE is satisfied by the supersymmetric generators $G$ themselves,
\begin{equation}
\label{G G OPE}
G(z)G(w) \simeq \frac{2c/3}{(z-w)^3} + \frac{2T(w)}{z-w}\,.
\end{equation}
Together, (\ref{T T OPE}), (\ref{T G OPE}), (\ref{G G OPE})
define the $N = 1$ superconformal algebra.

As a supercurrent
that determines supersymmetry transformations,
 the operator $G$ transforms  bosonic degrees of freedom via fermionic ones, and vice versa.
 This is expressed in the OPEs
 \begin{equation}
 \label{G j and G psi OPE}
 \begin{aligned}
 G(z) \mathtt{j}^a(w)&\simeq \frac{\psi^a(w)}{(z-w)^2} + \frac{\p \psi^a(w)}{z-w}\,,\\
 G(z) \psi^a(w) &\simeq \frac{\mathtt{j}^a(w)}{z-w}\,.
 \end{aligned}
 \end{equation}
Notice that the bosonic superpartner of the fermion $\psi^a$
is given by the full current $\mathtt{j}^a$ rather than by its bosonic-sector component $j^a$.
In components, the supercurrent is represented by the expression
\begin{equation}
G(z) = \frac{2}{k} \, \left(\psi^a\,j^a -
\frac{i}{3k}\, f_{abc} \, \psi^a\,\psi^b\,\psi^c\right)
= \frac{2}{k} \, \left(\psi^a\,\mathtt{j}^a +
\frac{2i}{3k}\, f_{abc} \, \psi^a\,\psi^b\,\psi^c\right)\,,
\end{equation}
where the products in r.h.s. are normally-ordered.

Let us now turn to the discussion of effective action in supersymmetric WZW models.
The starting point is supersymmetric action $k\,S_{\textrm{susy}}[g,\chi^a,\tilde\chi^a]$
at level $k$. By performing a change of fermionic degrees of freedom,
we can reformulate the theory in terms of decoupled free fermions $\psi^a$, $\tilde \psi^a$,
discussed above. This is a chiral transformation, that produces a non-trivial functional
determinant that can be reformulated as $e^{c_G\,{\cal S}[g]}$. The total action
is then given by $\mathtt{k}\,\mathtt{S}[g] +k\, S _f[\psi,\tilde \psi]$, where
${\cal S} _f[\psi,\tilde \psi]$ is the action of free two-dimensional
Majorana-Weyl fermions $\psi^a$, $\tilde \psi^a$  at 
level $k=1$\cite{Tseytlin:1992ri,Tseytlin:1993my}. Consistently with our discussion above,
we recognize the shifted level $\mathtt{k} = k - c_G$,
that appeared in the OPE (\ref{j j bos OPE}) of the bosonic sector Kac-Moody currents $j^a$.

Since the fermions $\psi^a$, $\tilde\psi^a$ are free and decoupled, the rest of derivation 
of the effective action proceeds similarly to the bosonic WZW case. 
Analogously to the effective action (\ref{Gamma g bosonic WZW}),
going from classical supersymmetric WZW action to the effective action
of supersymmetric WZW model,
the level $\mathtt{k}$ gets shifted as $\mathtt{k}\rightarrow \mathtt{k} + c_G = k$.
In other words, starting with the action $k\,S_{\textrm{susy}}[g,\chi^a,\tilde\chi^a]$ at  level $k$, first
the fermionic determinant due to the change of variables  $\chi,\tilde\chi\rightarrow \psi,\tilde\psi$
shifted the level $k$ prefactor of the bosonic sector action by $-c_G$, that subsequently 
was offset by the $+c_G$ shift in calculation of the effective action,
yielding at the end the total effective action 
\begin{equation}
\label{susy wzw effective action}
\Gamma_{\textrm{susy}}[g,\psi,\tilde\psi] = k\,\mathtt{S}[g] + k\,S _f[\psi,\tilde \psi]\,.
\end{equation}
Notice that the prefactor $k$ in front of $\mathtt{S}[g]$ in  r.h.s. of (\ref{susy wzw effective action})
is consistent with the normalization $1/k$ of the $j^aj^a$ terms in the stress-energy tensor 
expression (\ref{set susy wzw}). Of course the fermionic action $k\, S _f[\psi,\tilde \psi]$ 
at the level $k$ reproduces the fermionic sector contribution to (\ref{set susy wzw}).\\

In the main text, whenever we write down action of (gauged) WZW model, 
we usually have in mind supersymmetric model, considered at level $k$,
and write down the exact world-sheet effective action thereof. This becomes important, for instance,
when we write down the action on the coset (\ref{big coset}),
that combines contributions from the $SL(2,\mathbb{R})$ and $SU(2)$
sectors. Due to the $\pm c_G$ cancellations between the level $k$
shifts, the effective action of supersymmetric model on  (\ref{big coset})
is at level $k$  both in the  $SL(2,\mathbb{R})$ and $SU(2)$ sectors.
While such $1/k$ corrections are unimportant in the semiclassical
limit $k\rightarrow \infty$, our results are therefore in fact exact at finite $k$ in the full quantum
world-sheet theory.

\subsection{Some (anti-)commutation relations}

In this appendix we collect various  (anti-)commutation relations
satisfied by the amplitudes of the superconformal and the super-Kac-Moody currents.
Specifically, we are interested in the NS sector of superstring states,
and therefore all the fermionic operators are expanded in the half-integer modes.

From the $N = 1$ superconformal algebra OPEs (\ref{T T OPE}), (\ref{T G OPE}), (\ref{G G OPE})
one obtains the (anti-)commutation algebra relations,
\begin{align}
[L_m,\;L_n] &= (m-n)\,L_{m+n}+\frac{c}{12}\,m(m^2-1)\,\delta_{m+n,\,0}\,,\\
\{G_r,\;G_s\}&= 2L_{r+s}+\frac{c}{12}\,(4r^2 - 1)\,\delta_{r+s,\,0}\,,\\
[L_m,\; G_r] &= \frac{m-2r}{2}\,G_{m+r}\,.
\end{align}
Expanding the total Kac-Moody currents as\footnote{For abelian currents the total Kac-Moody current coincides with
its purely bosonic theory version, $\J^x = j^x$, $\J^w = j^w$, $\J^i = j^i$, $i=1,2,3,4$.}
\begin{equation}
\J^M = \sum_n\frac{\J^M_n}{z^{n+1}}\,,\qquad
\J^w = \sum_n\frac{\J^w_n}{z^{n+1}}\,,
\end{equation}
and the fermions as
\begin{equation}
\label{psi M w mode expansion}
\psi^{M}  = \sum_{r\in\mathbb{Z}+\frac{1}{2}}\frac{\psi^{M} _r}{z^{r+\frac{1}{2}}}\,,\qquad
\psi^{w}  = \sum_{r\in\mathbb{Z}+\frac{1}{2}}\frac{\psi^{w} _r}{z^{r+\frac{1}{2}}}\,,
\end{equation}
where $M=a,\,a',\,i$ labels the  $SL(2,\mathbb{R})$ polarizations $a=1,2,3$,
$SU(2)$ polarizations $a'=1,2,3$, and $\mathbb{T}^4$ polarizations $z^i$, $i=1,2,3,4$,
one obtains from the OPEs (\ref{T j general}), (\ref{T psi general})
\begin{equation}
\label{L j psi (anti)commutators}
\begin{aligned}
[L_m,\, \J_n^M] &= - n\, \J^M_{m+n}\,,\qquad &&[L_m,\, \J_n^w] = - n\, \J^w_{m+n}\,,\\
[L_m,\, \psi_r^M] &= - \frac{m+r}{2}\, \psi^M_{m+r}\,,\qquad &&[L_m,\, \psi_r^w] = - \frac{m+r}{2}\, \J^w_{m+n}\,,\\
\end{aligned}
\end{equation}
and from the OPEs (\ref{G j and G psi OPE}) we obtain
\begin{equation}
\label{G psi anticom}
\begin{aligned}
\{G_r,\, \J^M_n\} &= -n\,\psi^M_{n+r}\,,\qquad
&&\{G_r,\, \J^w_n\} = -n\,\psi^w_{n+r}\,,\\
\{ G_r,\,\psi^M_s \} &= \J^M_{r+s}\,,\qquad
&&\{ G_r,\,\psi^w_s \} = \J^w_{r+s}\,.
\end{aligned}
\end{equation}

Using the OPE  expressions (\ref{psi psi ope}), (\ref{psi x ope}), (\ref{psi w ope}), 
we obtain (here $\delta^{+-}=\delta^{-+} = 2$)
\begin{equation}
\label{psi r anticom}
\begin{aligned}
\{\psi^a_r,\,\psi^b_s\} &= \frac{k}{2}\delta^{ab}\delta_{r+s,\,0}\,,\qquad
&&\{\psi^{\prime a}_r,\,\psi^{\prime b}_s\} = \frac{k}{2}\delta^{ab}\delta_{r+s,\,0}\,,\\
\{\psi^x_r,\,\psi^x_s\} &= \frac{1}{2}\delta_{r+s,\,0}\,,\qquad
&&\{\psi^w_r,\,\psi^w_s\} = -\frac{1}{2}\delta_{r+s,\,0}\,,
\end{aligned}
\end{equation}
and similarly for the anti-holomorphic sector.

\section{GHY term in string frame}
\label{sec:GHY in string frame}

In this appendix, we are going to derive the GHY boundary term in string frame,
that needs to be added to the action (\ref{bulk gravity action}).
To derive it, we first re-write the action (\ref{bulk gravity action})
in Einstein frame, and identify the boundary terms necessary to allow for a
well-defined variational principal. We will identify two contributing boundary terms:
the usual GHY action in Einstein frame, and the extra term necessary to offset the boundary
contributions due to the dilaton total derivative term. Finally, we will combine these two
terms and re-write them in terms of metric in string frame.

 Expressing the Ricci scalar $\tilde R$ in the Einstein frame in terms of its counterpart $R$ in the string
 frame for the general transformation $ {\cal G}_{\mu\nu} = e^{2\Omega}\,\tilde{\cal G}_{\mu\nu}$
 we obtain in $D$-dimensional space-time
 \begin{equation}
  R = e^{-2\Omega} \, \left(\tilde R - 2(D-1)\nabla^2\Omega - (D-2)(D-1)(\p\Omega)^2\right)\,.
 \end{equation}
 Fixing $D = 10$, $\Omega = (\Phi - \Phi_0)/4$, we obtain
 \begin{equation}
\label{bulk gravity action Einstein frame}
S = \frac{1}{2\kappa^2}\,\int d^{6}x\,\tilde V_{\mathbb{T}^4}\,\sqrt{-\det\, \tilde {\cal G}}\,\left(\tilde R-\frac{9}{2}\,(\p\Phi)^2
-\frac{9}{2}\,\tilde \nabla^2\Phi-\frac{1}{12}\,e^{\Phi-\Phi_0}\,
{\cal H}_{\mu\nu\lambda}
{\cal H}^{\mu\nu\lambda}\right)\,,
\end{equation}
where now contraction of indices is done with the Einstein metric tensor, and 
 tilde on top of the covariant derivative indicates that it is calculated
in Einstein frame.
Here we can extract the pure boundary term, at $\chi = \Lambda$, using the Stokes's theorem,
 \begin{equation}
\label{boundary gravity action Einstein frame}
\begin{aligned}
S\supset -\frac{9}{4\kappa^2}\,\int d^{5}x\,\sqrt{-\det \tilde {\gamma}}\,
\tilde n_M\,\p^M\Phi\, |_{\chi = \Lambda}\,,
\end{aligned}
\end{equation}
where $\tilde \gamma_{\mu\nu}$ is the boundary metric in Einstein frame, and
\begin{equation}
\label{def of euclidean normal n}
\tilde n_M = \left(0,\;\sqrt{{\tilde {\cal G}}_{\chi\chi}},\;0,\;0,\;0,\;0,\;0,\;0,\;0,\;0\right)
\end{equation}
is an outward-pointing vector normal to the boundary.
The gravity action in string frame is to be supplemented with the GHY boundary term,  
\begin{equation}
S_{\textrm{GHY}} = \frac{1}{\kappa^2}\,\int d^{5}x\,\tilde V_{\mathbb{T}^4}\,\sqrt{-\det \tilde {\gamma}}\,
\tilde\nabla_M\tilde n^M\,,
\end{equation}
The GHY term in Einstein frame needs to be supplemented
with an extra contribution, to compensate for (\ref{boundary gravity action Einstein frame}).
In total, we obtain the boundary term in Einstein frame 
\begin{equation}
\label{tot Einstein bt}
S_b =  \frac{1}{\kappa^2}\,\int d^{5}x\,\sqrt{-\det\tilde \gamma}\,\tilde V_{\mathbb{T}^4}\,\left(\tilde \nabla_M \tilde n^M 
+\frac{9}{4}\tilde n^\chi\,\partial_\chi\Phi\right)\,|_{\chi = \Lambda}\,.
\end{equation}
On the other hand, in string frame we the outward-point unit normal to the boundary
\begin{equation}
\label{def of normal n}
n^M = \left(0,\;\sqrt{ {\cal G}^{\chi\chi}},\;0,\;0,\;0,\;0,\;0,\;0,\;0,\;0\right) = e^\frac{\Phi_0-\Phi}{4}\,\tilde n^M\,.
\end{equation}
Using the relation between Christoffel symbols,
\begin{equation}
\Gamma^M_{M\chi} = \tilde \Gamma^M_{M\chi} + \frac{5}{2}\,\partial_\chi \Phi\,,
\end{equation}
we obtain
\begin{equation}
\nabla_M n^M =  e^\frac{\Phi_0-\Phi}{4}\,\left(\tilde\nabla_M\tilde n^M+\frac{9}{4}\,\tilde n^\chi\partial_\chi\Phi\right)
\end{equation}
Introducing the boundary metric in string frame, $\gamma_{\mu\nu} = e^{\frac{\Phi-\Phi_0}{2}}\,\tilde\gamma_{\mu\nu}$,
and  combining everything together, we can re-write the total boundary term (\ref{tot Einstein bt}) as
\begin{equation}
\label{GHY in string frame}
S_b=\frac{1}{\kappa_0^2}\,\int d^{5}x\,e^{-2\Phi}\,\sqrt{-\det\gamma}\,V_{\mathbb{T}^4}\,\nabla_M n^M \,.
\end{equation}
The action (\ref{GHY in string frame}) therefore gives the GHY term in string frame.
In the main text we will denote (\ref{GHY in string frame}) as $S_{\textrm{GHY}}$,
having in mind that it is the full boundary term in the string frame.

\section{Thermodynamics of $SL(2,\mathbb{R})\times U(1) / U(1)$}
\label{sec: sl2r u1 over u1 thermodynamics}

In this appendix, we are going to demonstrate that our new results 
for  thermodynamics of the
coset geometry (\ref{big coset}) reproduce the known results for the
\begin{equation}
\label{un-gauged su2 coset}
\frac{SL(2,\mathbb{R})\times U(1) }{ U(1)} \times SU(2)\times \mathbb{T}^4
\end{equation}
background, that corresponds to the limit of un-gauged  $SU(2)$ sub-group.\footnote{See \cite{Chakraborty:2020swe,Apolo:2019zai} for discussion of related thermodynamic systems with microscopic description of NS5-F1 system and single-trace $T\bar{T}$ deformation.}
According to (\ref{U1 def}), this limit can be obtained by setting $a_2 = b_2 = 0$.
The anomaly-free conditions (\ref{time-like gauging}) then impose
two constraints $a_1^2+a_3^2 = 1$, $b_1^2+b_3^2 = 1$ on the four 
parameters of the model, $a_{1,3}$, $b_{1,3}$.
The remaining two independent parameters can be expressed in terms of the angles $\psi\in [0,\pi/2]$, $\chi\in [0,\pi/2]$,
as follows \cite{Giveon:2005mi}
\begin{equation}
\label{un-gauged su2 choice of parameters}
a_1 = \cos(\chi-\psi)\,,\quad b_1 = \cos\chi\,,\quad
a_3 = \sin(\chi-\psi)\,,\quad b_3 = \sin\chi\,.
\end{equation} 
The same parameters $\psi$, $\chi$ are also used in the conventions
of \cite{Chakraborty:2020yka}. In the particular case of \cite{Giveon:2003ge,Goykhman:2013oja,Giveon:2005jv}
one sets $\chi = 0$.\footnote{In conventions of \cite{Giveon:2005jv} one also sets $\alpha\rightarrow \psi$.}

We begin by writing down expression for the temperature (\ref{general T result})
for the choice of parameters (\ref{un-gauged su2 choice of parameters}),
\begin{equation}
\label{ungauged su2 T}
T = \frac{1}{\pi\sqrt{k}\,\ell_s}\frac{\cos\chi\cos(\chi-\psi)}{\cos\chi+\cos(\chi-\psi)}\,.
\end{equation}
In particular, setting $\chi = 0$, (\ref{ungauged su2 T}) is seen to agree with the
known result for the coset (\ref{un-gauged su2 coset}), that can be found, e.g., in eq. (2.26) of \cite{Giveon:2005jv}.

The entropy of (\ref{un-gauged su2 coset}) can be found in eq. (3.31) of \cite{Giveon:2005mi}\footnote{
The extra factor of $1/\sqrt{2}$ relative to that equation is necessary once we relax the convention of $\alpha ' =\ell_s^2 = 2$.}
\begin{equation}
\label{ungauged su2 S}
{\cal S} = \pi\,\sqrt{k}\,\ell_s\,\left(\sqrt{M^2 - Q_L^2}+\sqrt{M^2-Q_R^2}\right)\,,
\end{equation}
where $M$ is the ADM mass, and the $U(1)_{L,R}$ charges are given by (C.16) of \cite{Giveon:2005mi},\footnote{
In particular, in case of $\chi = 0$, we recover (2.30) of \cite{Giveon:2005jv}, with $Q_L=Q$, $Q_R=0$. In the latter, we 
substitute the mass $M$ and charge $Q$ of the black hole, determined by eqs. (34), (36) of \cite{Giveon:2003ge}.
Specifically, one plugs $M = me^{-2\phi_0}$, $Q=qe^{-2\phi_0}$ into ${\cal S} = \pi\,\sqrt{k}\,\ell_s(M+\sqrt{M^2-Q^2})$,
where $m=\frac{1}{2}\frac{1+p^2}{1-p^2}$, $q=\frac{p}{1-p^2}$, $p=\tan(\psi/2)$, and $\phi_0 = \Phi_0
+\frac{1}{2}\log\frac{1+p^2}{2(1-p^2)}$, expressed in terms of asymptotic value of the flat space dilaton $\Phi_0=
\log(g_s)$
via (29), (35) of \cite{Giveon:2003ge}.}
\begin{equation}
\label{ungauged su2 QLR}
Q_L = M\sin(\psi-\chi)\,,\qquad Q_R = M\sin\chi\,.
\end{equation}
Combining (\ref{ungauged su2 T}), (\ref{ungauged su2 S}), (\ref{ungauged su2 QLR}), we then obtain
\begin{equation}
T\,{\cal S} = M\cos\chi\cos(\chi-\psi)\,.
\end{equation}
This agrees with (\ref{BH entropy result}) for the choice of parameters 
(\ref{un-gauged su2 choice of parameters}), provided $M =  \frac{4\pi^2}{\kappa^2}\,
k^\frac{3}{2}\ell_s^2\,V_{\mathbb{S}^1}\,V_{\mathbb{T}^4}$.
The latter indeed matches  our result for the energy, (\ref{energy gen result}).

The free energy is  determined by 
\begin{equation}
{\cal F} = M - T{\cal S}-\mu_LQ_L-\mu_RQ_R\,,
\end{equation}
where $\mu_{L,R}$ are the left- and right-moving components of the chemical potential.
As pointed out in \cite{Giveon:2005jv}, the free energy of the theory (\ref{un-gauged su2 coset})
vanishes, which is in agreement with our result (\ref{vanishing grand potential}) for (\ref{big coset}) 
obtained independently of the choice of the anomaly-free gauging parameters $a_i$, $b_i$.
We can in fact match expressions for the free energy term by term.
We have already matched $M$ and $T\,{\cal S}$ with our results in the limit of
un-gauged $SU(2)$.
Now let us match $\mu_LQ_L+\mu_RQ_R$.
The chemical potential can be found from the asymptotic behavior of the gauge field components\footnote{
Again, the  factor of $1/\sqrt{2}$ appears when we relax the convention of $\alpha ' =\ell_s^2 = 2\rightarrow \ell_s=\sqrt{2}$
used in \cite{Giveon:2005mi}. This in turn rescales the time coordinate measured in units of $\ell_s$ by the factor 
of $1/\sqrt{2}$.}
\begin{equation}
A_t^{G,B} =\pm \sqrt{2} \mu^{G,B}+\dots\,,
\end{equation}
where the gauge fields can be found in eq. (C.9) of \cite{Giveon:2005mi}.
Using eq. (C.9) of \cite{Giveon:2005mi}, we
first determine the relation $y = \frac{t\,\sqrt{2}}{\sec\chi+\sec(\chi-\psi)}$
between the time coordinate $y$, used in that equation, and the time
coordinate $t$, featuring asymptotic behavior of the metric $g_{tt} = -1$.
We then obtain $\mu_G = \sec\left(\chi-\frac{\psi}{2}\right)\sin\left(\frac{\psi}{2}\right)$,
$\mu_B = \cos\chi\tan\left(\frac{\psi}{2}\right)-\sin\chi$.
Consequently
\begin{equation}
\begin{aligned}
\mu_L &= \frac{\mu_G+\mu_B}{2} = -\frac{1}{2} \cos (\chi ) \sec \left(\frac{\psi }{2}\right) \sin (\chi -\psi ) \sec \left(\chi -\frac{\psi }{2}\right)\,,\\
\mu_R &= \frac{\mu_G-\mu_B}{2} =
\frac{1}{2} \sin (\chi ) \sec \left(\frac{\psi }{2}\right) \cos (\chi -\psi ) \sec \left(\chi -\frac{\psi }{2}\right)\,.
\end{aligned}
\end{equation}
We then obtain
\begin{equation}
\label{mu Q for un-unaged su2}
\mu_LQ_L+\mu_RQ_R = \frac{M}{2}\,(2-\cos(2\chi-\psi)-\cos\psi)\,.
\end{equation}
At the same time, from (\ref{OmegaJ gen result}), using (\ref{energy gen result}), (\ref{un-gauged su2 choice of parameters})
we obtain
\begin{equation}
 \Omega\cdot {\cal J} = {\cal E}(1-\cos(\chi-\psi) \cos\chi)\,,
\end{equation}
that indeed agrees with (\ref{mu Q for un-unaged su2}) for ${\cal E} = M$.
This completes the term-by-term match of our result for the free energy
for the choice of parameters (\ref{un-gauged su2 choice of parameters}) with the known result
for the  (\ref{un-gauged su2 coset}).

\end{document}